%% file: main.tex
\documentclass[aps,prl,twocolumn,superscriptaddress]{revtex4-2}
\usepackage{blindtext}
\usepackage{cancel}
\usepackage{mathtools,mathalfa,amssymb,amsmath,amsfonts,amsthm, bm}
\usepackage{graphicx}
\usepackage{multirow}
\usepackage[table, dvipsnames]{xcolor}
\usepackage{color}
\definecolor{292}{rgb}{0.3843, 0.6588, 0.8980}
\usepackage[unicode, colorlinks=true, 
linkcolor=blue, 
urlcolor=blue, 
citecolor=292]{hyperref}
\usepackage{soul}
\usepackage{orcidlink}
\usepackage{float}

\newcommand{\re}{\mathrm{Re}}
\newcommand{\im}{\mathrm{Im}}
\newcommand\numberthis{\addtocounter{equation}{1}\tag{\theequation}}

\usepackage{appendix}
\usepackage{float}
\usepackage{subfigure}
\usepackage{xcolor}
\usepackage{tabularray}
\usepackage{multirow}

\colorlet{green}{teal}
\graphicspath{{./images/Linear_Plots/}{/images/}{/images/new/}}

\begin{document}
	%\title{Nonlinearities at Black Hole Horizons}
	\title{Nonlinear ringdown at the black hole horizon}
    \author{Neev Khera \orcidlink{0000-0003-3515-2859}}
    \email{nkhera@uoguelph.ca }
    \affiliation{University of Guelph, Guelph, Ontario N1G 2W1, Canada}
	
    \author{Ariadna Ribes Metidieri \, \orcidlink{0009-0001-4084-2259}}
	\email{ariadna.ribesmetidieri@ru.nl}
	\affiliation{Institute for Mathematics, Astrophysics and Particle Physics, Radboud University, Heyendaalseweg 135, 6525 AJ Nijmegen, The Netherlands}

	\author{B\'eatrice Bonga \, \orcidlink{0000-0002-5808-9517}}
	\email{bbonga@science.ru.nl}
	\affiliation{Institute for Mathematics, Astrophysics and Particle Physics, Radboud University, Heyendaalseweg 135, 6525 AJ Nijmegen, The Netherlands}

    \author{Xisco Jim\'enez Forteza \, \orcidlink{0000-0002-8158-5009}}
	\email{frjifo@aei.mpg.de}
	\affiliation{Albert-Einstein-Institut, Max-Planck-Institut f{\"u}r Gravitationsphysik, Callinstra{\ss}e 38, 30167 Hannover, Germany}
	\affiliation{Leibniz Universit{\"a}t Hannover, 30167 Hannover, Germany}
 \affiliation{Nikhef, Science Park 105, 1098 XG Amsterdam, The Netherlands}
\affiliation{Institute for Gravitational and Subatomic Physics (GRASP),
Utrecht University, Princetonplein 1, 3584 CC Utrecht, The Netherlands}

    \author{Badri Krishnan\, \orcidlink{0000-0003-3015-234X}}
	\email{badri.krishnan@ru.nl}
	\affiliation{Institute for Mathematics, Astrophysics and Particle Physics, Radboud University, Heyendaalseweg 135, 6525 AJ Nijmegen, The Netherlands}
	\affiliation{Albert-Einstein-Institut, Max-Planck-Institut f{\"u}r Gravitationsphysik, Callinstra{\ss}e 38, 30167 Hannover, Germany}
	\affiliation{Leibniz Universit{\"a}t Hannover, 30167 Hannover, Germany}
    \author{Eric Poisson \, \orcidlink{0000-0001-7679-1265}}
    \email{epoisson@uoguelph.ca}
    \affiliation{University of Guelph, Guelph, Ontario N1G 2W1, Canada}
	
    \author{Daniel Pook-Kolb \, \orcidlink{0000-0001-7317-1087}}
	\email{daniel.pook.kolb@aei.mpg.de}
	\affiliation{Institute for Mathematics, Astrophysics and Particle Physics, Radboud University, Heyendaalseweg 135, 6525 AJ Nijmegen, The Netherlands}
	\affiliation{Albert-Einstein-Institut, Max-Planck-Institut f{\"u}r Gravitationsphysik, Callinstra{\ss}e 38, 30167 Hannover, Germany}
	\affiliation{Leibniz Universit{\"a}t Hannover, 30167 Hannover, Germany}

    \author{Erik Schnetter\,\orcidlink{0000-0002-4518-9017}}
    %\email{eschnetter@perimeterinstitute.ca}
    \affiliation{Perimeter Institute for Theoretical Physics, Ontario, N2L 2Y5, Canada}
    \affiliation{Department of Physics and Astronomy, University of Waterloo, Ontario, Canada}
    \affiliation{Center for Computation \& Technology, Louisiana State University, Baton Rouge, LA, USA}
	
    \author{Huan Yang\, \orcidlink{0000-0002-9965-3030}}
    \email{hyang@perimeterinstitute.ca}
    \affiliation{Perimeter Institute for Theoretical Physics, Ontario, N2L 2Y5, Canada}
\affiliation{University of Guelph, Guelph, Ontario N1G 2W1, Canada}
	   
\date{2023-06-19}

	\begin{abstract}
 The gravitational waves emitted by a perturbed black hole ringing down are well described by damped sinusoids, whose frequencies are those of quasinormal modes. Typically, first-order black hole perturbation theory is used to calculate these frequencies. Recently, it was shown that second-order effects are necessary in binary black hole merger simulations to model the gravitational-wave signal observed by a distant observer. Here, we show that the horizon of a newly formed black hole after the head-on collision of two black holes also shows evidence of non-linear modes. 
 %Specifically, we identify quadratic modes for $l=2,4,6$ in simulations with varying mass ratio and boost parameter. 
 Specifically, we identify one quadratic mode for the $l=2$ shear data, and two quadratic ones for the $l=4,6$ data in simulations with varying mass ratio and boost parameter.
 The quadratic mode amplitudes display a quadratic relationship with the amplitudes of the linear modes that generate them.

\end{abstract}
	
	\maketitle

\section{Introduction}

Gravitational wave observations are increasing rapidly and with them the science we can extract from these observations. 
Some examples are the statistical inference of the mass distribution of stellar mass black holes in our universe (see e.g. \cite{LIGOScientific:2016vpg,KAGRA:2021duu,Libanore:2023ovr,Nitz:2021zwj})
, lessons on the formation of heavy elements in the merger of binary neutron stars \cite{Diehl:2022jnq} and tests of strong-field gravity and black holes (see e.g. \cite{Berti:2018vdi,LIGOScientific:2021sio,LIGOScientific:2019fpa,Kastha:2021chr,Isi:2019aib,Isi:2020tac,Forteza:2022tgq}). 
For the latter, black hole spectroscopy is a valuable tool \cite{Berti:2016lat,Yang:2017zxs,Berti:2018vdi,Ma:2023cwe,Detweiler:1980gk,Dreyer:2003bv,Berti:2006cc}. This method relies on the fact that after the merger of two black holes, the newly formed object settles down to a new stationary black hole by emitting gravitational waves with a discrete set of complex frequencies called quasinormal modes (QNMs). These QNMs depend only on the two parameters describing black holes: their mass $M$ and spin $J$. If more than one QNM can be observed, one can test for consistency of these modes (as has been done in~\cite{Nee:2023osy,Baibhav:2023clw}). 
Black hole spectroscopy requires the observation of multiple QNMs, which in turn depends not only on the strength of the gravitational wave signal but also on our ability to model these modes accurately. Linear perturbation theory on a Kerr spacetime can be used to calculate the linear frequencies of the QNMs \cite{Teukolsky:1973ha,Leaver:1985ax,Berti:2009kk,Yang:2012he,Yang:2013uba}, analyze gravitational wave observations \cite{Chandrasekhar:1975zza,LIGOScientific:2016lio}, and make forecasts for the detectability of QNMs \cite{Capano:2021etf,Dreyer:2003bv,LIGOScientific:2021sio}.
Studies of numerical waveforms have shown the importance of various effects of the linear modes, including higher overtones, mirror modes, mode mixing, and the influence of the Bondi-Metzner-Sachs frames~\cite{Giesler_2019, Dhani:2020nik, Li:2021wgz, MaganaZertuche:2021syq}.

However, nonlinearities are naturally expected in the ringdown stage~\cite{Campanelli:1998jv,Yang:2014tla,Yang:2015jja,Ripley:2020xby,Sberna:2021eui,Redondo-Yuste:2022czg}. In particular, it has been shown that modes with a frequency expected from perturbation theory at quadratic order fit the ringdown phase better than higher overtones in the linear theory \cite{Cheung:2022rbm,Mitman:2022qdl,Baibhav:2023clw,Nee:2023osy} (see also pioneering work in \cite{London:2014cma}).
This is an important result both conceptually, as general relativity is a non-linear theory after all, and practically as these quadratic
QNMs may be detectable in observations and thereby improve our strong-field tests of general relativity and black holes. 

The source emitting gravitational radiation---in the case of QNMs, the time-dependent merger object that settles down to a Kerr black hole---emits waves that go out to infinity and fall into the horizon. The waves at infinity are the ones we observe and interpret as QNMs, but numerical simulations have indicated that the shear modes at the horizon are also accurately described by a superposition of modes with frequencies matching those of the signal at infinity \cite{Mourier:2020mwa}. Given that the horizon is in the strong field regime, one would naturally expect that the signal at the horizon should also show evidence of non-linearities. In particular, one would expect the shear modes to be better fitted by a model that takes the next-to-leading order QNM frequencies into account than a model based on frequencies derived solely from linear perturbation theory. We investigated this for simulations of head-on collisions of two black holes. %We find evidence for the presence of quadratic tones in the shear modes $l=2,4$ and 6. 
We find evidence for the presence of a single quadratic tone in the shear mode $l=2$, and of two quadratic tones in the shear mode $l=4$ and 6.

Hence, non-linearities are present both at infinity and near the horizon. This invalidates the idea that non-linearities may be hidden behind the horizon or are even absent~ \cite{Cheung:2022rbm,Giesler_2019,Prasad_2020,okounkova2020revisiting,jaramillo2022airyfunction,Chen_2022}. 

\section{Set up}
We study the head-on collision of non-spinning black holes with different boost parameters and mass ratios, as summarized in Tab.~\ref{tab:simulations-summary}. In particular, we investigate two sets of simulations. The first set (S1-S4) describes black holes initially at rest with Brill-Lindquist's bare masses $m_1$ and $m_2$~\cite{BrillLindquist}. The second set (S5-S9) are equal-mass black holes with Bowen-York initial data~\cite{BowenYork}, in which both black holes have equal and opposite Bowen-York momentum parameter of magnitude $P$ expressed in units of $M_\circ=m_1+m_2$, the total bare mass~ \footnote{ 
For initial data corresponding to a large separation of the black holes, this parameter can be interpreted as the individual momenta of the black holes. However, these simulations have an initial separation of $0.4 M_\circ$, so this interpretation is not applicable. Nonetheless, increasing $P$ corresponds to increasing the coordinate velocities of the black holes.}.

The boosted simulations have larger linear amplitudes than the unboosted ones, typically by a factor of ten. Consequently, the quadratic amplitudes are also larger in these boosted simulations and it is easier to confidently establish their presence for larger $l$ modes.

 \begin{table}[h]
     \centering
     \begin{tabular}{c||c c c c c c c c c}
     \hline
     Simulation & S1 & S2 & S3 & S4 &S5 & S6 & S7 & S8 & S9\\
     \hline
     \hline
     $\mu = m_1/m_2$ & 1 & 1.6 & 2 & 3 & 1 &1 & 1 &1 &1 \\
    $P$ & \; 0 \; & \; 0 \; & \; 0 \; & \; 0 \; & 0.90& 1.20 & 1.52 & 1.80 & 2.10\\
    \hline
     \end{tabular}
     \caption{Mass ratio $\mu$ and momentum parameter $P$ of nine simulations of the  head-on collision of two Schwarzschild black holes.}
     \label{tab:simulations-summary}
 \end{table}

We track the evolution of the outermost marginally outer trapped surface (MOTS), which traces out a 2+1-dimensional world-tube $\mathcal{H}$ and is the dynamical black hole horizon of the newly formed black hole at $t=0$ ($t=1.06M$) for the boosted (unboosted) simulations. Initially, this surface is highly distorted and dynamic, but it quickly settles down to a nearly spherical MOTS as the black hole approaches a Schwarzschild solution. 
We define the onset of the ringdown as the time during which the change in the area of the MOTS becomes oscillatory and below $1\%$.  
In practice, we take $t_{\rm rd}= 8.2 M$, where $M$ is the mass of the remnant, such that all simulations have reached this ringdown regime. This makes it easier to compare the simulations.  Having access to the evolution of the horizon area allows us to avoid the contamination in our data from the merger phase, a common issue in this type of analysis, discussed in depth in
~\cite{Nee:2023osy,Baibhav:2023clw}. 
We follow the horizon evolution until $t_\text{f}\approx 40 M$ for the unboosted simulations and $t_\text{f}\approx 32 M$ for the boosted ones.

The shear of the outward null normal $\ell$ to the MOTS is a measure of the gravitational waves going into the horizon \cite{HawkingHartle72,Prasad_2020}, but also a geometric quantity measuring the deformation of the horizon surface. In the following, we focus uniquely on the shear, although a complementary study using the mass multipole moments shows qualitatively similar results.  
We fix a unique $\ell$ via $\ell^a \nabla_a t = 1$, where
$\nabla_a$ is the spacetime covariant derivative
and $t$ is the coordinate time of the simulation.
The shear $\sigma$ of $\ell$ is then calculated similarly to~\cite{Mourier:2020mwa}, but we also multiply by the remnant mass to make it dimensionless. 

During the ringdown, we can decompose $\sigma$ at the horizon 
as a sum of damped sinusoids~\cite{Mourier:2020mwa}, namely,
\begin{align*}\label{eq:model}
\sigma(t,\theta, \varphi)& =\sum_{l \geq s,  m,  n,\pm} {\mathcal{A}}_{lmn}  e^{- i \; \omega_{lmn}^\pm (t-t_{\rm rd}) + i\phi_{lmn}}  \,_{2}Y_{lm}(\theta,\varphi)\\
&\equiv \sum_{l \geq s,  m} \sigma_{lm} \,_{2}Y_{lm}(\theta,\varphi) \numberthis 
\end{align*}
where the $(l,m)$ indices describe the angular decomposition of the
modes (with $m = -l, \dots, l$), ${\mathcal{A}}_{lmn}$ are the constant (dimensionless) amplitudes, $\phi_{lmn}$ the phases, ${}_{2\!} Y_{lm}$  the spin-weighted
spherical harmonics with spin-weight $s=2$ and $\omega_{lmn}^{\pm}$ the complex frequencies corresponding to the co-rotating $\re [\omega_{lmn}^+]>0$ and counter-rotating $\re [ \omega_{lmn}^-]<0$ modes. The $n = 0, 1, 2, \dots$ denote the $n$-tone excitation of a given $(l, m)$ mode, with $n = 0$ being the fundamental tone and
$n \ge 1$ correspond to overtones.
Due to the rotational symmetry of the head-on collision, the shear is fully described by the $m=0$ modes and $\omega_{lmn}^+=|\omega_{lmn}^-|$, so we work with $\omega_{ln} = \omega_{l0n}^+=-\omega_{l0n}^-$. Hence, we set $m=0$ and drop the $m$ subindex in both the shear modes $\sigma_{l0}=\sigma_{l}$   and the complex frequencies. Additionally, all odd $l$ modes vanish for the boosted simulations since the mass-ratio is one.  
Further, by the symmetries of the problem, $\sigma_l$ is a real-valued function, so the positive and negative frequencies combine to provide a manifestly real expansion of the shear modes 
\begin{align}
    \label{eq:model-real}
    \sigma_l &= \sum_{n=0}^{n_\text{max}} \left(C_{ln} e^{-\im [\omega_{ln}](t-t_{\rm rd})} \cos\left[\re [\omega_{ln}](t-t_{\rm rd})\right] \right. \nonumber\\ &\phantom{=} \left. + S_{ln} e^{-\im [\omega_{ln}](t-t_{\rm rd})} \sin\left[\re [\omega_{ln}](t-t_{\rm rd})\right] \right)\,.
\end{align} Here $C_{ln}$ and $S_{ln}$ are real amplitudes with $2 \mathcal{A}_{ln}= \sqrt{C_{ln}^2 + S_{ln}^2}$,  and $\tan\phi_{ln}=-S_{ln}/C_{ln}$ \cite{Mourier:2020mwa}.

Using black hole perturbation theory, one can calculate the values of the QNM frequencies rather straightforwardly to linear order in the metric perturbations (see the efficient open software routine in \cite{Stein:2019mop}). The next order in perturbation theory is rather involved (see \cite{Brizuela:2006ne,Brizuela:2007zza,Brizuela:2009qd,spiers2023secondorder,Ioka:2007ak,Nakano:2007cj}) 
 but for each pair of linear QNM frequencies $\omega_{ln}$ and $\omega_{l'n'}$, we expect a corresponding quadratic QNM frequency $\omega_{ln\times l' n'}= \omega_{ln}+\omega_{l'n'}$. We find evidence of quadratic frequencies in the shear modes $l=2,4$ and 6, which we show in Tab.~\ref{tab:Quadratic-frequencies}.  Our analysis is inconclusive for the $l=3$ mode in the unboosted simulations and the $l=8$ in the boosted ones due to the signal's weak amplitude. 

Here, we only present the detailed analysis for the $l=2$ shear modes of the boosted simulation S7 with a spatial discretization, $\Delta x = M_\circ/\mathrm{res}$, where $M_\circ$ is the total bare mass and `$\mathrm{res}$' refers to the resolution of the grid spacing. The results presented here for $S7$ use $\mathrm{res}=120$. We also briefly discuss the results for the unboosted simulation S2 with $\mathrm{res}=312$.
 The details for the remaining simulations in Tab.~\ref{tab:simulations-summary} are not discussed explicitly, since they are completely analogous to the ones presented here. The results for higher $l$ modes can be found in the Supplementary Material.   

\section{Mismatch and stability}
\label{sec:stability}
When fitting the data, several combinations of linear and quadratic tones are possible.  To minimize the risk of overfitting as discussed in~\cite{Baibhav:2023clw,Nee:2023osy,Forteza:2021wfq}, we consider a model with the lowest possible number of tones for which the quadratic mode is resolved~\footnote{We follow a similar 'bootstrap' strategy as in~\cite{Nee:2023osy}, where we first confirm the presence of the fundamental tone at late times, and the first fundamental tone, before searching for higher tones.}. For the boosted simulations, a model with three tones suffices ($n_\text{max}=2$ in Eq.~\eqref{eq:model-real}), while for the unboosted ones, we need a model with at least four tones ($n_\text{max}=3$). We analyze the boosted and unboosted simulations independently, so we only compare models with the same number of modes (and thus the same number of free parameters). Using that the QNM model~\eqref{eq:model-real} is linear in $C_{\ell n}$ and $S_{\ell n}$, we use a linear least square fitting algorithm to minimize the $L^2$ norm of the residual. A nonlinear fitting algorithm, such as the one used in~\cite{Mourier:2020mwa}, yields completely analogous results.

\begin{figure}
    \centering
\includegraphics[width=0.48\textwidth]{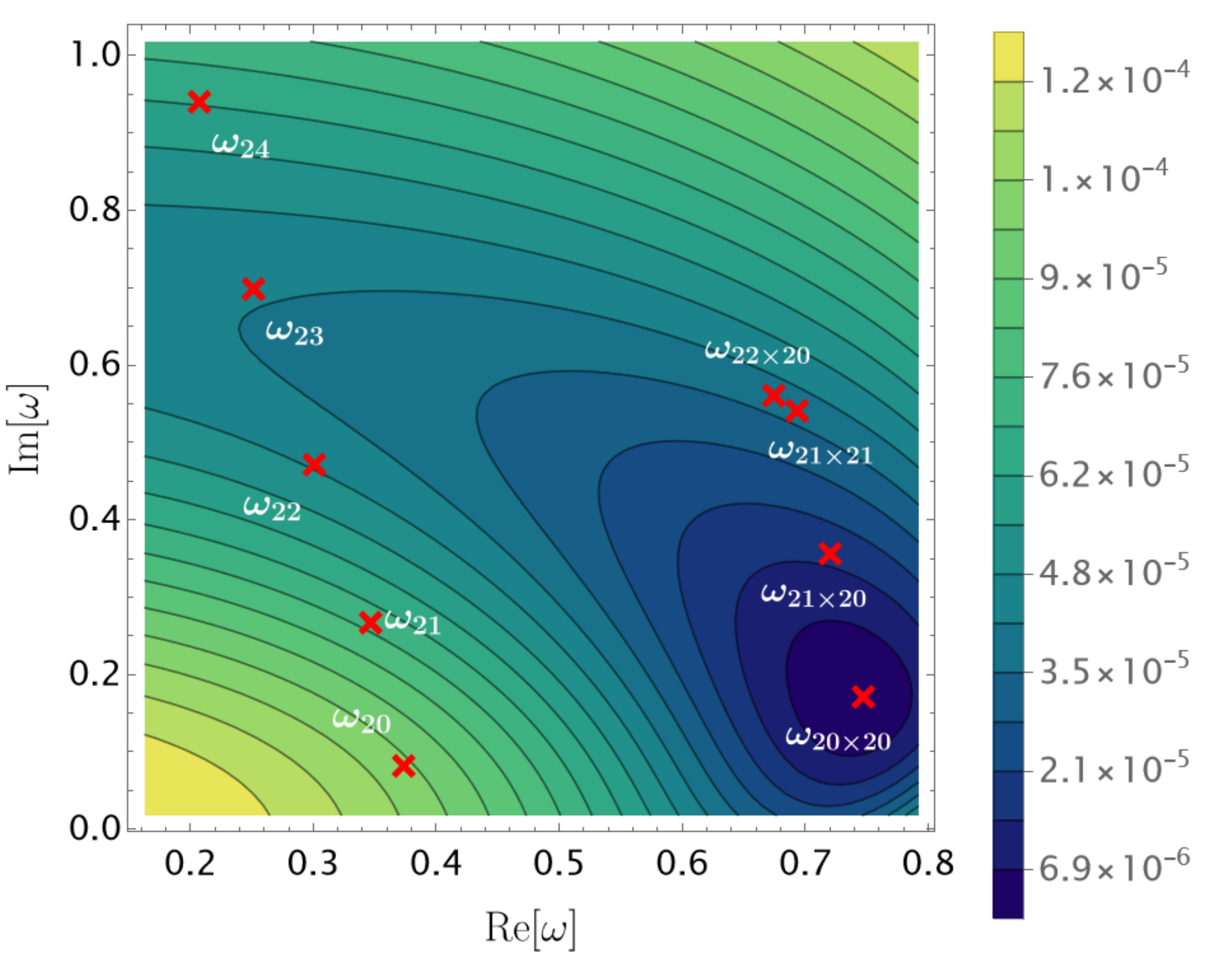}
    \caption{Mismatch between the  $l=2$ data ($t\in [t_\text{rd},t_f]$) of the boosted simulation S7 starting at time $t_{\rm rd}$ and a model for $\sigma_2$ with three tones, in which two frequencies are fixed to the GR predictions $\omega_{20}$ and $\omega_{21}$, and the third one is varied.
  }
    \label{fig:l2n2Mismatch}
\end{figure}

We first explore which quadratic modes could be present in our ringdown dataset $t\in [t_{\rm rd},t_{\rm f}]$ by scanning over the complex frequency of the last overtone in the model~\eqref{eq:model-real}. We compute the mismatch (see~\cite{Mourier:2020mwa} for its definition) for each possible frequency of the last overtone. The model with the smallest mismatch is considered the best model. Fig.~\ref{fig:l2n2Mismatch} shows that the quadratic frequencies $\omega_{20\times 20}$ and $\omega_{21\times 20}$ are favored over the linear overtone $\omega_{22}$. The unboosted simulations show the same trend. 

\begin{figure*}
    \centering
  \subfigure[Mismatch of the three models with 
 the third tone's frequency $\omega_{22}$, $\omega_{20\times 20}$ or $\omega_{21\times 20}$.]{       
            \includegraphics[width=0.33\textwidth, height=0.22\textwidth]{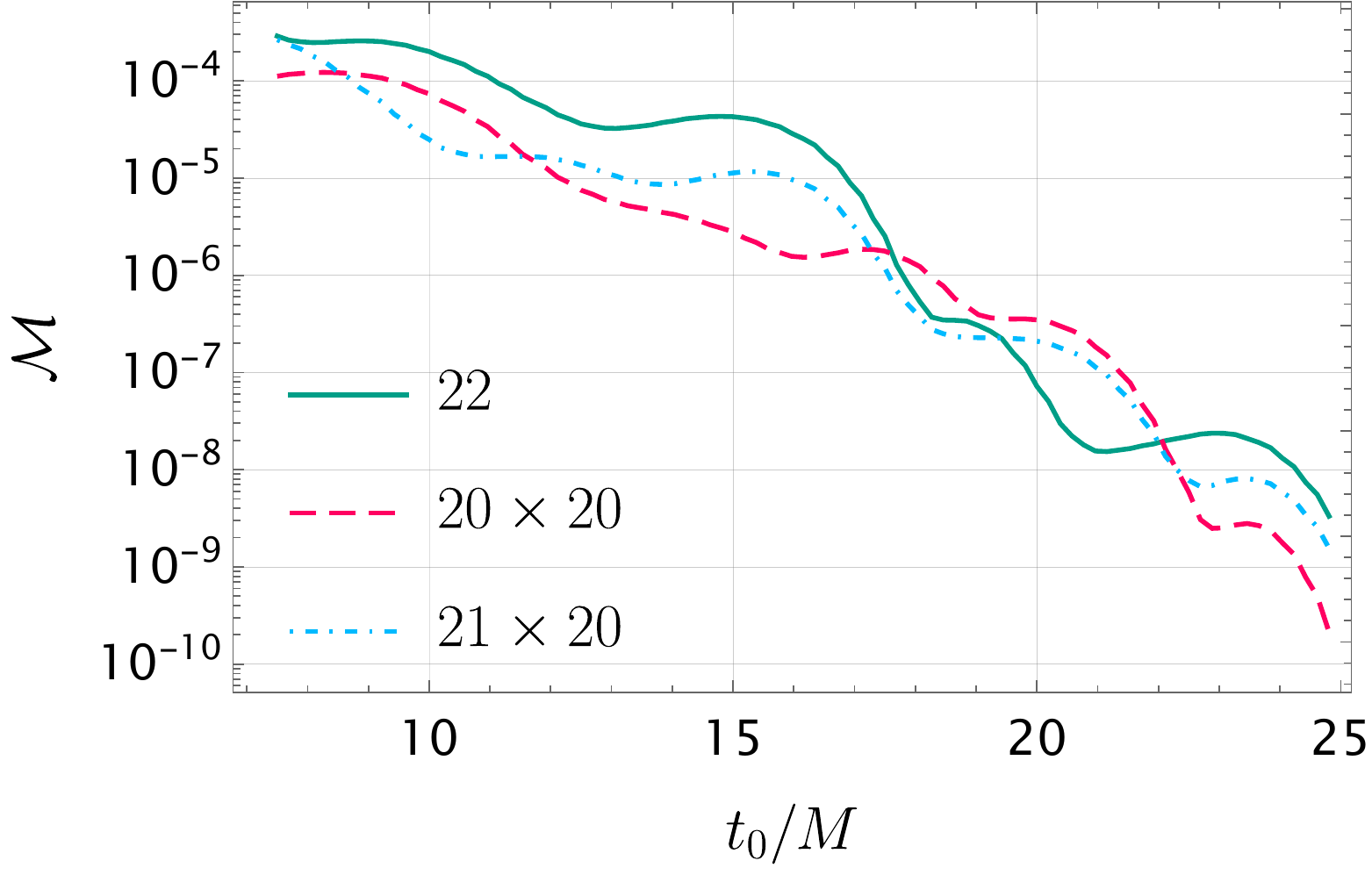}
               \label{fig:Mismatchl2n2-boost}
        }~\subfigure[Amplitudes of the three QNMs for the linear (no superindex) and quadratic models.]{       
            \includegraphics[width=0.33\textwidth, height=0.25\textwidth]{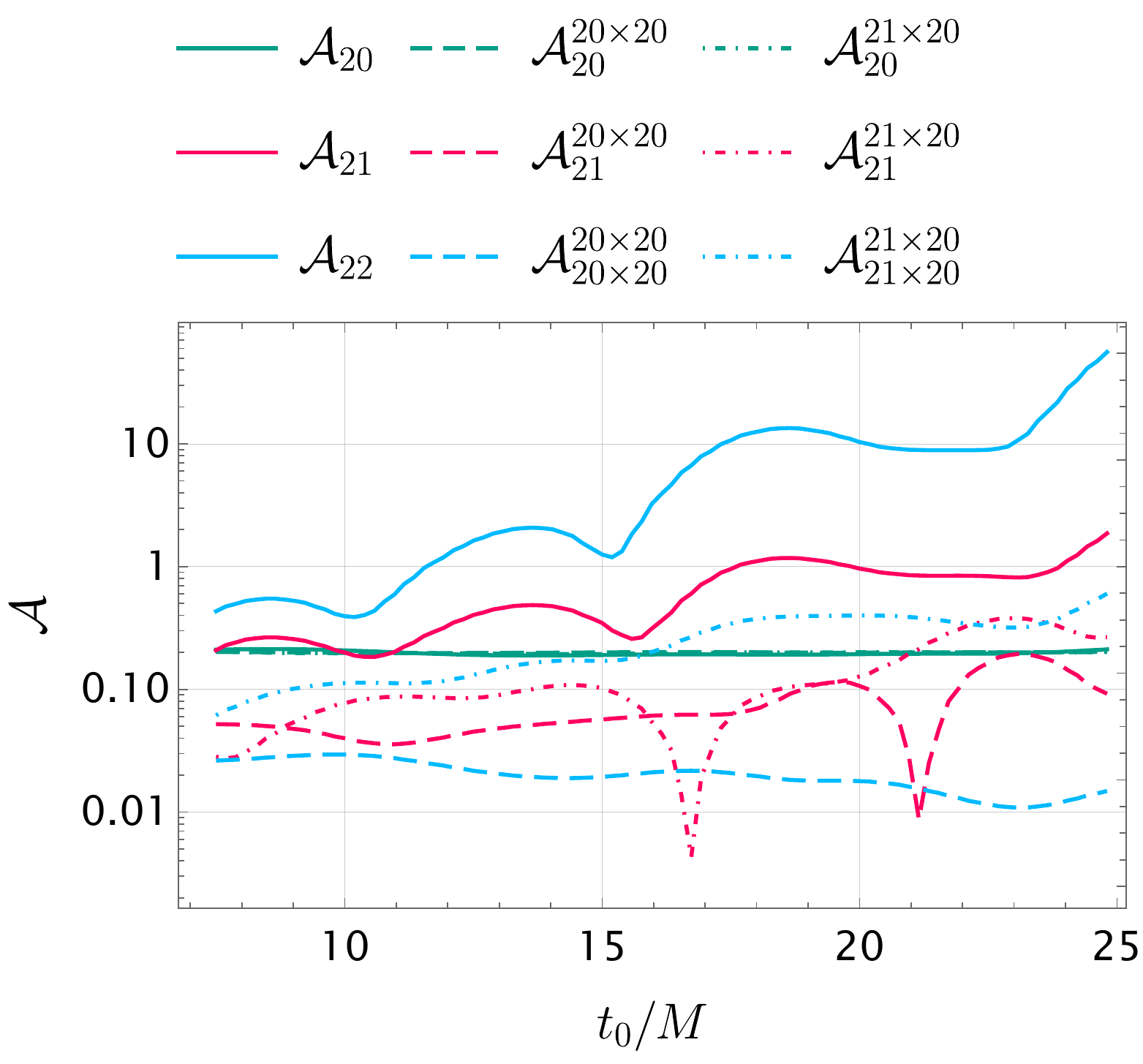}
             \label{fig:Ampl2n2-S7}
        }~\subfigure[Relative variation of the optimal frequency with respect to $\omega_{22}$, $\omega_{20\times 20}$ and $\omega_{21\times 20}$.]{     
    \includegraphics[width=0.33\textwidth, height=0.22\textwidth]{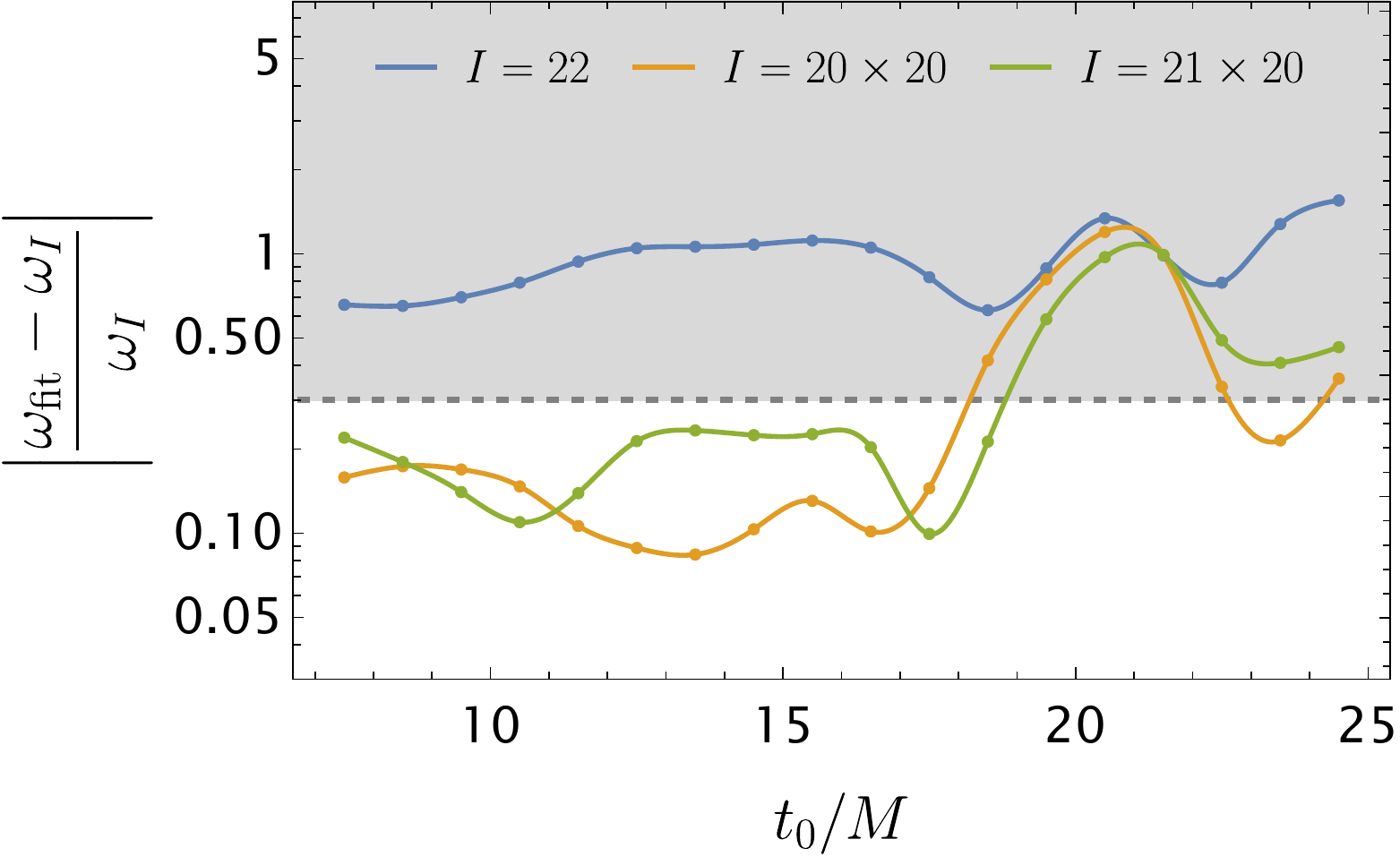}
             \label{fig:free-frequency-l2n2}
        }
   \label{fig:Qualityl4}
    \caption{
    Mismatch, amplitudes, and best-fitted frequency between the $l=2$ data of the boosted simulation S7 and a model for $\sigma_2$ with three tones as a function of the starting time $t_0\in[t_{\rm rd},25 M]$. }
    \end{figure*}

 We then assess the presence of these quadratic tones by ensuring that they persist when narrowing our dataset. Specifically, we compute the mismatch for the different models while varying the starting time of the dataset to later times $t_0\in [t_{\rm rd}, 25M]$.  Fig.~\ref{fig:Mismatchl2n2-boost} shows that the models containing the quadratic frequencies $\omega_{20\times 20}$ and $\omega_{21\times 20}$ have lower mismatch at earlier times, when these modes are expected to be resolvable. In Fig.~\ref{fig:Ampl2n2-S7}, we track the stability of the fit in this process, i.e., the evolution of the mode's amplitudes with the starting time $t_0$. 
 The model including the quadratic frequency $\omega_{20\times 20}$ is the most stable, with a maximum relative variation at early times $t\in[t_{\rm{rd}}, 15M]$ of $\sim 2\%$ for the fundamental tone's amplitude and $\sim 30\%$ for the amplitudes of the first overtone and the quadratic tone.  As already noticed in~\cite{Mourier:2020mwa}, the model with two linear overtones has amplitudes varying over several orders of magnitude and is therefore unstable: the maximum relative variation of the fundamental tone's amplitude is $\sim 10\%$ while for the first and second overtones, it is $\sim 60\%$ and $\sim 80\%$ respectively. 

We finally consider the model~\eqref{eq:model-real} not only with free amplitudes and phases but also with \textit{the frequency of the last overtone free}.
We then implement an algorithm of mismatch minimization to find the frequency for which the fit over the dataset $t\in[t_0,t_\text{f}]$ is optimal. In other words, we effectively track the frequency in  Fig.~\ref{fig:l2n2Mismatch} for which the mismatch is minimal as we vary the starting time. Fig.~\ref{fig:free-frequency-l2n2} shows the relative variation of the optimal frequency $\delta\omega= (\omega_{\rm fit}-\omega_{\rm known})/\omega_{\rm known}$ with respect to known possible frequencies (with $\omega_{\rm known}=\omega_{ln\times l' n'}\,,\omega_{ln}$). The advantage of this procedure is that it sets an absolute lower bound to the mismatch by finding the optimal numerical frequency, and consequently, it enables us to discard possible tones in our model. In fact, Fig.~\ref{fig:free-frequency-l2n2} shows that a linear model is not favored, not even when the quadratic modes have already decayed since the deviation of the linear overtone with respect to the optimal frequency remains above 50\% at all starting times.  Further, Fig.~\ref{fig:free-frequency-l2n2} also  shows that both quadratic frequencies $\omega_{20\times 20}$ and $\omega_{21\times 20}$ have a minimum deviation with respect to the optimal frequency of about 7\% and only surpass the 30\% deviation once the $l=2$ shear mode can be accurately described by the fundamental tone (around $t =20 M$). This deviation is consistent with the criteria used in Fig.~1 in~\cite{Cheung:2022rbm}, and therefore the quadratic frequencies $\omega_{20\times 20}$ and $\omega_{21\times 20}$ are good candidates to be in our model. The amplitude relation detailed in the next section confirms the presence of the quadratic tone $\omega_{20 \times 20}$.

\section{Amplitude relations}
\label{sec:Amplitude-relations}

\begin{figure*}
    \centering
  \subfigure[Boosted simulations with constant mass ratio $\mu=1$.]{       
            \includegraphics[width=0.45\textwidth,height=0.3\textwidth]{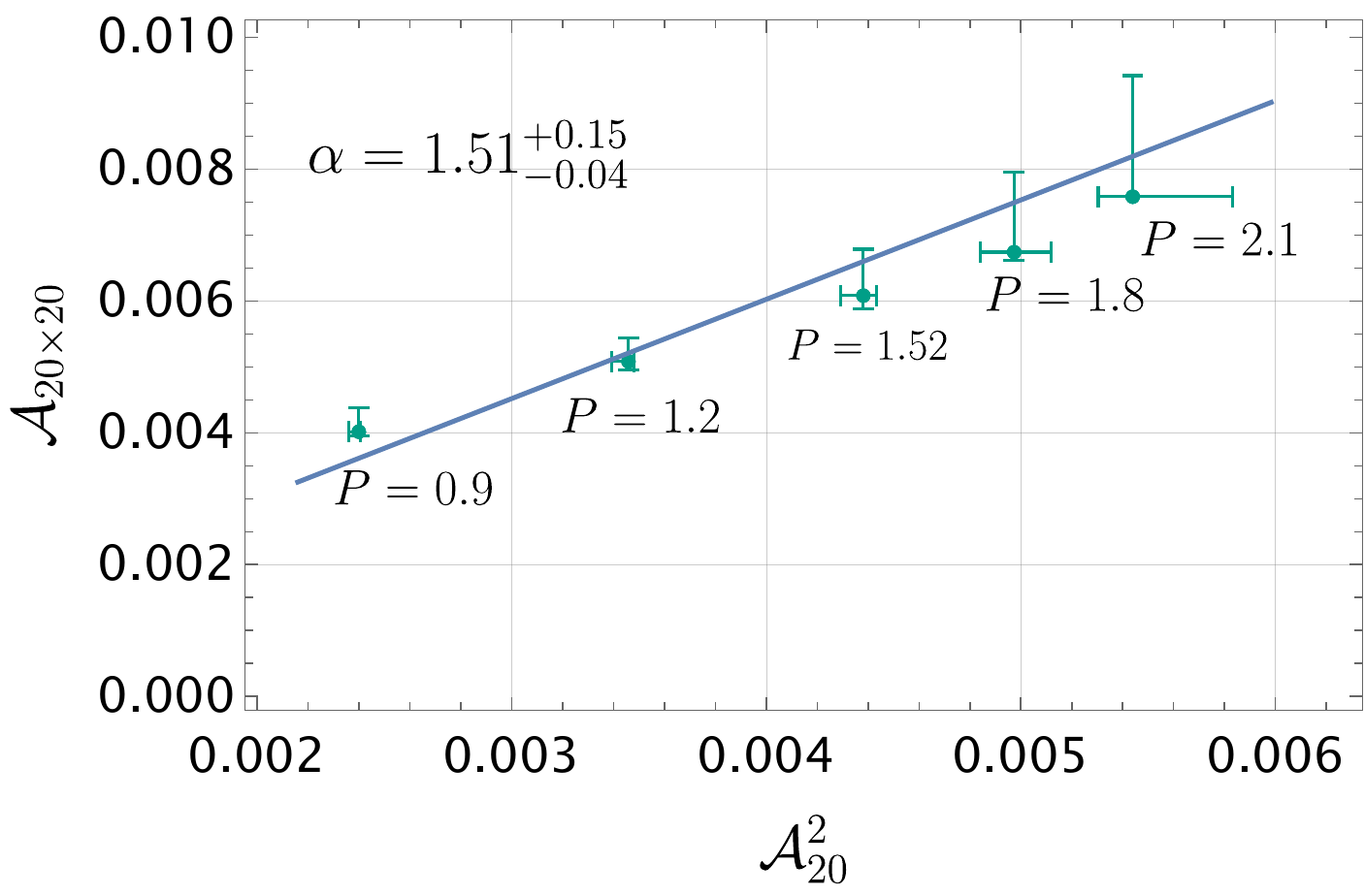}
               \label{fig:Ampl-gamma-l2n2t20x20}
        }~\subfigure[Unboosted simulations with different mass ratios (and momentum parameter $P=0$).]{       
            \includegraphics[width=0.45\textwidth,height=0.3\textwidth]{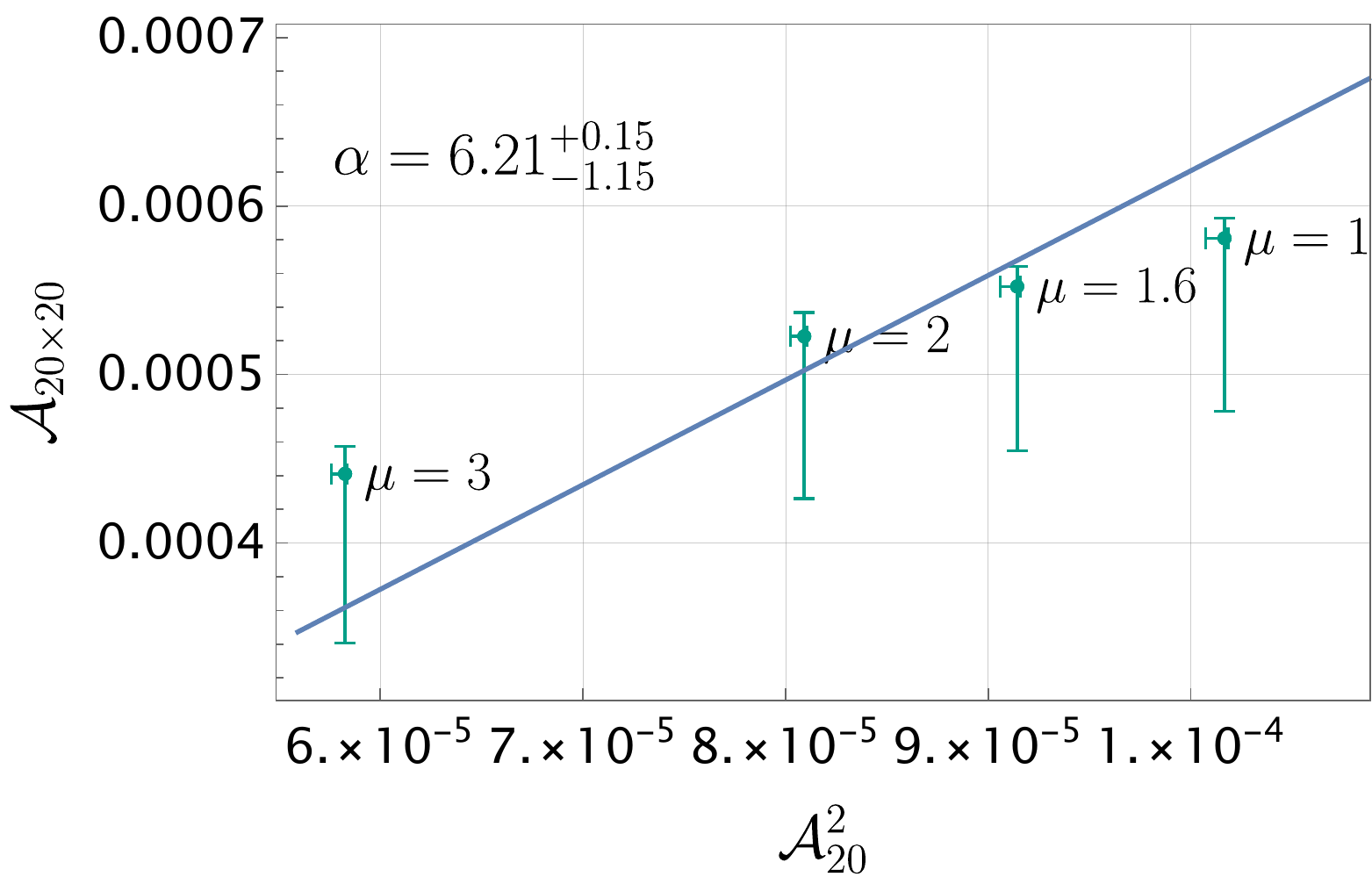}
             \label{fig:Ampmul2n3}
         }
    \caption{Amplitude relation for the quadratic tone $\omega_{20\times 20}$ in the shear mode $l=2$ when considering both the boosted and unboosted simulations. In both cases the dataset initial time $t_0$ is fixed to $t_0=15 M$. }
     \label{fig:Ampl2}
    \end{figure*}

The amplitudes of the quadratic modes are related to the ones of the linear modes through 
 \begin{equation}\label{eq:amplitude-rel}
     \mathcal{A}_{ln\times l'n'}=\alpha \mathcal{A}_{ln} \mathcal{A}_{l'n'}\,,
 \end{equation} 
 where $\alpha$ is the slope of the line passing through the origin. If a quadratic mode is present in our data, then we should be able to confirm its presence by fitting Eq.~\eqref{eq:amplitude-rel} across different simulations. Otherwise, the presence of the quadratic mode could be a consequence of overfitting or an artifact of the mode mixing.  We analyze the non-boosted (S1-S4) and boosted (S5-S9) simulations independently. In this part of the analysis, we used lower resolution simulations, as these were accurate enough 
 (in particular, we used a resolution of $120$ for the boosted and at least $90$ for the unboosted simulations).

 Fig.~\ref{fig:Ampl2} shows the amplitude relation for the $l=2$ shear mode. In both sets of simulations, such a relation is found at late times $t_0\geq 15 M$ within the uncertainty bars~\footnote{We evaluate the amplitudes on an interval of $4M$ centered around the time of the fit $t_0$. The range of amplitudes in the interval is the uncertainty bars in Fig.~\ref{fig:Ampl2}.}. At earlier times, the presence of higher overtones blurs this relationship. This analysis confirms the presence of the quadratic frequency $\omega_{20\times 20}$ in the $l=2$ shear mode. The slope in Eq.~\eqref{eq:amplitude-rel} is reported in Tab.~\ref{tab:Quadratic-frequencies}.  
  
Given that the remnant black hole in both sets of simulations is Schwarzschild, one might have expected the slopes in the two sets of simulations to be comparable.  We see instead that the slopes in the two sets of simulations are inconsistent.  This could be because the error bars do not capture systematic modeling errors, for instance, due to the finite separation of the holes in our initial data, inherent to this type of numerical simulations. Additionally, initial conditions are more important than one may naively think. In \cite{Redondo-Yuste:2022czg}, the authors showed that the slope of the amplitude relation when solving the second-order Teukolsky equation is three times smaller than the slope obtained from the non-linear numerical simulations in~\cite{Cheung:2022rbm} and~\cite{Mitman:2022qdl}. A detailed study of such systematic errors, including spin effects, will be presented elsewhere.  
 
 In Tab.~\ref{tab:Quadratic-frequencies}, we collect the slopes for the quadratic relation Eq.~\eqref{eq:amplitude-rel} of the quadratic tones that we found for the $l=2, 4$, and $6$ shear modes. For the $l=2$ shear mode, we only include the quadratic tone $\omega_{20\times 20}$, since the amplitude relation for the quadratic frequency $\omega_{21\times 20}$ is not satisfied for the set of boosted simulations within the uncertainty bars (see Fig.~7 in the Supplementary Material), and therefore we conclude this mode is not present in the $l=2$ shear mode given our resolution. For the $l=4$ shear mode we find that a model with the fundamental tone and two quadratic tones, the $\omega_{20\times 20}$ and $\omega_{20\times 40}$, is the most favored. Both quadratic tones satisfy the amplitude relation over several boosted simulations, and their slopes have been therefore included in Tab.~\ref{tab:Quadratic-frequencies}. Finally, for the $l=6$ shear mode, we also find that a model with the fundamental mode and two quadratic ones is the most suitable to fit the shear data. In this case, models including the quadratic frequencies $\omega_{20\times 40}$ and $\omega_{20\times 60}$ or $\omega_{20\times 40}$ and $\omega_{40\times 40}$ are both possible, and the small difference in the fit residuals between the two prevents us from opting for one or the other. The figures and the full discussion for the $l=4$ and $6$ shear modes can be found in the Supplementary Material. We would like to highlight that we find the same combination of relevant modes as Cheung et al.~\cite{Cheung:2022rbm} for the $l=4$ mode, which provides us with new and novel evidence of arising correlations between the horizon dynamics and the gravitational wave radiation.

 \begin{table}[h]
     \centering
     \begin{tblr}{c|c|c|c}
     \hline
       Mode                    & $\omega_{ln\times l' n'}$     & Boosted ($\alpha$)        & Unboosted ($\alpha$) \\
       \hline
       \hline
        $l=2$                  & $\omega_{20\times 20}$        & $1.51_{-0.04}^{+0.15}$    &  $6.21^{+0.15}_{-1.15}$ \\
         \hline
         \SetCell[r=2]{c} $l=4$ & $\omega_{20\times 20}$        &   $0.73_{-0.33}^{+0.06}$  & - \\
                                &  $\omega_{20\times 40}$       &   $2.6_{-0.26}^{+0.26}$    & - \\ 
          \hline
          \SetCell[r=4]{c} $l=6$ ${}^*$          &  $\omega_{20\times 40}$        &  $1.78_{-0.74}^{0.53}$                     & -\\
                               & $\omega_{20\times 60}$        &  $2.52_{-0.59}^{+1.29}$    & -\\
         \cline{2-4}
                               &  $\omega_{20\times 40}$        &  $1.78_{-0.65}^{0.44}$                          & -\\
                               & $\omega_{40\times 40}$        &  $2.82_{-0.62}^{+1.5}$     &  -\\
        \hline
     \end{tblr}
     \caption{The slope in Eq.~\eqref{eq:amplitude-rel} for different quadratic modes present for different $l$ modes. For the boosted simulations, we detect the presence of the quadratic tones when using models with the fundamental mode, one overtone and one quadratic mode ($l=2$) or the fundamental tone and two quadratic modes ($l=4$ and $l=6$), while for the unboosted $l=2$ shear mode we need to include at least one additional linear overtone. We marked with an asterisk the results for the shear mode $l=6$ since the models with either the quadratic frequencies $\omega_{20\times 60}$ or $\omega_{40\times 40}$ are possible, as explained in the Supplementary Material. We present both possibilities separated by a dashed line. }
     \label{tab:Quadratic-frequencies}
     
 \end{table}

\section{Conclusion}
While black hole horizon simulations have existed for a while and one naturally expects the non-linear nature of general relativity to be important in this regime, this is the first demonstration of non-linear effects at the horizon. In particular, we have shown that the $l=2,4,6$ shear modes at the horizon---a strong field regime---soon after a head-on collision of two black holes are better fitted with a model that includes next-to-leading order effects in perturbation theory than a purely linear model. These quadratic modes are in agreement with those found by~\cite{Cheung:2022rbm} at infinity. Finding the presence of quadratic modes was subtle, it required (1) high-accuracy numerical data, and/or (2) a signal with large linear amplitudes so that the corresponding quadratic amplitudes are also large (as is the case for a boosted signal). 

The excitement of observing electromagnetic signals often stems from their origin in ``interesting'' objects, which allows us to gain insights into the emitter's properties. While the initial detection of gravitational waves was inherently thrilling, gravitational waves increasingly become a tool to investigate the sources emitting them.
Black holes are ideal sources to investigate with gravitational waves given their blackness. However, to maximize our understanding of black holes and their horizons, it is crucial to establish a clear connection between the gravitational wave observed at infinity and the horizon geometry. This work is a small step in that direction by showing that just as the wave at infinity, also the horizon geometry of black holes requires non-linear effects to accurately describe it. It is worth noting the intriguing possibility of a connection with the Kerr/CFT correspondence, as suggested in \cite{Kehagias:2023ctr}.

\section{Acknowledgement}
The authors are grateful {to Mark Ho-Yeuk Cheung, Thomas Helfer, Emanuele Berti, and Gregorio Carullo for useful discussions on the GRChombo head-on data, and to Gregorio Carullo for his useful comments on the manuscript}. N.K, E.P., and H.Y are supported by the Natural Science and Engineering Council of Canada. H.Y. is also supported by Perimeter Institute for Theoretical Physics.
Research at Perimeter Institute is supported in part by the Government of Canada through the Department of Innovation, Science and Economic Development Canada and by the Province of Ontario through the Ministry of Colleges and Universities.

\bibliographystyle{apsrev4-1}
\bibliography{nonlinearities.bib, einsteintoolkit.bib}
\include{supplementary_material}

\end{document}

%% file: supplementary_material.tex
\setcounter{page}{0}
\section{SUPPLEMENTARY MATERIAL}

\subsection{Numerical simulations}

As explained in the main text, we simulate the head-on collision of two (boosted) black holes.
We use the Einstein Toolkit \cite{Loffler:2011ay, EinsteinToolkit:2022_05, EinsteinToolkit:web} to simulate coalescing binary black holes while explicitly imposing axisymmetry on a uniform grid spanning the $x\ge0, y=0$ half-plane. We used the TwoPunctures thorn \cite{Ansorg:2004ds} to set up initial conditions, and used an axisymmetric variant of the McLachlan code \cite{Brown:2008sb} to evolve the Einstein equations. We ensured that the outer boundaries were causally disconnected from the observed horizons.

\begin{figure}[h]
    \centering
    \includegraphics[width=0.5\textwidth]{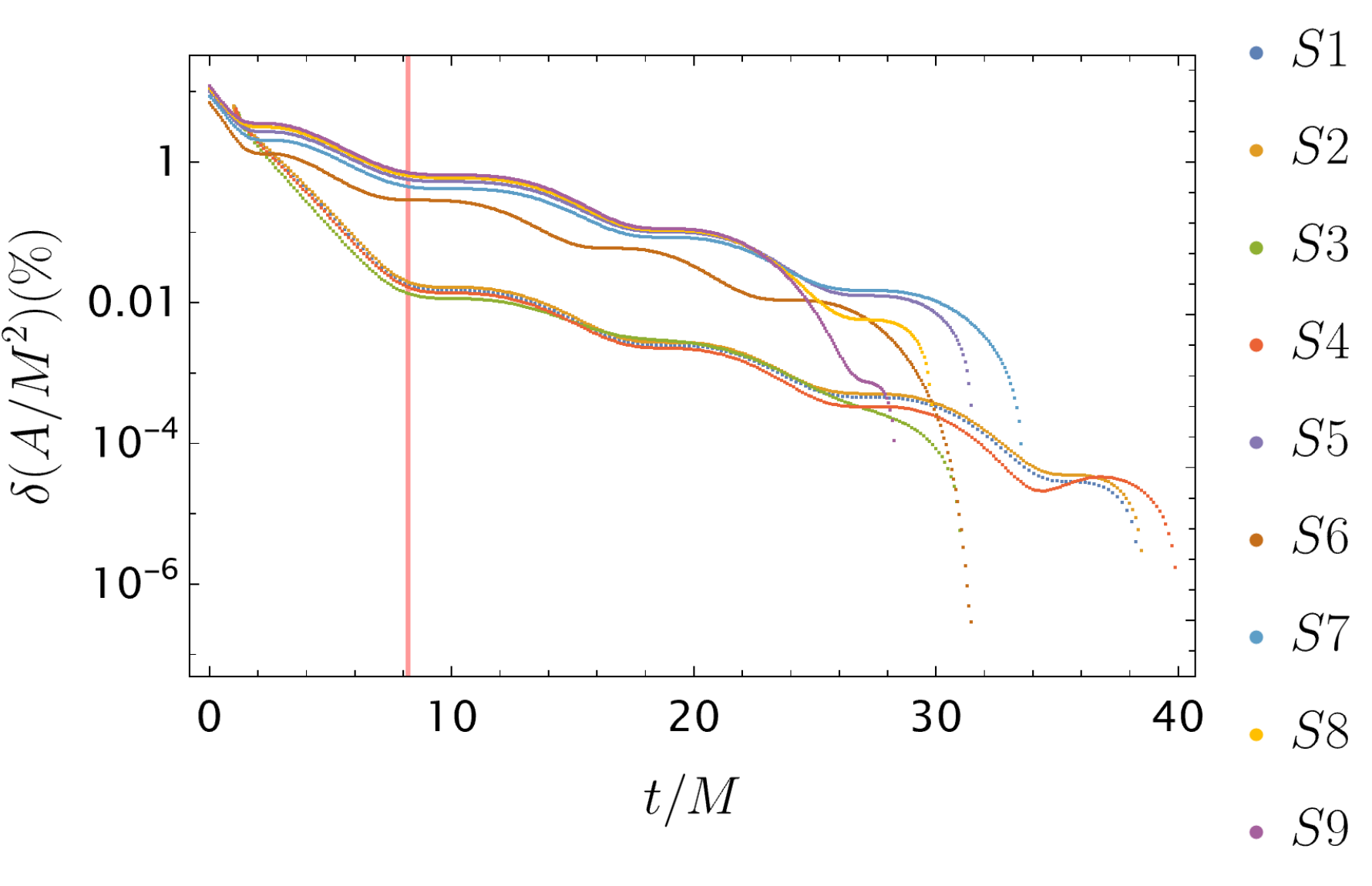}
    \caption{Relative variation of the area with respect to its stationary value at $t_\text{f}$ in \%. The total variation, from $t_\text{bifurcate}$ to $t_\text{f}$, amounts to $\sim$ 6\% for the unboosted simulations and $\sim$ 10\% for the boosted ones. Two regimes are distinguished. During the first, dynamic interval (until $t_{\rm rd} =8.2 M$, in red), approximately 5\% (9\% for the boosted simulations) of the total area variation occurs. 
    During the second regime, the variation of the area remains below 1\% and we observe an oscillatory behavior. The mass ratio and momentum parameter of the simulations S1-S9 are summarized in Tab.~\ref{tab:simulations-summary}. 
    }
    \label{fig:ringdown}
\end{figure}

At the start of the evolution, there are two disjoint marginally outer trapped surfaces (MOTS) $\mathcal{S}_1, \mathcal{S}_2$ representing the horizons of the two individual black holes. (The location of the MOTS is determined using tools from \cite{Pook-Kolb:2018igu}). 
As time evolves, $\mathcal{S}_1$ and $\mathcal{S}_2$ approach each other, touch, and go through each other~\cite{Gupta:2018znn}. But before the two MOTS touch, an additional common MOTS forms (at $t_{\rm bifurcate} = 1.06 M$ in the unboosted simulations and is already present at $t=0$ in the boosted ones), and immediately bifurcates into an inner and outer branch. The evolution of the outer MOTS traces out a 2+1-dimensional world-tube $\mathcal{H}$, which is the dynamical horizon of the newly formed black hole. Initially (until $\sim t=2.5M$ in the boosted simulations and  $\sim t=7.6M$ in the unboosted ones), this surface is highly distorted and dynamic, but it quickly settles down to a nearly spherical MOTS as the black hole asymptotically approaches a Schwarzschild black hole. The relative area increase with respect to its asymptotic value amounts to a 10.04\% for the boosted simulations and a 6.06\% in the unboosted ones, of which a 2.66\% and a 6.03\% is reached by the end of this first dynamical regime respectively (see Fig.~\ref{fig:ringdown}). In the second regime, the area derivative oscillates 
as the area reaches its asymptotic value.  This second regime would correspond to what qualitatively we understand as the ringdown. However, the relative area increase is still too high (of around a $\sim$ 7\%) in the case of the boosted simulations. Hence, we define the onset of the ringdown at a later time, as explained in the main text. We highlight this time in red in Fig.~\ref{fig:ringdown}.

Notice that by identifying the ringdown regime in Fig.~\ref{fig:ringdown}, we are assuming that a decomposition in QNMs exists (as it was already done in~\cite{Mourier:2020mwa}).  However, the equations that determine the frequencies of the QNMs rely on the separability of the wave-like equation for the metric perturbations on a Schwarzschild spacetime. This separability is guaranteed in the standard Schwarzschild coordinates, but if the simulation time coordinate is a function of the Schwarzschild time \emph{and} one or multiple of the other coordinates, then this separability no longer applies and one would therefore no longer expect the QNMs to be a good description~\footnote{If $t_{\rm simulation} = t_{\rm Sch} + f(r,\theta, \varphi)$, then the wave equation remains separable---we have in mind a more complicated relation such as $t_{\rm simulation} = f(t_{\rm Sch},r,\theta, \varphi)$.
}. However, this possibility is not likely given that the gauge conditions used in the numerical simulation are ``symmetry seeking'' in the sense that they attempt to find a timelike Killing vector if there is one, and therefore should find a coordinate that is related to the Schwarzschild time coordinate in a simple way that should not break the separability. In this sense, it is sensible to define the ringdown onset as we did in Fig.~\ref{fig:ringdown}.  

\subsection{Extra material for the shear mode $l=2$}

Here we complement the results in the main text with the figures for the unboosted simulation S2 and the amplitude relation of the quadratic mode $\omega_{21\times 20}$. The discussion is completely analogous to the one in the main text, so we shall be brief. 

\begin{figure}[ht!]
    \centering
\includegraphics[width=0.5\textwidth]{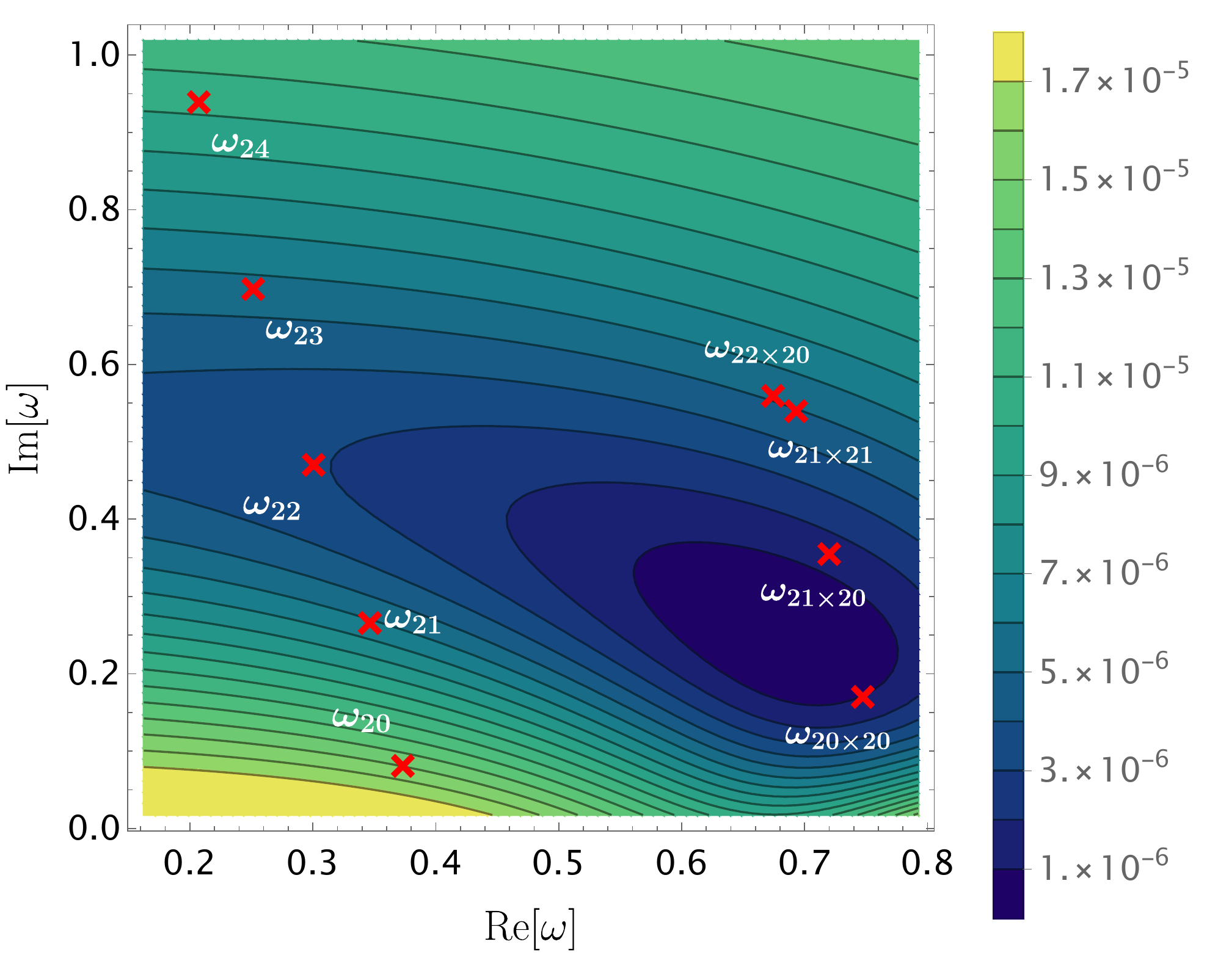}
    \caption{  
    Mismatch between the  $l=2$ data ($t\in [t_\text{rd},t_f]$) of the  unboosted simulation S2 and a model for $\sigma_2$ with four tones, in which three frequencies are fixed to the GR predictions $\omega_{20}$ and $\omega_{21}$, $\omega_{22}$, and the fourth one is varied. }
    \label{fig:Misml2n3S1}
\end{figure}

In Fig.~\ref{fig:Misml2n3S1}, we show the lowest mismatch contours for the frequency of the overtone $n_{\rm max}=3$ in the model~\eqref{eq:model-real}. As in Fig.~\ref{fig:l2n2Mismatch}, the quadratic frequencies $\omega_{20\times 20}$ and $\omega_{21\times 20}$ are favored over the linear frequency $\omega_{23}$.

\begin{figure*}
    \centering
  \subfigure[Mismatch of the three models with 
 the last tone's frequency $\omega_{23}$, $\omega_{20\times 20}$, and  $\omega_{21\times 20}$.]{       \includegraphics[width=0.33\textwidth, height=0.22\textwidth]{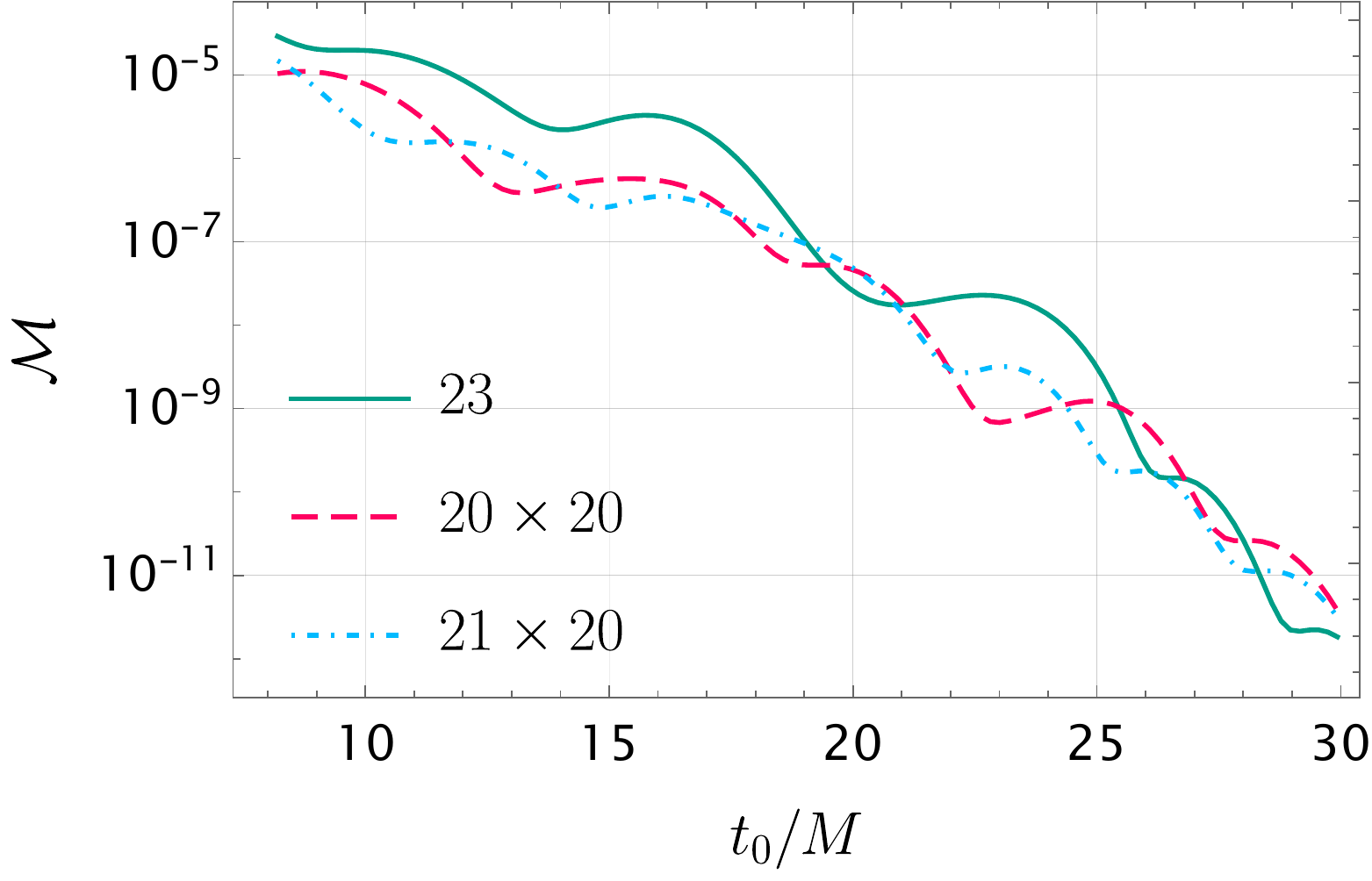}
               \label{fig:Mismatchl2n3-boost}
        }~
        \subfigure[Amplitudes for the linear (no superindex) and quadratic models with  $\omega_{20\times 20}$ versus $\omega_{21\times 20}$. ]{       
            \includegraphics[width=0.33\textwidth,height=0.25\textwidth]{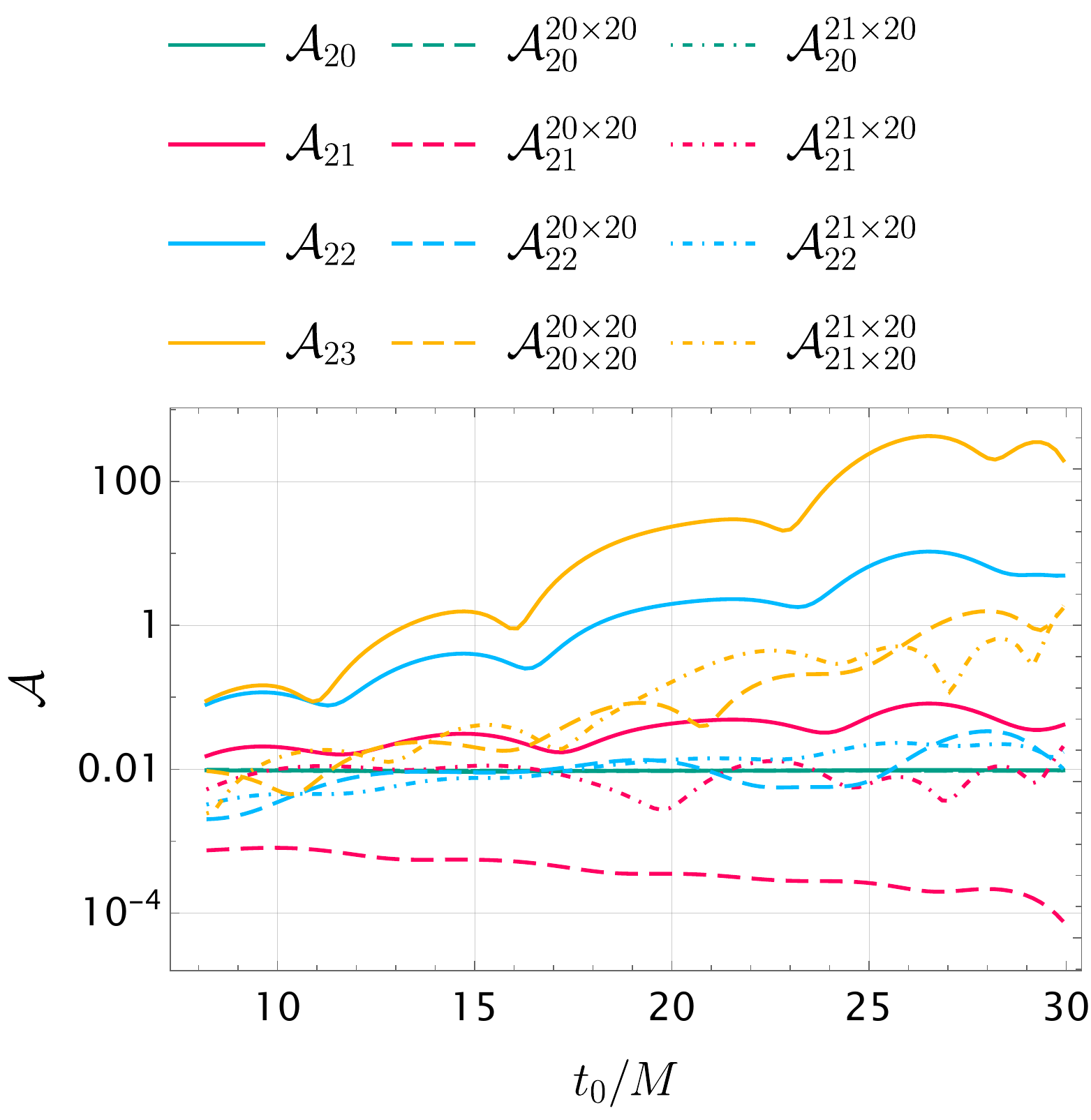}
             \label{fig:Ampl2n3}
        }~\subfigure[Relative variation of the optimal frequency with respect to $\omega_{23}$, $\omega_{20\times 20}$ and $\omega_{21\times 20}$.]{     
    \includegraphics[width=0.33\textwidth, height=0.22\textwidth]{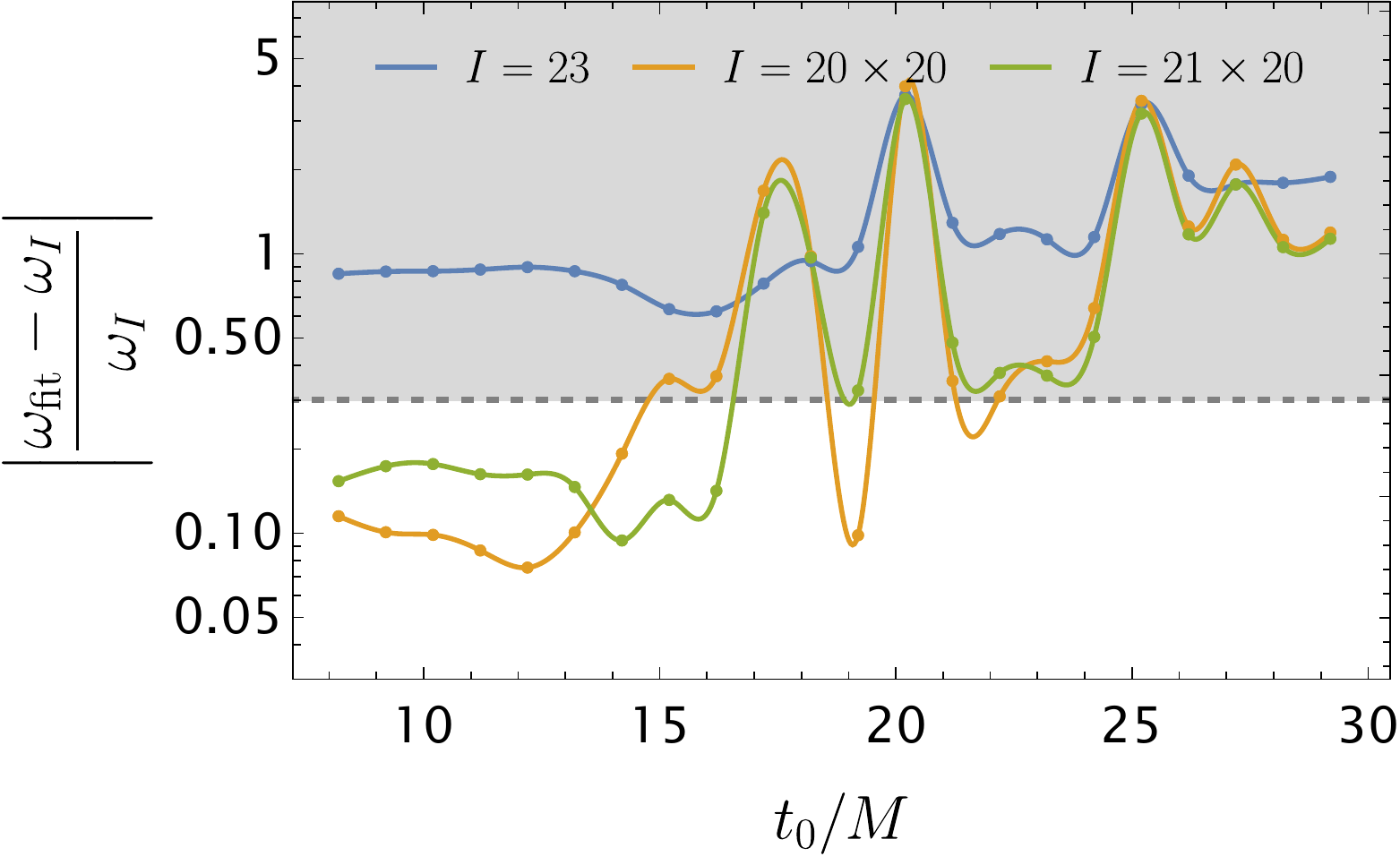}
             \label{fig:free-frequency-l2n3}
        }
   
    \caption{Mismatch and tone's amplitudes as a function of the starting time $t_0\in[t_{\rm rd},30 M]$ for the $l=2$ shear mode  in the unboosted simulation S2 with $n_{\rm max}=3$. The amplitude variation in the linear model is less than two orders of magnitude, while for the quadratic models the amplitudes vary less than one order of magnitude.}
    \label{fig:Qualityl2S1}
    \end{figure*}

We confirm the presence of the quadratic modes while varying the starting time of the fit in Fig.~\ref{fig:Qualityl2S1}, where we show the mismatch and amplitude stability of the three models.  Again, the quadratic models minimize the mismatch and provide a more stable fit than the linear model. Analogously to Fig.~\ref{fig:free-frequency-l2n2}, the optimal frequency is best approximated by the quadratic modes than the linear one, thus suggesting the presence of (at least one of) these modes in our data (see the discussion in the main text for the boosted simulations).

\begin{figure*}
    \centering
  \subfigure[Boosted simulations with constant mass ratio $\mu=1$.]{       
            \includegraphics[width=0.42\textwidth,height=0.25\textwidth]{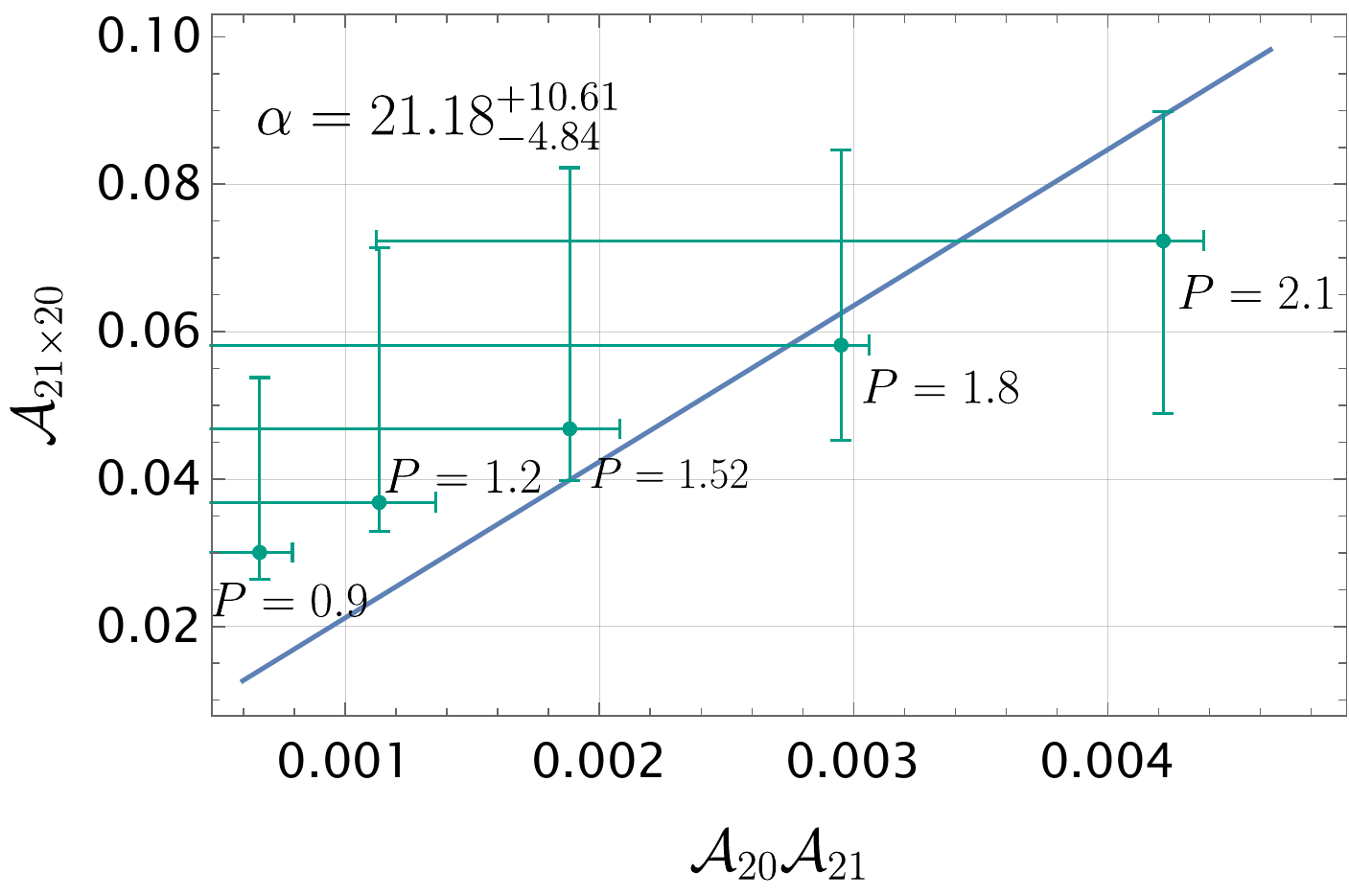}
               \label{fig:Ampl-gamma-l2n2t21x20}
        }~\subfigure[Unboosted simulations with different mass ratios.]{       
            \includegraphics[width=0.42\textwidth,height=0.25\textwidth]{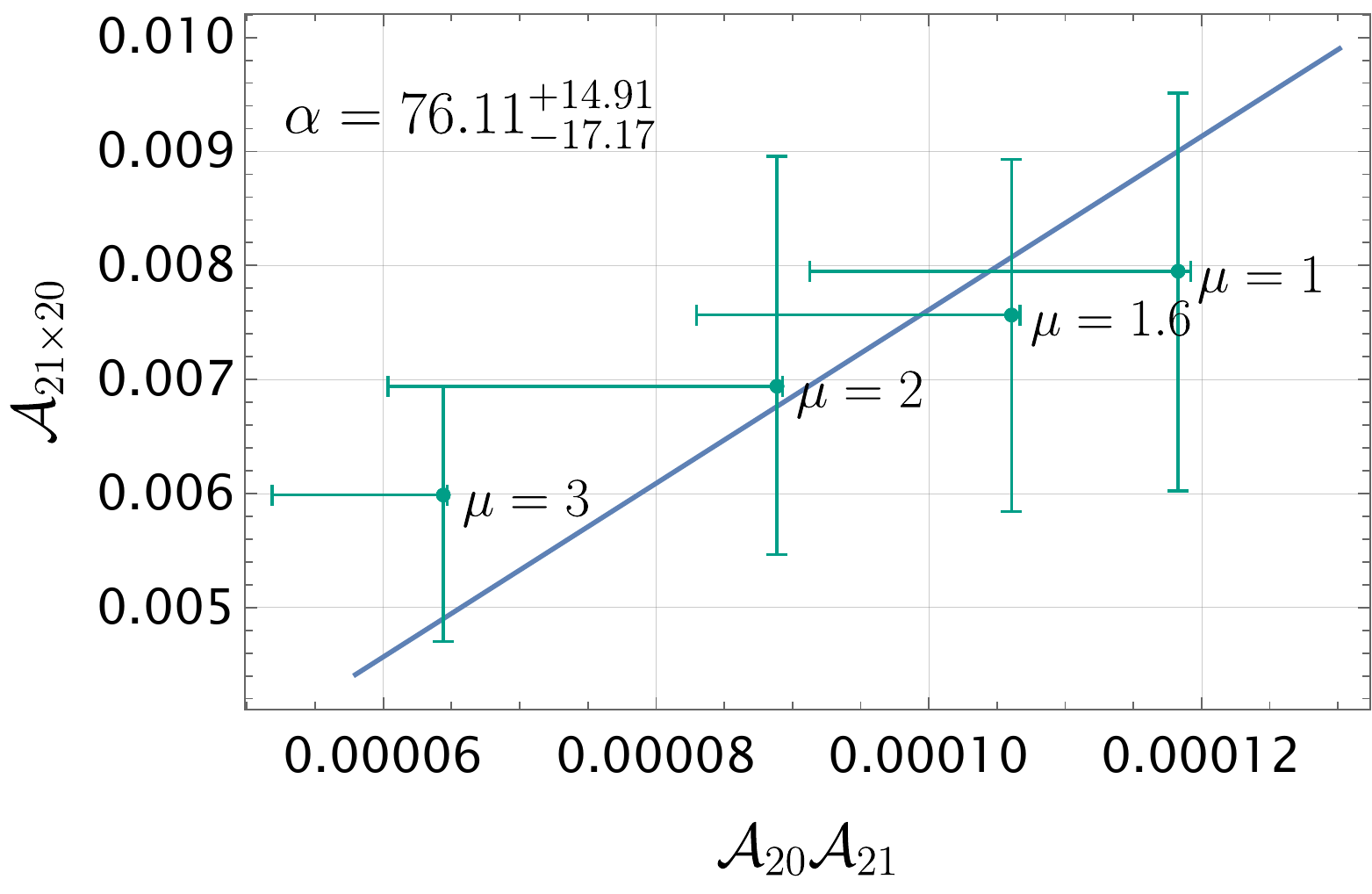}
             \label{fig:Ampmul2n3t21x20}
        }
    \caption{Amplitude relation for the quadratic tone $\omega_{21\times 20}$ in the $l=2$ shear mode for the boosted (left figure) and unboosted simulations (right figure). In all cases, $t_0$ is fixed to $15 M$. The boosted simulations clearly show a trend that deviates from the blue fitted line.}
     \label{fig:Ampl2t21x20}
    \end{figure*}

   Finally, the amplitude relation for the frequency $\omega_{21\times 20}$ is shown in Fig.~\ref{fig:Ampl2t21x20} for the sets of boosted (left) and unboosted simulations (right). The amplitude relation~\eqref{eq:amplitude-rel} could not be obtained within the uncertainty bars for the set of boosted simulations (see Fig.~\ref{fig:Ampl-gamma-l2n2t21x20}), while the large error bars in the unboosted simulations Fig.~\ref{fig:Ampmul2n3t21x20}~ make the amplitude relation unreliable, so the presence of the quadratic frequency $\omega_{21\times 20}$ cannot be confirmed up to our resolution. 

\subsection{The shear mode $l=3$ for the unboosted simulations}
\begin{figure}[ht!]
    \centering
\includegraphics[width=0.5\textwidth]{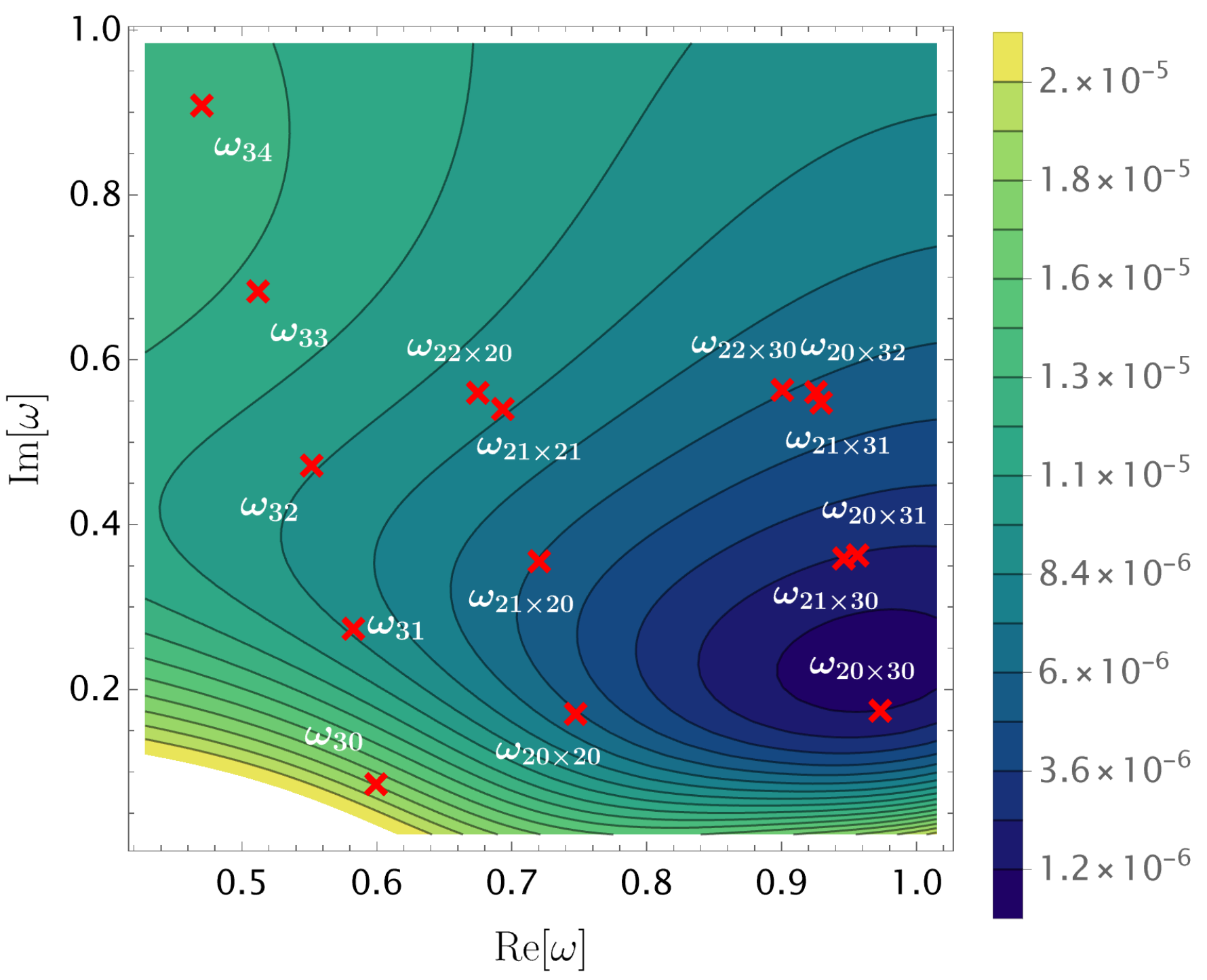}
    \caption{
     Mismatch between the  $l=3$ data ($t\in [t_\text{rd},t_f]$) of the  unboosted simulation S2 and a model for $\sigma_3$ with four tones, in which three frequencies are fixed to the GR predictions $\omega_{30}$ and $\omega_{31}$, $\omega_{32}$, and the fourth one is varied. }
    \label{fig:Misml3n3S1}
\end{figure}
\begin{figure*}
    \centering
        \subfigure[Mismatch of the four models with 
 the last tone's frequency $\omega_{33}$, $\omega_{20\times 20}$ or $\omega_{20\times 30}$. ]{       
            \includegraphics[width=0.33\textwidth, height=0.22\textwidth]{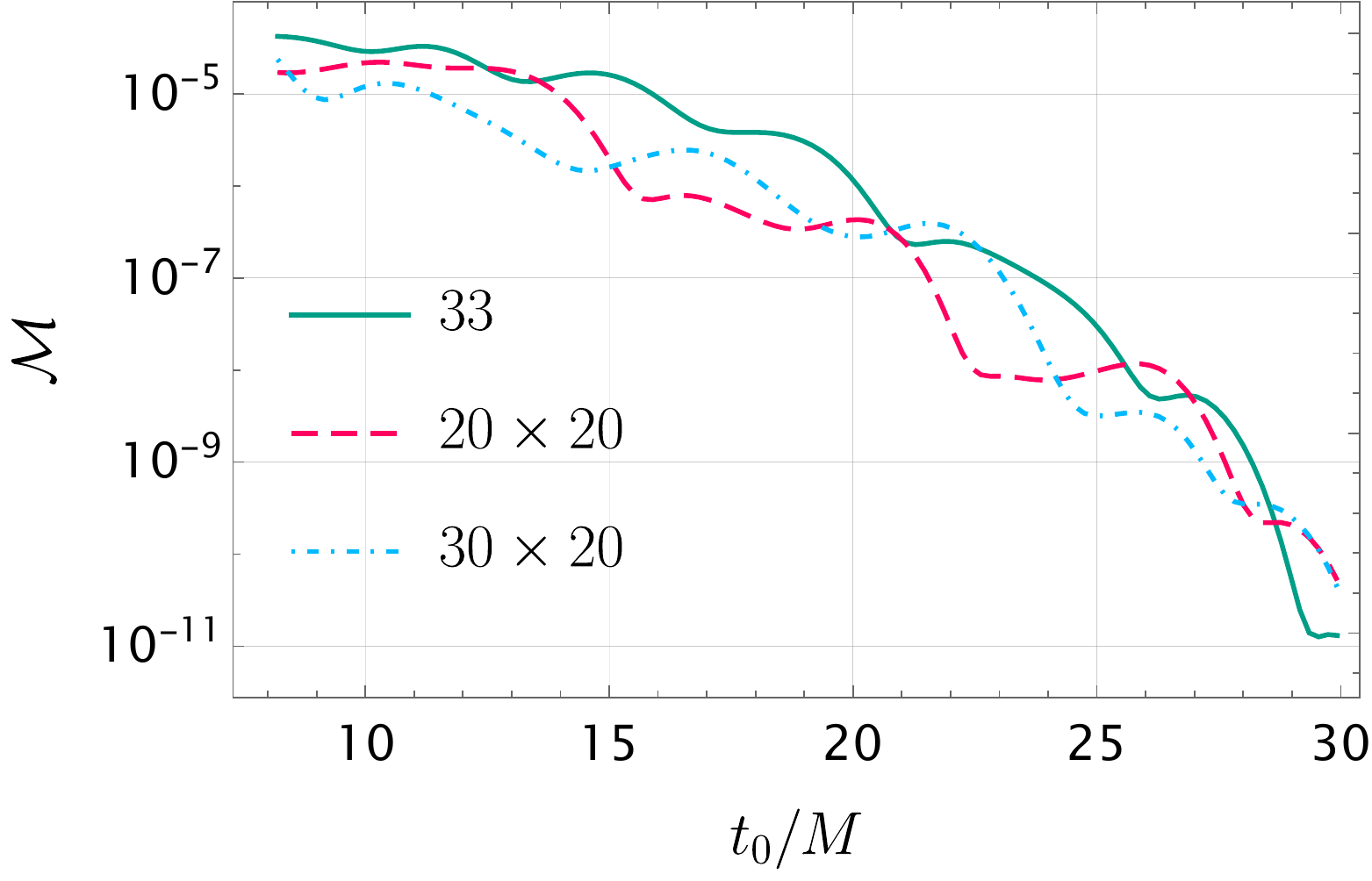}
               \label{fig:Mismatchl3n3-boost}
        }~\subfigure[Amplitudes for the linear (no superindex) and quadratic models with $\omega_{20\times 20}$ versus $\omega_{20\times 30}$.  ]{       
\includegraphics[width=0.33\textwidth, height=0.25\textwidth]{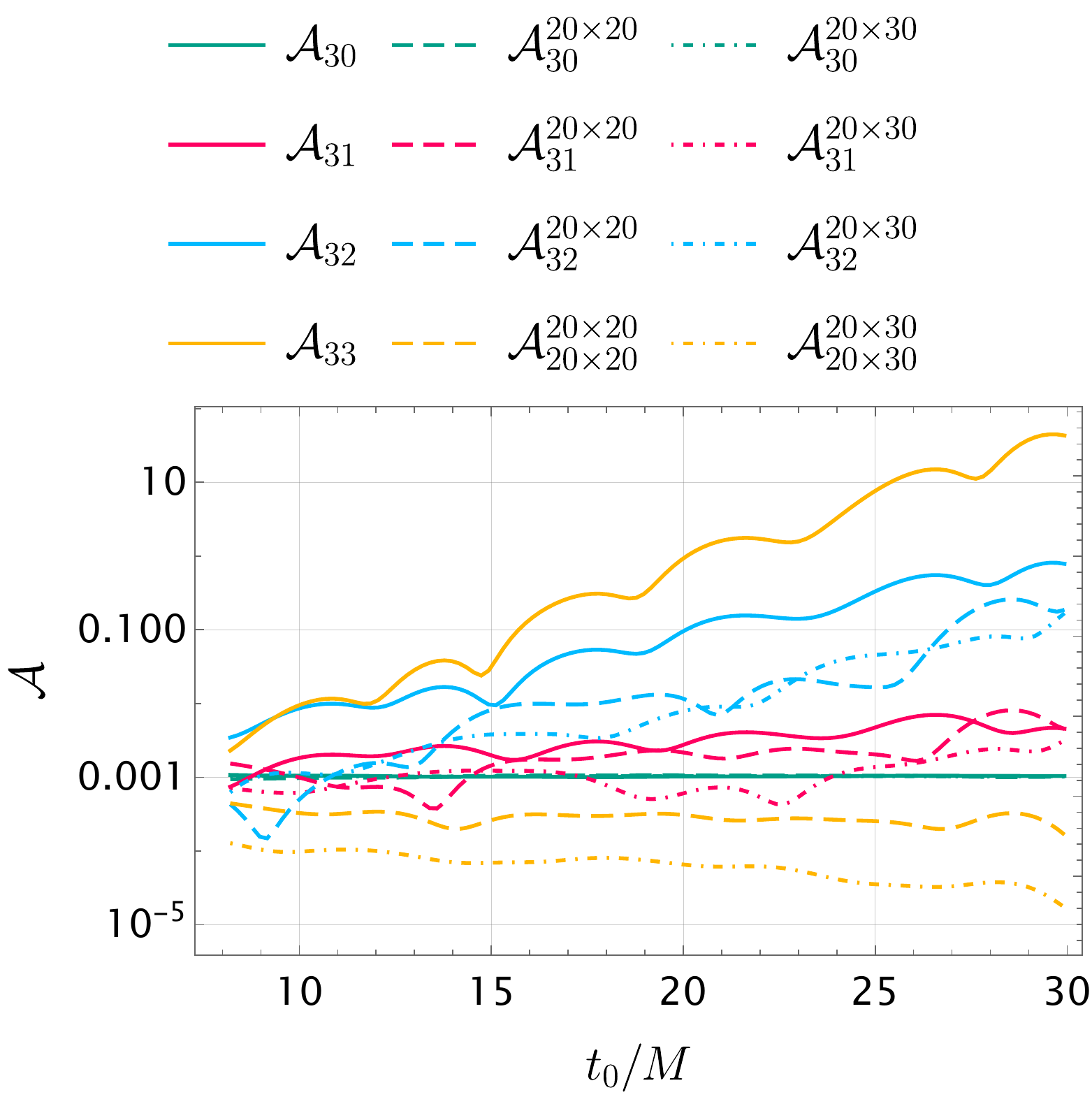}
             \label{fig:Ampl3n3}
        }~\subfigure[Relative variation of the optimal frequency with respect to $\omega_{33}$, $\omega_{20\times 20}$ and $\omega_{20\times 30}$.]{     
\includegraphics[width=0.33\textwidth, height=0.22\textwidth]{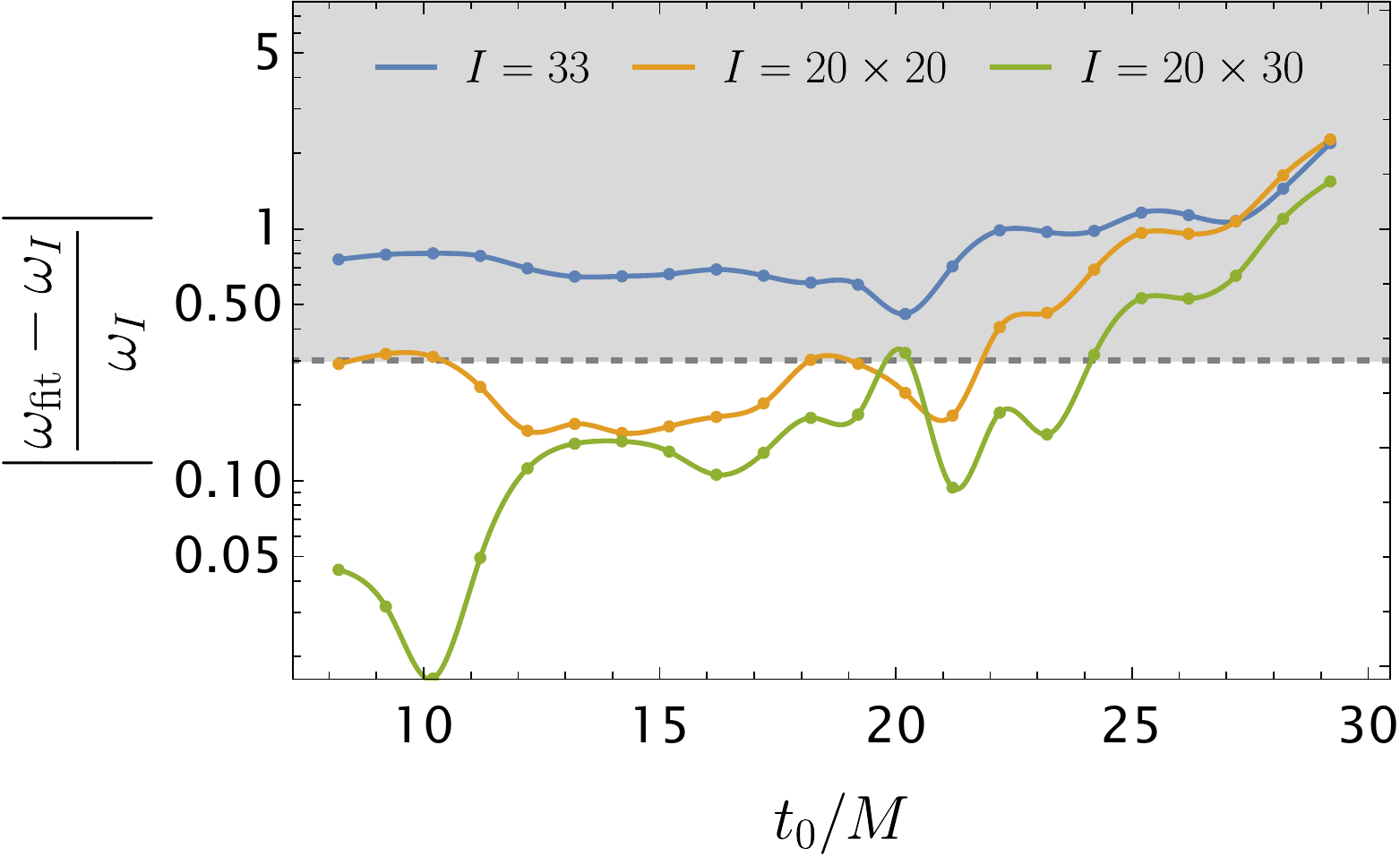}
             \label{fig:free-frequency-l3n3}
        }  
    \caption{Mismatch and tone's amplitudes as a function of the starting time $t_0\in[t_{\rm rd},30 M]$ for the $l=3$ shear mode  in the boosted simulation S2 with $n_{\rm max}=3$. While the mismatch is reasonable, the amplitudes vary significantly in all models. The variation of the amplitudes is largest for the linear model: $\mathcal{A}_{33}$ varies even by three orders of magnitude.}
    \label{fig:Qualityl3S2}
    \end{figure*}
\begin{figure*}
    \centering
 \subfigure[Tone $20\times 20$ in the shear mode $l=3$.]{       
            \includegraphics[width=0.42\textwidth,height=0.25\textwidth]{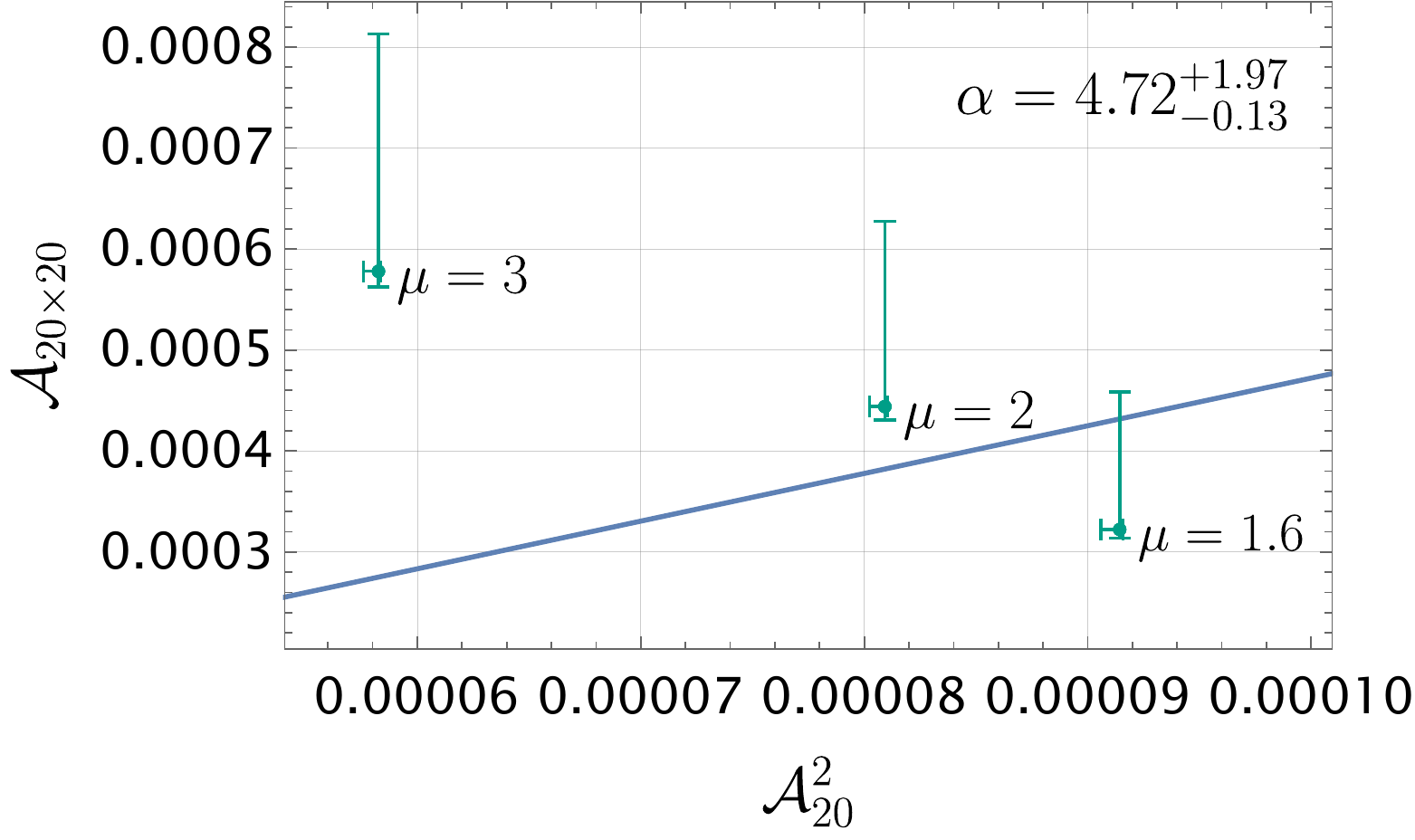}
             \label{fig:l3n3t21x20}
        }~\subfigure[Tone $20\times 30$ in the shear mode $l=3$. ]{       
            \includegraphics[width=0.42\textwidth,height=0.25\textwidth]{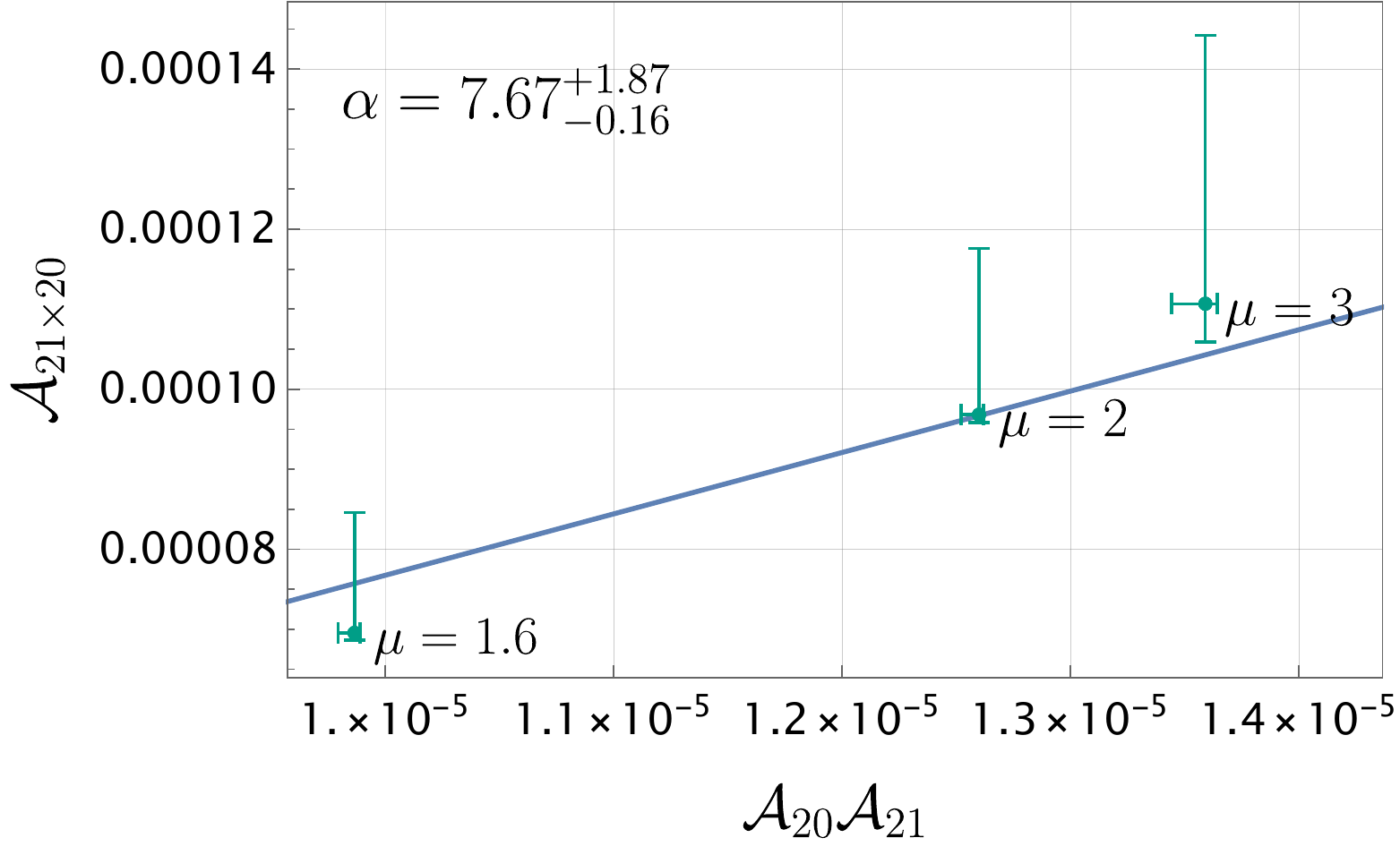}
             \label{fig:l3n3t20x30}
        }
   
    \caption{Amplitude relation for different quadratic modes in the  $l=3$ shear mode: $\omega_{20\times 20}$ (left), $\omega_{21\times 20}$ (middle) and $\omega_{22\times 20}$ (right). In all three cases, the dataset initial time $t_0$ is fixed to $t_0=15 M$. The quadratic relation is not conclusive given the large error bars. This relates to the large variation of the amplitudes in Fig.~\ref{fig:Ampl3n3}.}
    \label{fig:Amplitudel3}
    \end{figure*}

The set of boosted simulations was generated with mass ratio $\mu=1$, so the odd modes vanish. Hence, we can only use simulations S1, S2, and S4 to test the presence of quadratic modes in the odd shear modes. Here we show the mismatch and stability plots using the data of simulation S2, and the amplitude relation figures using all three simulations mentioned above. In Fig.~\ref{fig:Misml3n3S1}, we see that the quadratic frequencies $\omega_{20\times 20}$ and  $\omega_{20\times 30}$ could be present in the $l=3$ shear mode. Fig.~\ref{fig:Mismatchl3n3-boost} shows that the quadratic models fit the data with a lower mismatch than that of the linear model. 
 The stability study in Fig.~\ref{fig:Ampl3n3}, discards the presence of the linear mode $\omega_{33}$, and the optimization of the frequency in Fig.~\ref{fig:free-frequency-l3n3} shows a clear preference for the quadratic frequency $\omega_{20\times 30}$ over $\omega_{20\times 20}$, as it should, given the selection rules.
 The quadratic mode $\omega_{20\times 30}$ seems to satisfy a linear relation. However, the large error bars in Fig.~\ref{fig:l3n3t20x30} and the relatively small improvement of the mismatch in Fig.~\ref{fig:Mismatchl3n3-boost} (comparable to that of the mode $\omega_{20\times 20}$ for some range of the initial time) prevent us from reaching a conclusion regarding the presence of the quadratic modes in the $l=3$ shear data.

\subsection{The $l=4$ and $l=6$ shear modes for the boosted simulations}
\begin{figure*}
    \centering
  \subfigure[Shear mode $l=4$.]{       
            \includegraphics[width=0.44\textwidth]{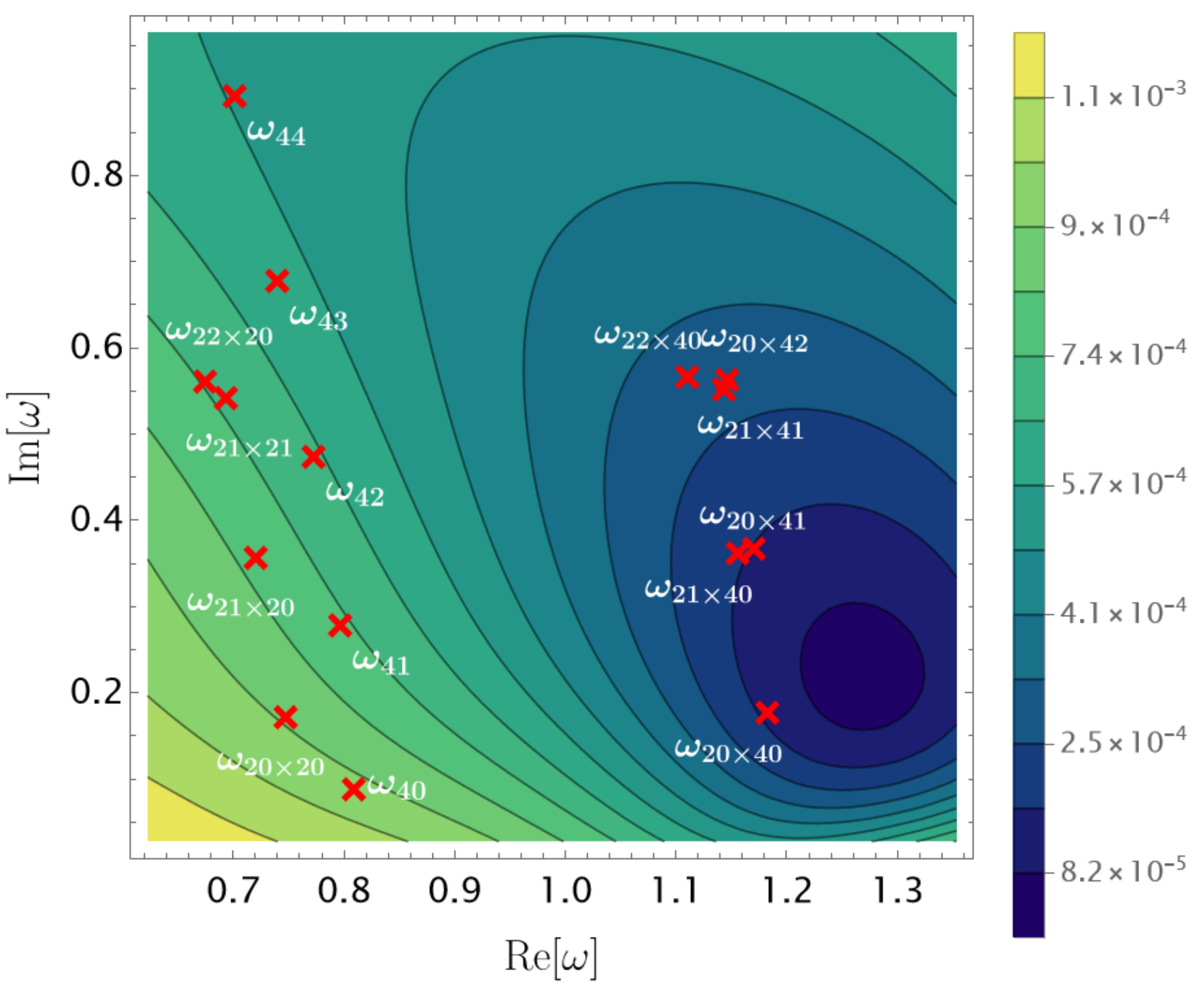}
               \label{fig:l4n2Mismatch}
        }~\subfigure[Shear mode $l=6$.]{       
            \includegraphics[width=0.44\textwidth]{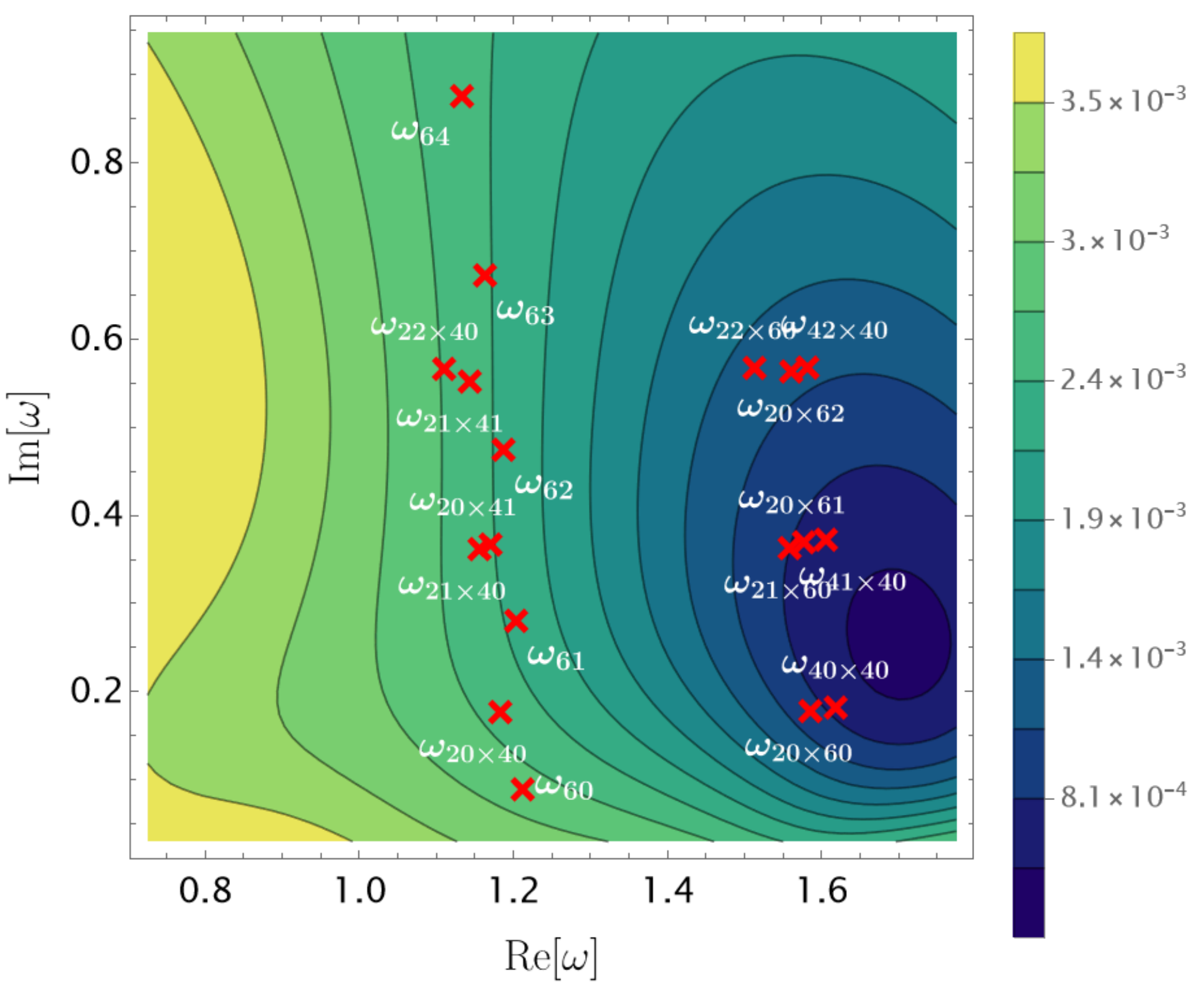}
             \label{fig:l6n2Mismatch}
        }
   
    \caption{
     Mismatch between the $l=4$ and $l=6$ data ($t\in [t_\text{rd},t_f]$) of the boosted simulation S7 and the models $\sigma_4$ and $\sigma_6$ with four tones each, in which three frequencies are fixed to the GR predictions $\omega_{l0}, \omega_{l1}$ and $\omega_{l2}$ with $l=4,6$ respectively, and the fourth one is varied.
    }
    \label{fig:l4&l6n2Mismatch}
\end{figure*}

\begin{figure*}
    \centering
  \subfigure[ {Mismatch of the four models with 
 the second and third tone's frequencies $\omega_{41}$ and $\omega_{42}$; $\omega_{41}$ and $\omega_{20\times 20}$; $\omega_{41}$ and $\omega_{20\times 40}$, and $\omega_{20\times 20}$ and $\omega_{20\times 40}$.}]{       
        \includegraphics[width=0.33\textwidth, height=0.20\textwidth]{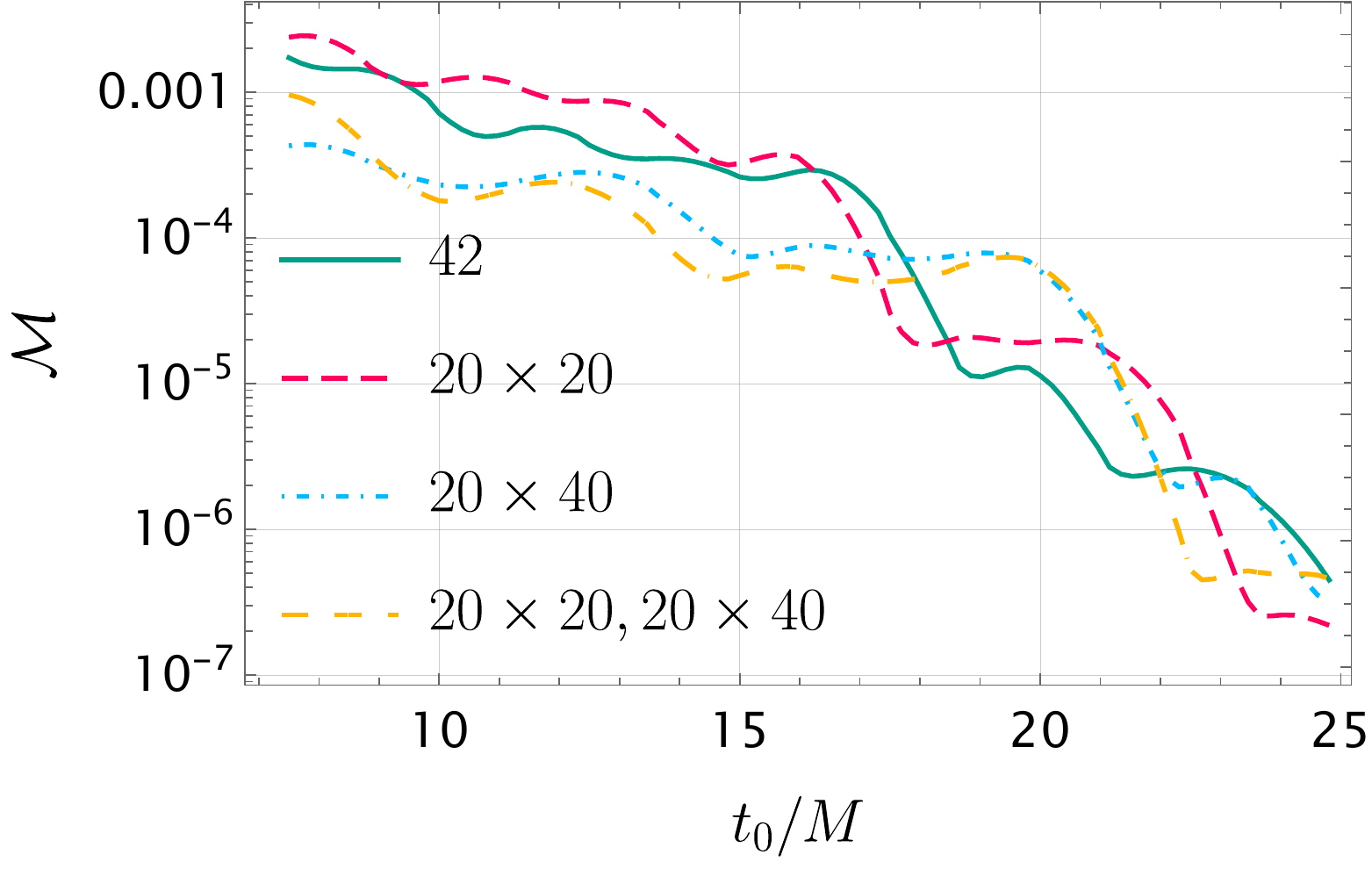}
               \label{fig:mismatch-l4n2}
        }~\subfigure[ {Amplitudes for the linear (no superindex) and quadratic models with the frequencies $\omega_{41}$ and $\omega_{20\times 40}$, or $\omega_{20\times 20}$ and $\omega_{20\times 40}$ in addition to the fundamental mode.
        %Overtone's amplitudes for the linear (no superindex) and quadratic models with the quadratic frequencies $\omega_{20\times 40}$, and $\omega_{20\times 20}$ and $\omega_{20\times 40}$.
        } ]{   
            \includegraphics[width=0.33\textwidth, height=0.20\textwidth]{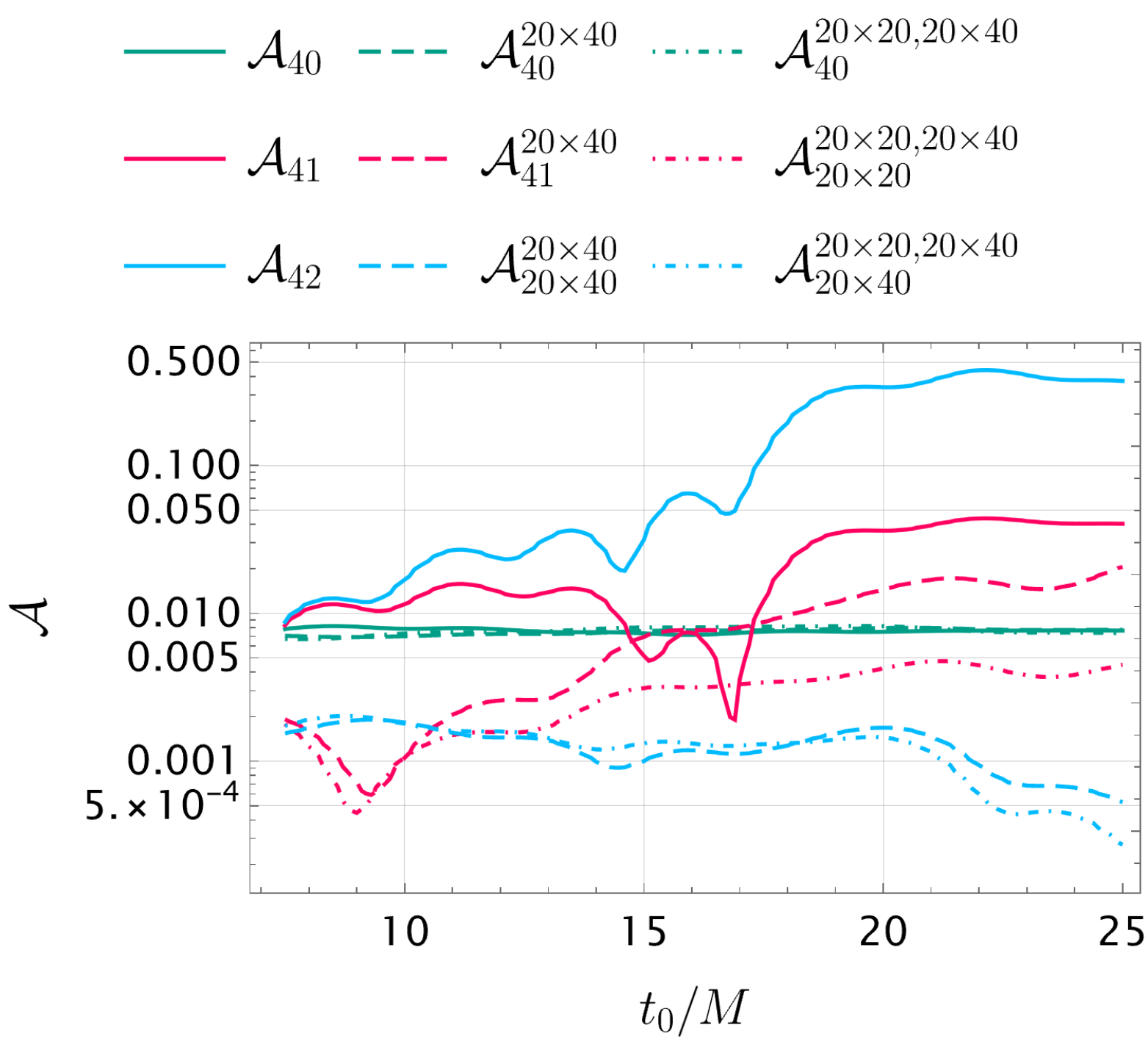}
             \label{fig:Ampl4n2-S7}
        }~\subfigure[ { Relative variation of the free fitted frequency with respect to known candidates. We have fixed the frequencies of the first tone to $\omega_{40}$ and of the second one to $\omega_{20\times 20}$. }]{       
            \includegraphics[width=0.33\textwidth, height=0.2\textwidth]{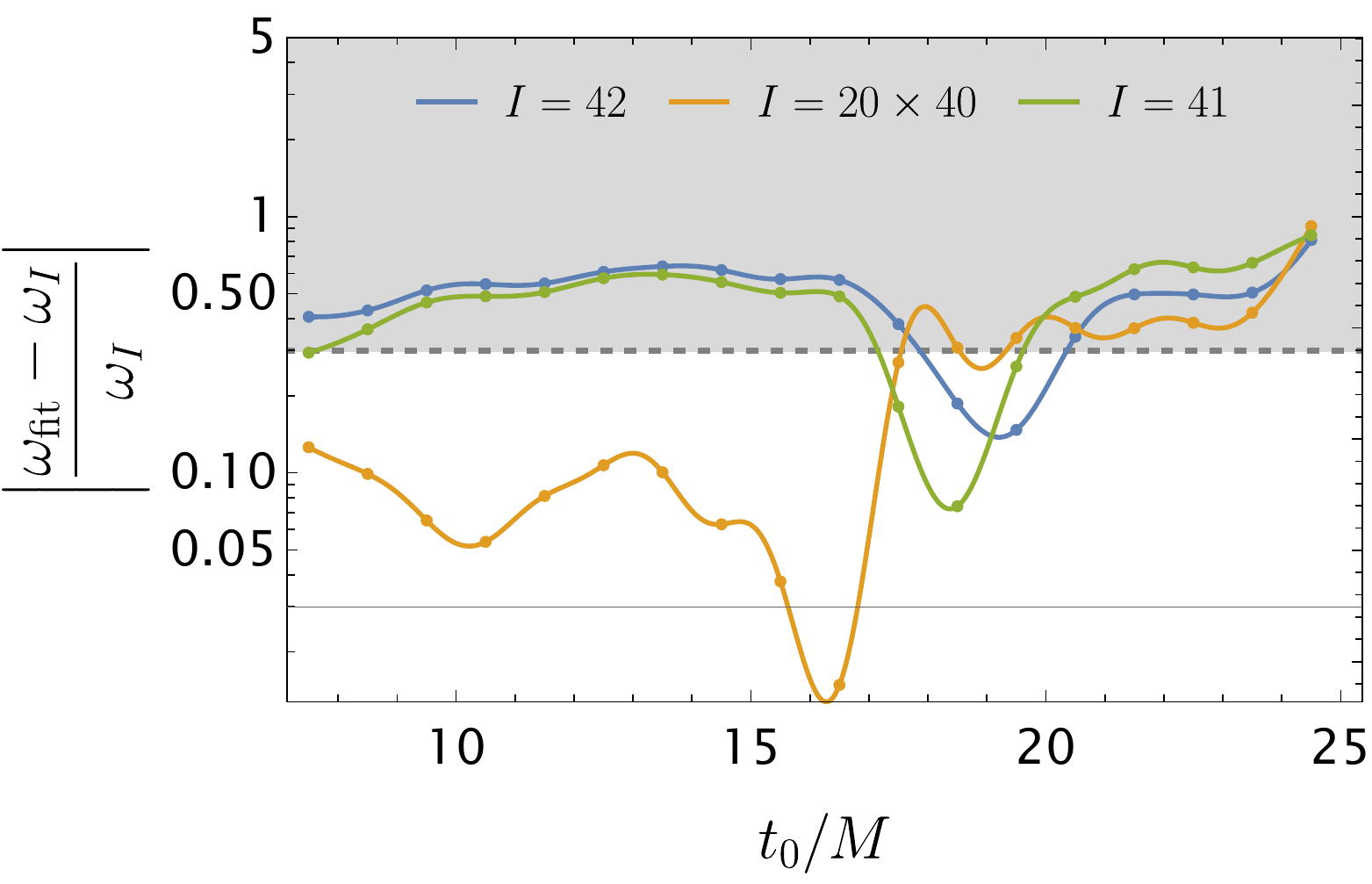}
             \label{fig:free-frequency-l4n3}
        }
    \caption{
    { Mismatch, tone's amplitudes, and best-fitted frequency between the $l=4$ data of the boosted simulation S7 and a model for $\sigma_4$ with three tones as a function of the starting time $t_0\in[t_{\rm rd},25 M]$. We label the different models through the quadratic tones that they contain (if only one quadratic mode labels the model, the other two tones are the fundamental one and the first overtone).} }
    \label{fig:Qualityl4}
    \end{figure*}

\begin{figure*}
    \centering
  \subfigure[ {Mismatch of the four models with 
 the second and third tone's frequencies $\omega_{61}$ and $\omega_{62}$, $\omega_{61}$ and $\omega_{20\times 60}$ or $\omega_{40\times 40}$;  $\omega_{20\times 40}$ and $\omega_{20\times 60}$, and $\omega_{20\times 40}$ and $\omega_{40\times 40}$.}]{       
        \includegraphics[width=0.33\textwidth, height=0.20\textwidth]{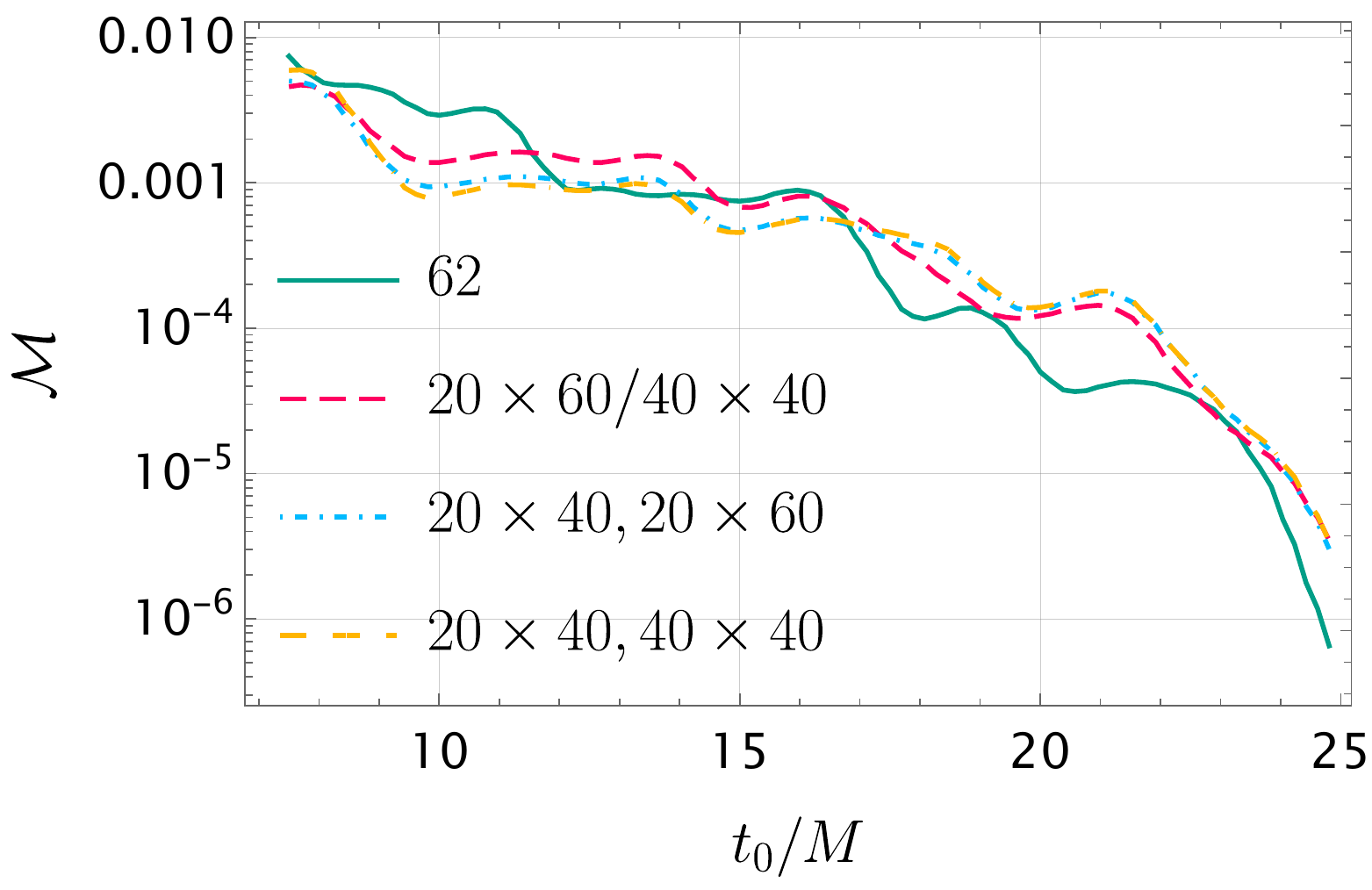}
               \label{fig:mismatch-l6n2}
        }~\subfigure[ {Amplitudes for the linear (no superindex) and quadratic models with the frequencies $\omega_{61}$ and $\omega_{40\times 40}$, or $\omega_{20\times 40}$ and $\omega_{40\times 40}$ in addition to the fundamental mode.
        } ]{   
            \includegraphics[width=0.33\textwidth, height=0.20\textwidth]{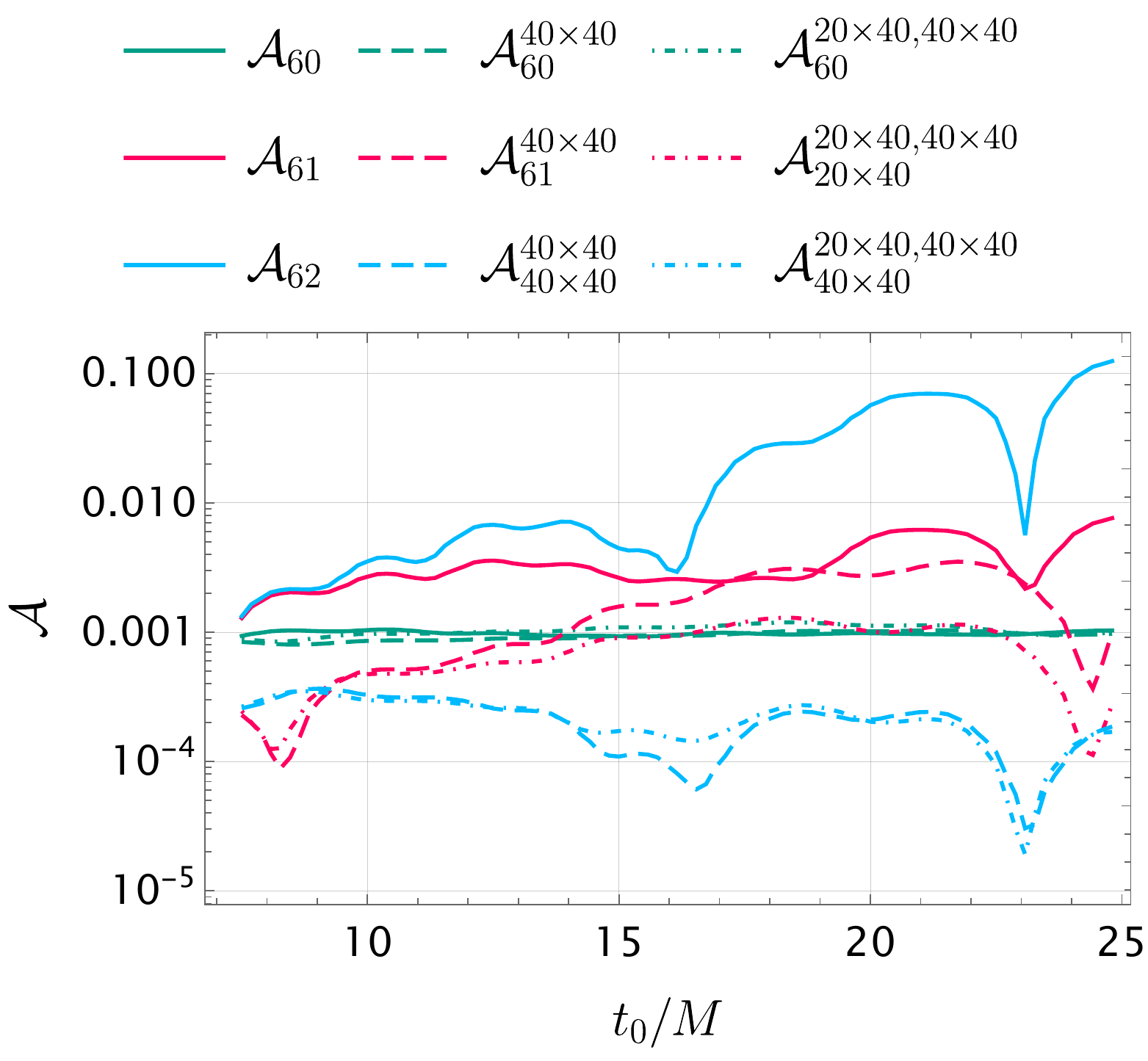}
             \label{fig:Ampl6n2-S7}
        }~\subfigure[ { Relative variation of the free fitted frequency with respect to known candidates. We have fixed the frequencies of the first tone to $\omega_{60}$ and of the second one to $\omega_{20\times 40}$. }]{       
            \includegraphics[width=0.33\textwidth, height=0.2\textwidth]{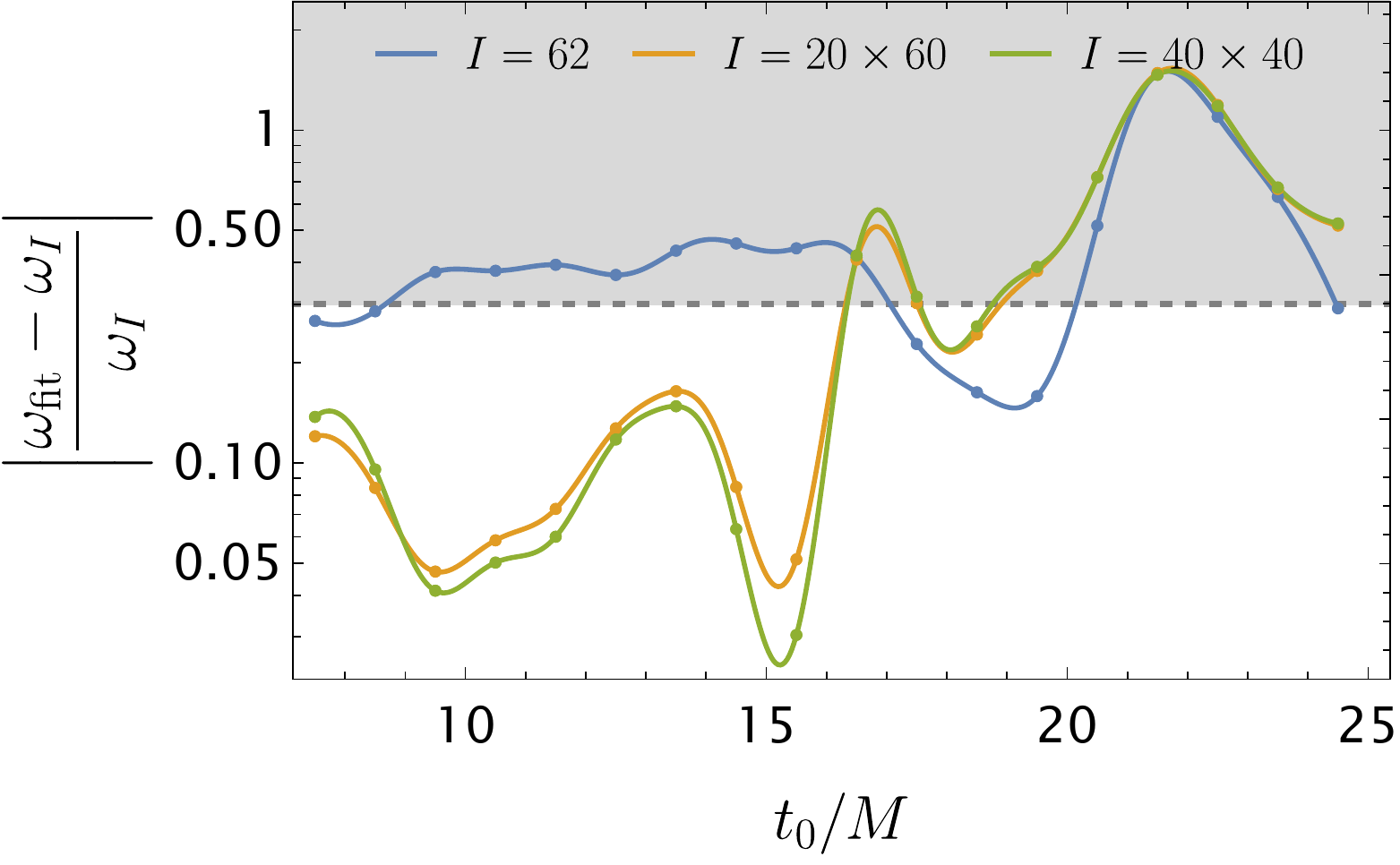}
             \label{fig:free-frequency-l6n3}
        }
    \caption{
    { Mismatch, tone's amplitudes, and best-fitted frequency between the $l=6$ data of the boosted simulation S7 and a model for $\sigma_6$ with three tones as a function of the starting time $t_0\in[t_{\rm rd},25 M]$. We label the different models through the quadratic tones that they contain (if only one quadratic mode labels the model, the other two tones are the fundamental one and the first overtone).} }
    \label{fig:Qualityl6}
    \end{figure*}

\begin{figure*}
    \centering
  \subfigure[  Amplitude relation for the $20\times 20$ quadratic mode.]{       
        \includegraphics[width=0.42\textwidth,height=0.25\textwidth]{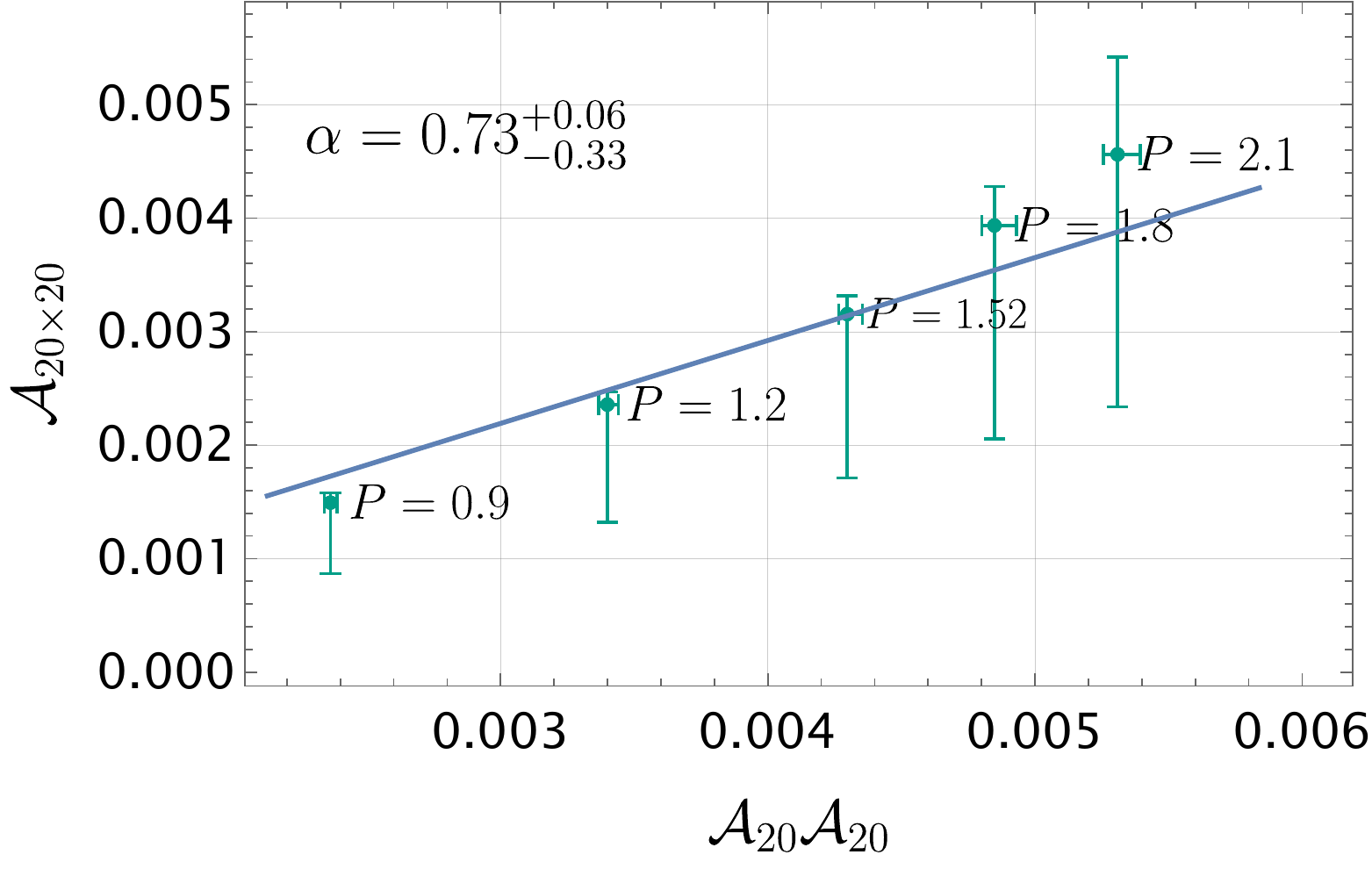}
               \label{fig:Amp-t20x20&t20x40-t20x20}
        }~\subfigure[  Amplitude relation for the $20\times 40$ quadratic mode.]{       
            \includegraphics[width=0.42\textwidth,height=0.25\textwidth]{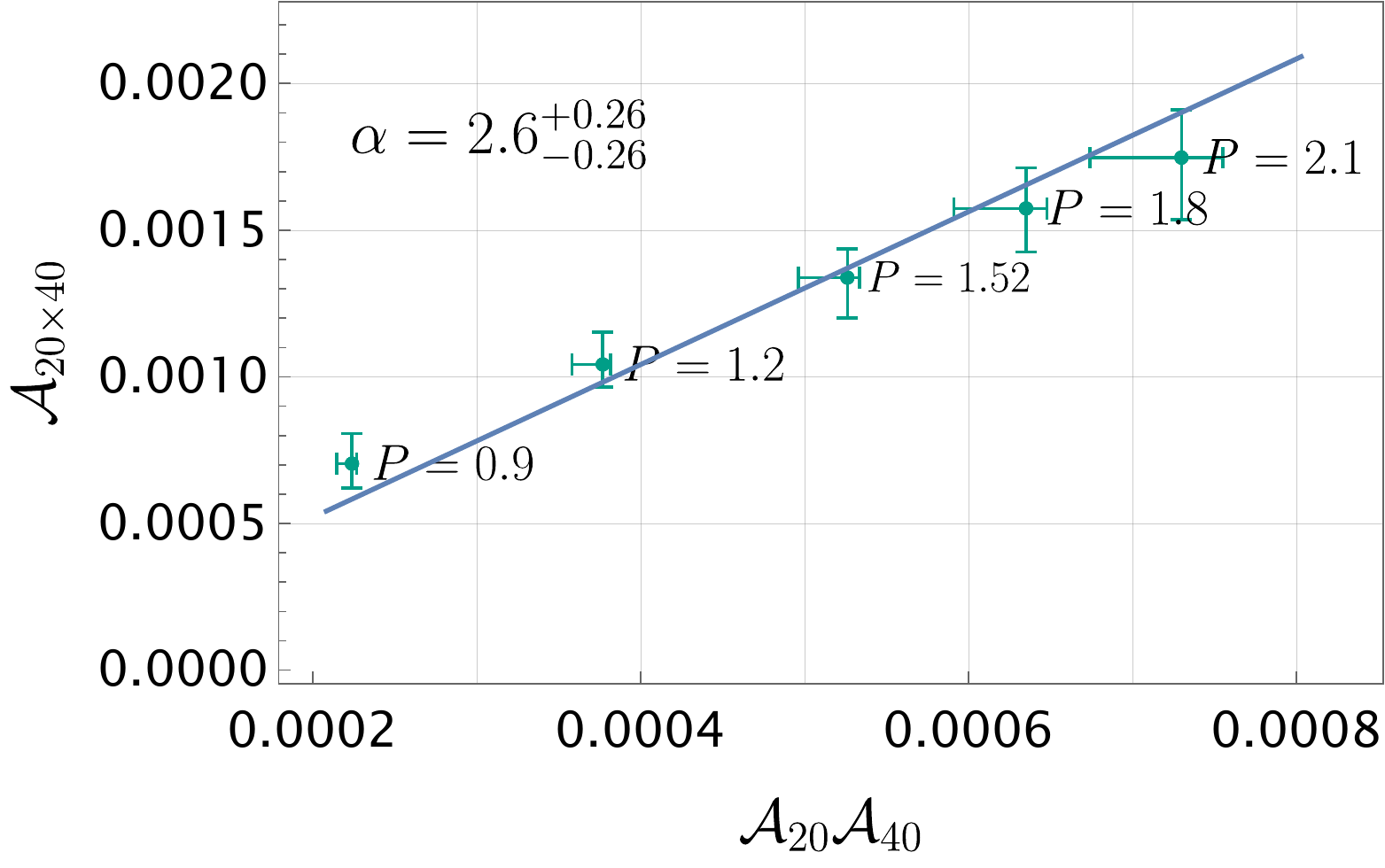}
             \label{fig:Amp-t20x20&t20x40-t20x40}
        }
    \caption{  Amplitude relation for the quadratic tones $\omega_{20\times 20}$ and $\omega_{20\times 40}$ in a model with the fundamental tone and two quadratic ones for the $l=4$ shear mode. The dataset initial time $t_0$ is fixed to $t_0=15 M$. 
   }
   \label{fig:Amp-t20x20&t20x40}
    \end{figure*}

\begin{figure*}
    \centering
  \subfigure[  Tone $20\times 40$ in the model with the frequencies $\{\omega_{60}, \omega_{20\times 40},\omega_{20\times 60}\}$. ]{       
        \includegraphics[width=0.42\textwidth,height=0.25\textwidth]{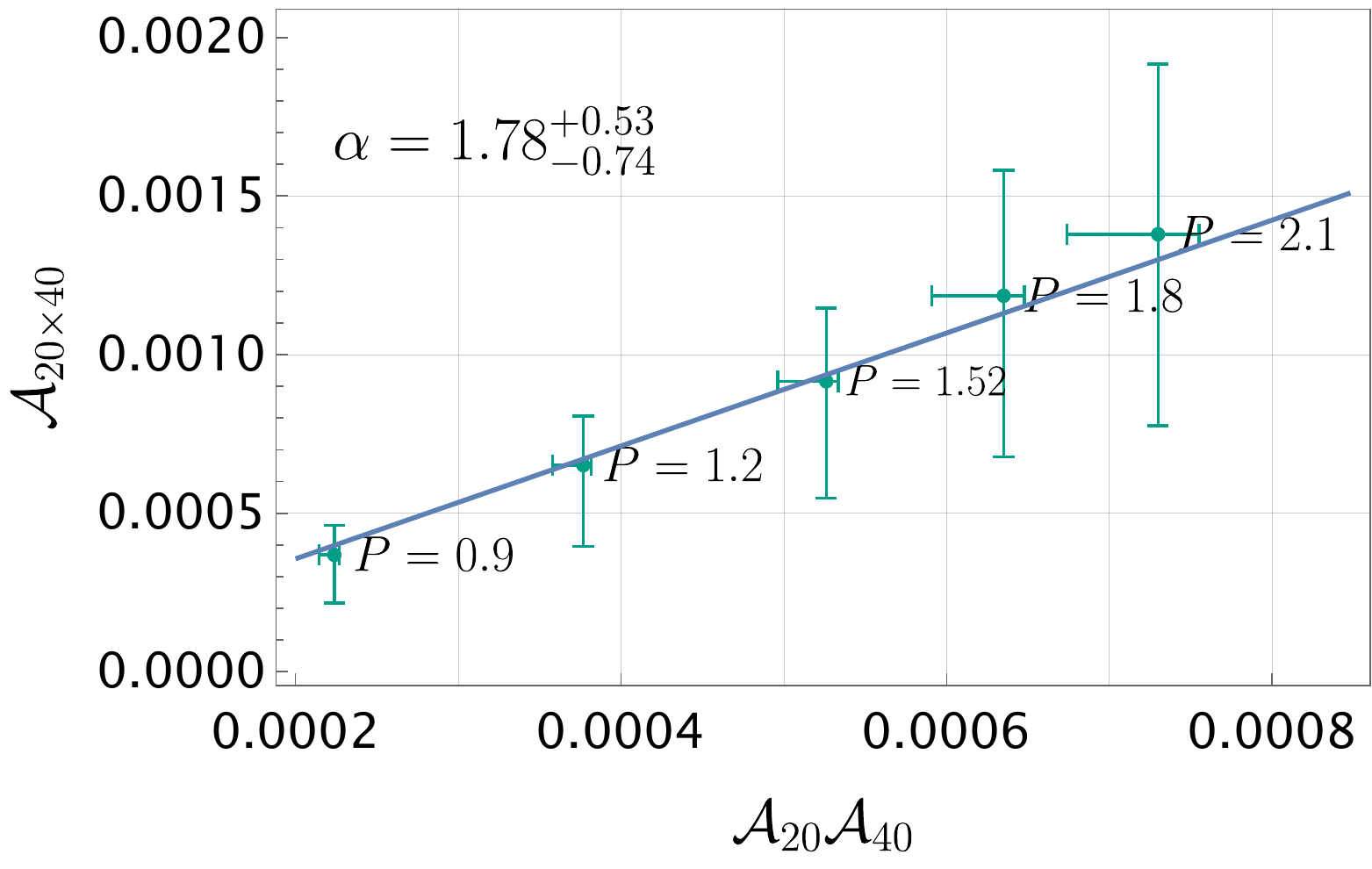}
               \label{fig:Amp-t20x40&t20x60-t20x40}
        }~\subfigure[  Tone $20\times 60$ in the model with the frequencies $\{\omega_{60}, \omega_{20\times 40},\omega_{20\times 60}\}$. ]{       
            \includegraphics[width=0.42\textwidth,height=0.25\textwidth]{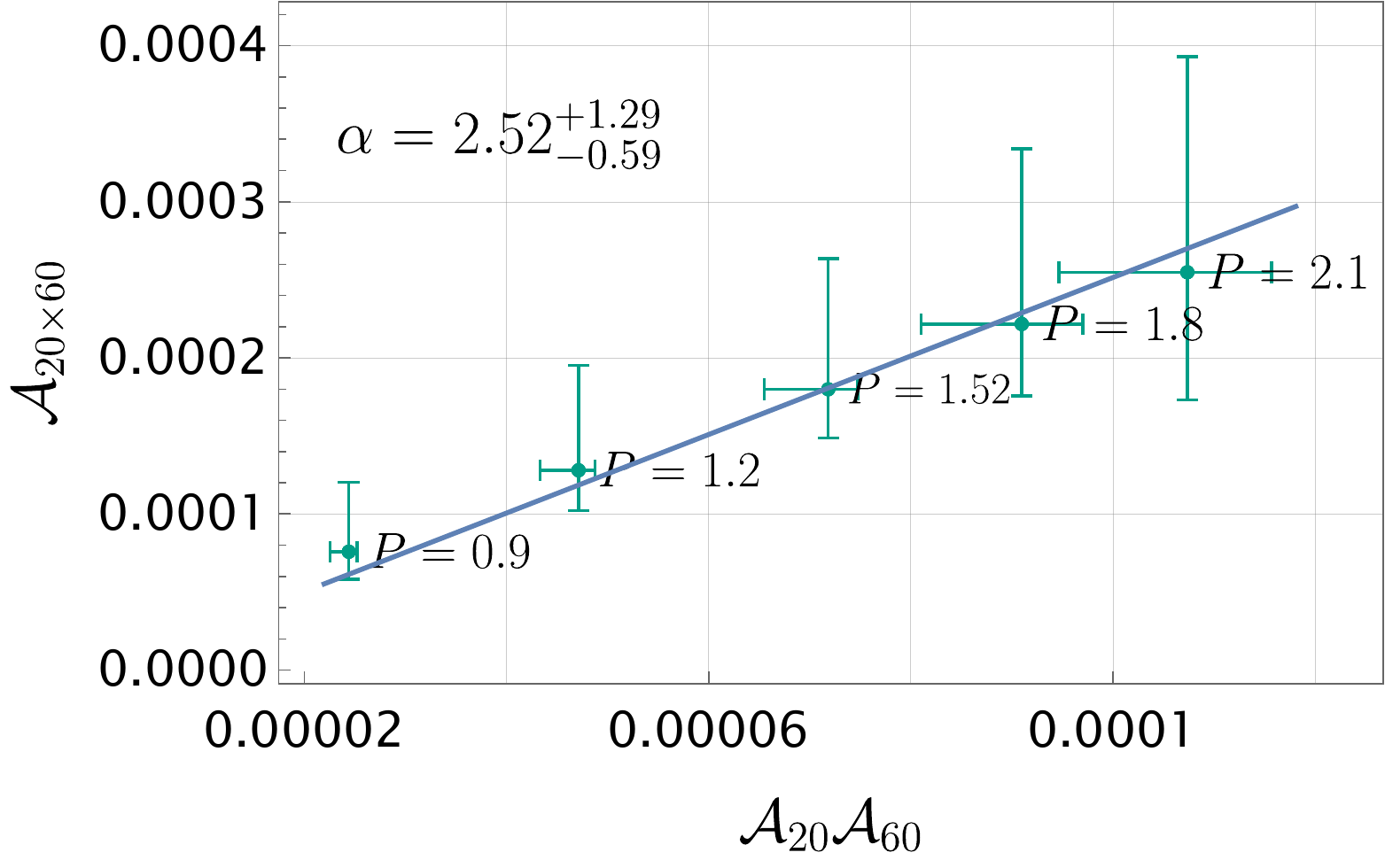}
             \label{fig:Amp-t20x40&t20x60-t20x60}
        }
         \subfigure[  Tone $20\times 40$ in the model with the frequencies $\{\omega_{60}, \omega_{20\times 40},\omega_{40\times 40}\}$. ]{       
        \includegraphics[width=0.42\textwidth,height=0.25\textwidth]{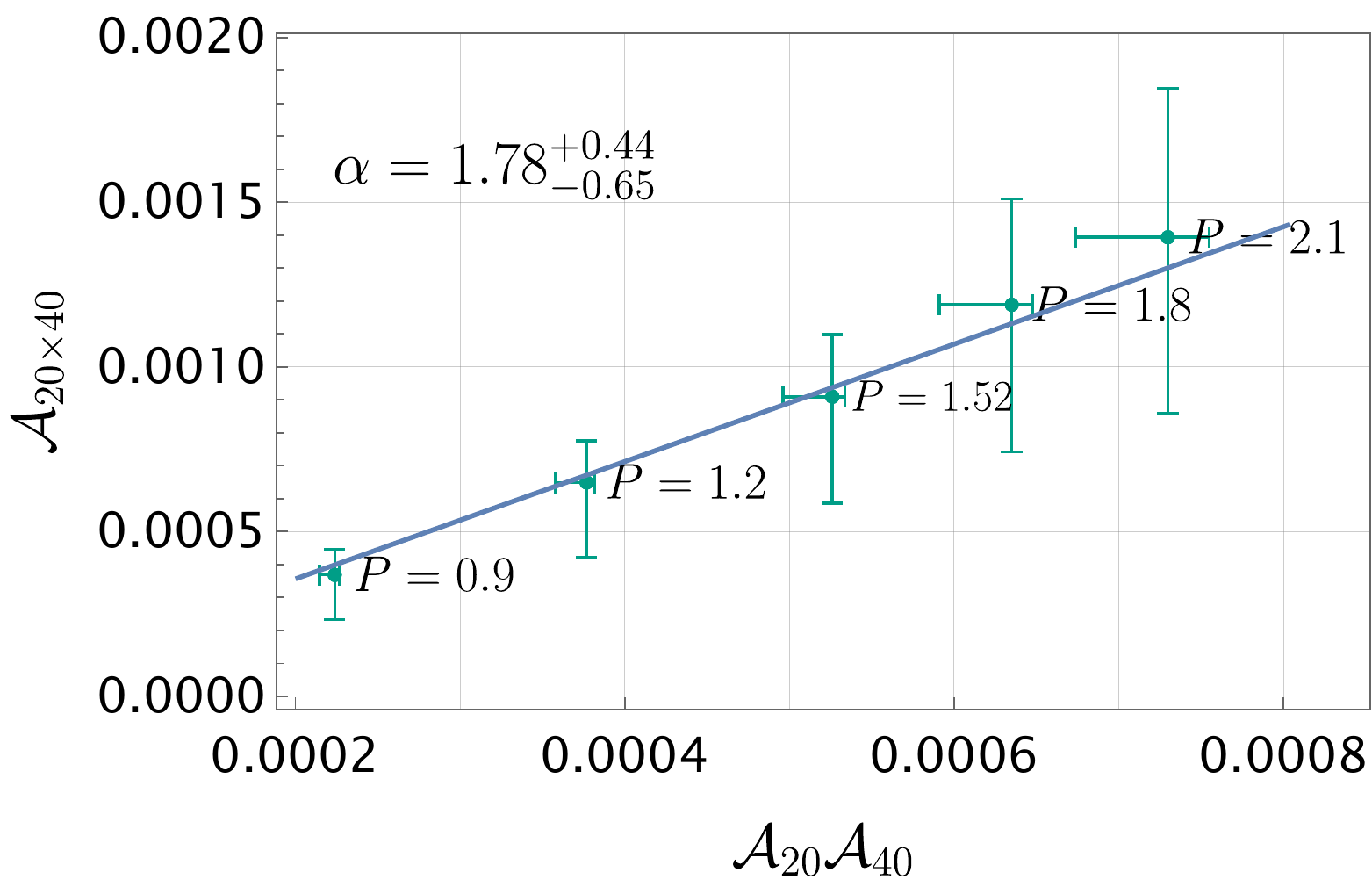}
               \label{fig:Amp-t20x40&t40x40-t20x40}
        }~\subfigure[  Tone $40\times 40$ in the model with the frequencies $\{\omega_{60}, \omega_{20\times 40},\omega_{40\times 40}\}$. ]{       
            \includegraphics[width=0.42\textwidth,height=0.25\textwidth]{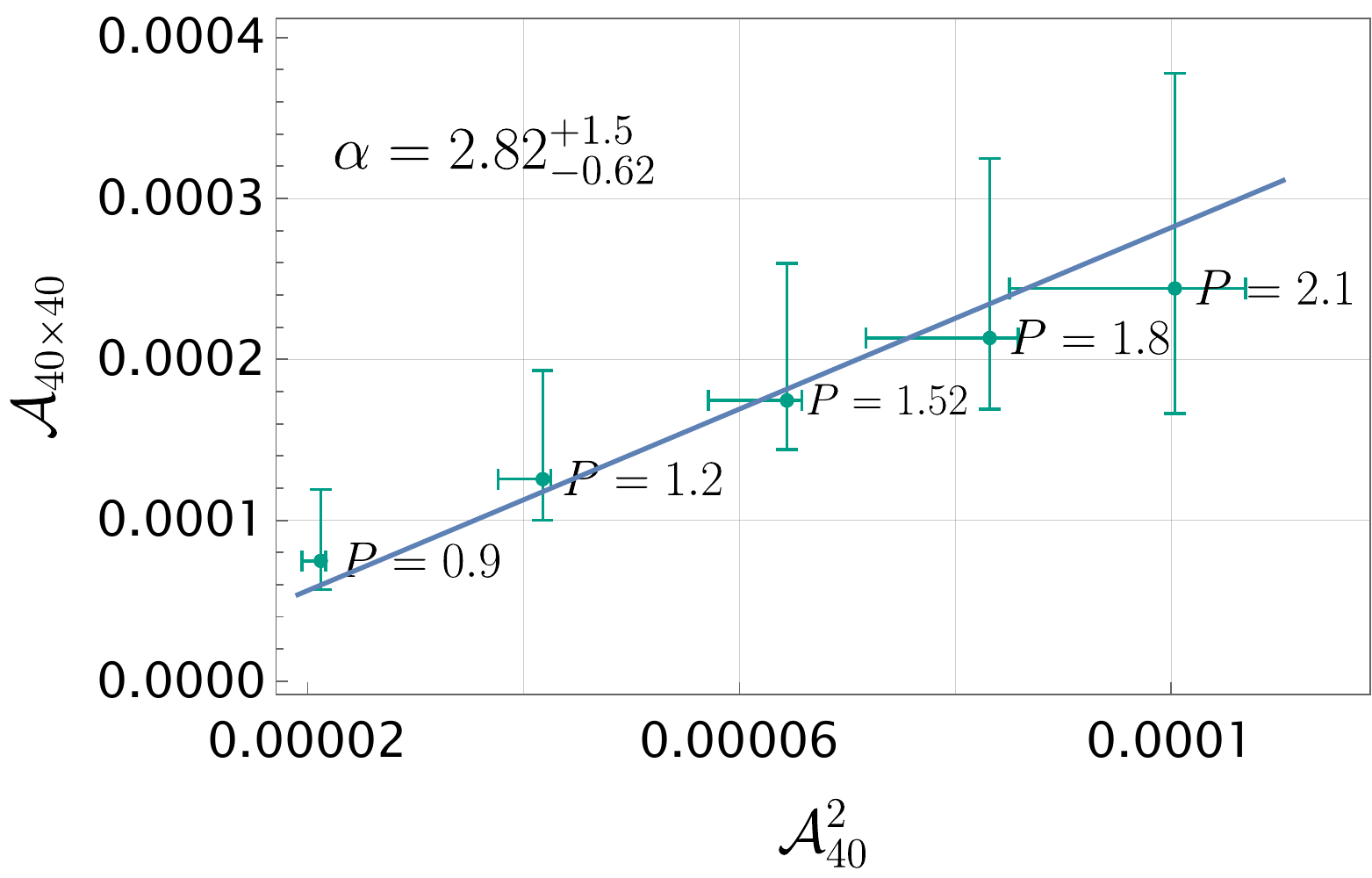}
             \label{fig:Amp-t20x40&t40x40-t40x40}
        }
    \caption{  Amplitude relation for the quadratic tones $\omega_{20\times 40}$ and $\omega_{20\times 60}$, and $\omega_{20\times 40}$ and $\omega_{40\times 40}$, in a model with the fundamental tone and two quadratic ones for the $l=6$ shear mode. The dataset initial time $t_0$ is fixed to $t_0=15 M$. 
   }
   \label{fig:Amp-l6n2-two_quadratics}
    \end{figure*}

We could not determine the presence of quadratic modes in the unboosted simulations for the shear modes $l=4$ and higher because the amplitudes were too weak (see the discussion on the restriction of the number of overtones). However, we could determine the presence of quadratic modes for the $l=4$ and $6$ shear modes in the boosted simulations. Completely analogously to the main text, we use the data of the simulation S7. In Fig.~\ref{fig:l4&l6n2Mismatch}, we screen over the complex frequency plane of the $n_{\rm max} =2$ overtone in Eq.~\eqref{eq:model-real} to determine the most likely modes in our model at $t_0=t_{\rm rd}$. For the shear mode $l=4$ we find that the frequency $\omega_{20\times 40}$ is favored, while for the $l=6$ mode, the tones $20\times 60$ and $40\times 40$ prevail.  {Given the similar spectrum of the linear frequency $\omega_{41}$ and the quadratic $\omega_{20\times 20}$ in Fig.~\ref{fig:l4n2Mismatch}, and that of the linear overtone $\omega_{61}$ with the quadratic frequency $\omega_{20\times 40}$ in Fig.~\ref{fig:l6n2Mismatch}, we also test the models with two quadratic tones for the $l=4$ and $l=6$ shear modes, i.e., the models with the frequencies $\{\omega_{40},\omega_{20\times 20},\omega_{20\times 40}\}$ and $\{\omega_{60},\omega_{20\times 40},\omega_{20\times 60}\}$ or $\{\omega_{60},\omega_{20\times 40},\omega_{40\times 40}\}$. In Figs.~\ref{fig:Qualityl4} and ~\ref{fig:Qualityl6}, we show the mismatch, amplitude stability, and frequency minimization of these models (all with $n_{\rm max}=2$) for the $l=4$ and $l=6$ shear modes respectively. Notice that in Fig.~\ref{fig:mismatch-l6n2} the similarity of the quadratic frequencies  $\omega_{20\times 60}$ and $\omega_{40\times 40}$ makes the single quadratic models indistinguishable from each other in terms of the fit residuals.
We see that for both the $l=4$ and $l=6$ shear modes, the multiple quadratic models reduce the fit residuals as compared to the single quadratic ones (including the linear overtones $41$ and $61$), and improve the stability of the fitted amplitudes as a function of the starting time. The presence of the quadratic modes in the $l=4$ and $l=6$  shear modes is confirmed by the amplitude relations in Figs.~\ref{fig:Amp-t20x20&t20x40} and~\ref{fig:Amp-l6n2-two_quadratics}. We find the same combination of relevant modes as Cheung et al.~\cite{Cheung:2022rbm} for the $l=4$ modes, which assesses the correlation between the outgoing gravitational wave radiation and the horizon dynamics. For the $l=6$ mode, we identify the $\omega_{20\times 40}$ quadratic frequency  (also reported in~\cite{Cheung:2022rbm}), in combination with either the quadratic frequency $\omega_{40\times 40}$ or $\omega_{20\times 60}$. Identifying these  extra quadratic tones is possible due to the high resolution and low noise level of our simulations. } 
%Their presence is further supported by the results in Fig.~\ref{fig:Qualityl4&6} as well as through the amplitude relations in Fig.~\ref{fig:Amplitudel4&6}.

%\begin{figure*}
%    \centering
%  \subfigure[Tone $20\times 40$ in the $l=4$ shear mode .]{     
            %\includegraphics[width=0.4\textwidth,height=0.23\textwidth]{figures/Amplitude_relations_l4n2_t15_boosted_20x40.pdf}
               %\label{fig:l4n2t20x40}}
        %~\subfigure[Tone $20\times 60$ in the $l=6$ shear mode .]{       
            %\includegraphics[width=0.4\textwidth,height=0.23\textwidth]{figures/Amplitude_relations_l6n2_t15_boosted_20x60.pdf}
             %\label{fig:l6n2t20x60}
        %}~\subfigure[Tone $40\times 40$ in the $l=6$ shear mode .]{       
            %\includegraphics[width=0.4\textwidth,height=0.23\textwidth]{figures/Amplitude_relations_l6n2_t15_boosted_40x40.pdf}
             %\label{fig:l6n2t40x40}
        %}
    %\caption{Amplitude relation for the quadratic tone $\omega_{20\times 40}$ in the shear mode $l=4$ and the tones $\omega_{20\times 60}$ and $\omega_{40\times 40}$ in the shear mode $l=6$. In all cases, the dataset initial time is set to $t_0=15 M$.
    %}
    %\label{fig:Amplitudel4&6}
%\end{figure*}

\subsection{Restrictions on the number of overtones}

The analysis for the $l=3$ (and higher) shear modes in the unboosted simulations, and the $l=8$ shear modes (and higher) in the boosted ones were inconclusive. In other words, we were not able to conclude whether quadratic modes were present in the shear data confidently due to either the similar mismatch obtained for the linear and quadratic models or the large error bars in the amplitude relation analysis (for an explicit example, see the results for the $l=3$ mode below). Further, increasing the number of overtones in our models doesn't change this outcome. Here we comment on why including higher overtones doesn't help the analysis.

Generally, including a higher number of overtones will make the amplitudes very unstable. Namely, the best-fit amplitudes can vary over orders of magnitude for a short change in starting time, making the amplitude measurement very inaccurate. Thus, the quadratic relation between the linear and quadratic amplitudes cannot be established. This instability can be understood to be stemming from the non-orthogonality of the quasinormal modes. Consider the minimization of $\lVert A^i h_i - d \rVert^2$, where $h^i(t)$ are some real QNM models described in Eq.~\eqref{eq:model-real}, $A^i$ are their real amplitude, $d$ is the data, and $\lVert \cdot \rVert$ is the $L^2$ norm. Define the Fischer-matrix by the $L^2$ inner product $\Gamma_{ij} = \langle h_i,h_j\rangle$. Given the QNM frequencies and a range of time, $\Gamma_{ij}$ can be calculated analytically. The amplitude that minimizes the least-squared norm is given by 
\begin{equation}
    \label{eq:Amplitude-Fit}
    A^i = (\Gamma^{-1})^{ij}\langle d, h_j\rangle
\end{equation}
Because the QNMs are not orthogonal, the matrix $\Gamma_{ij}$ can become nearly singular. Consequently, small errors in $d$ can be magnified and lead to large errors in the amplitudes. Indeed, for $d = \sigma_{\mathrm{GR}} + N$, where $\sigma_{\mathrm{GR}}$ is the exact GR shear and $N$ is the numerical error, the error in the amplitude that minimizes the residual compared to GR,  $\delta A^i$, satisfies 
\begin{equation}
    \label{eq:condition}
    \frac{\lVert \delta A^i \rVert}{\lVert A^i_{GR} \rVert} \leq  \kappa \frac{\lVert N \rVert}{\lVert \sigma_{GR} \rVert}\,,
\end{equation}
where $\kappa$ is the condition number of $\Gamma_{ij}$, given by $\kappa = |\lambda_{\mathrm{max}}/\lambda_{\mathrm{min}}|$, the absolute value of the ratio of the maximum and minimum eigenvalues of $\Gamma_{ij}$. Here the norm on the amplitudes is the Euclidean norm on the space of amplitudes. Hence, the left-hand side of Eq.~\eqref{eq:condition}  measures the relative error in amplitudes, while the right-hand side provides an upper bound for this quantity.

The condition number grows rapidly with the number of overtones, thus increasing the errors in the best-fit amplitudes. We use Eq.~\eqref{eq:condition} to estimate the uncertainty in the fitted amplitudes due to the numerical resolution only (e.g., we do not consider the errors coming from the systematic effects of the model, among others). Thus, we take the error $N$ to be a purely numerical error, which is estimated by comparing two numerical resolutions (for example, we denote by $\epsilon_{240-120}$ the error obtained comparing the data at $\text{res}=120$ and $\text{res}=240$), and we approximate $\lVert \sigma_{\mathrm{GR}}\rVert$  by the norm of the numerical shear modes. While this procedure gives an estimate for the upper bound in Eq.~\eqref{eq:condition}, in practice, the errors can be much smaller. Contrarily, we could be in a regime where systematic errors dominate over the resolution error, making the `true' error much larger. Further, notice that the error in amplitudes will typically be dominated only in certain modes. Nonetheless, despite being a rough estimate, Eq.~\eqref{eq:condition} provides a useful measure of how quickly the errors in the amplitudes can grow with increasing the number of overtones in our model. In Tab.~\ref{tab:amplitude-error} we show an example of the dependence of the relative error in the tone's amplitudes with the number of overtones for the simulation S7. We see that including 2 linear overtones for the modes of S7 can give significant errors in the amplitudes. However, adding a well-separated frequency such as a quadratic tone does not increase the condition number substantially. For instance, the bound for the $l=2$ shear mode of S7 with one linear overtone, and the quadratic  $(2,0)\times (2,0)$ tone is $5.5\times10^{-4}$, significantly smaller than the linear model including the $n=2$ overtone, as highlighted in Tab.~\ref{tab:amplitude-error}. 

The rapidly growing error bounds show how increasing the number of linear overtones to two or more increases the errors in our amplitudes. Consequently, even when the numerical resolution is many orders of magnitude below the mismatches of the model, adding more overtones gives us very large errors in the amplitudes. Therefore, we cannot try to dig deeper into the signal by continuing to increase the number of linear overtones. This prevents us from looking for the quadratic modes in the $l=8$ modes of the boosted simulations or the $l=3$ modes of the unboosted simulations.

\begin{table}[h]
     \centering
     \begin{tabular}{c|c|c c c c}
     \hline
       $l$ &$\frac{\epsilon_{240-120}}{\lVert \sigma_l \rVert}$  & & $\mathrm{Max}  \frac{\lVert \delta A^i \rVert}{\lVert A^i_{\mathrm{GR}} \rVert} $ & & \\
         & & $n=0$ & $n=1$ & $n=2$ & $n=3$\\
       \hline
       
        2 & $8.1\times 10^{-9}$ & $1.3\times 10^{-8}$ & $5.5\times 10^{-4}$   & 619 & $3.5\times 10^8$  \\
        4 & $7.2\times 10^{-8}$ & $9.1\times 10^{-8}$ & 0.012& $4.2\times 10^3$  & $3.0\times 10^9$\\
        6 & $5.0\times 10^{-7}$ & $5.8\times 10^{-7}$ & 0.11 & $2.7\times10^4$  & $1.2\times 10^{10}$ \\
        8& $2.3\times 10^{-6}$ & $2.6\times 10^{-6}$ & 0.59 & $1.2\times 10^5$ & $4.2\times 10^{10}$\\
        \hline
     \end{tabular}
     \caption{Table of estimated maximum relative errors in the norms of amplitude for different shear modes of S7, coming from the numerical resolution errors only. Systematic errors coming from the model are not taken into account. This bound is estimated using \eqref{eq:condition}. Here $\epsilon_{240-120}$ is the $L^2$ norm of the difference of shear $\sigma_l$ in the simulations with resolution 240 and 120 respectively, and $\lVert \sigma_l \rVert$ is the $L^2$ norm of the shear mode. $n$ is the number of linear overtones we include in the model. These values are for a starting time of $t_0=8.2M$.}
     \label{tab:amplitude-error}
 \end{table}

%% file: main.bbl
\begin{thebibliography}{69}%
\makeatletter
\providecommand \@ifxundefined [1]{%
 \@ifx{#1\undefined}
}%
\providecommand \@ifnum [1]{%
 \ifnum #1\expandafter \@firstoftwo
 \else \expandafter \@secondoftwo
 \fi
}%
\providecommand \@ifx [1]{%
 \ifx #1\expandafter \@firstoftwo
 \else \expandafter \@secondoftwo
 \fi
}%
\providecommand \natexlab [1]{#1}%
\providecommand \enquote  [1]{``#1''}%
\providecommand \bibnamefont  [1]{#1}%
\providecommand \bibfnamefont [1]{#1}%
\providecommand \citenamefont [1]{#1}%
\providecommand \href@noop [0]{\@secondoftwo}%
\providecommand \href [0]{\begingroup \@sanitize@url \@href}%
\providecommand \@href[1]{\@@startlink{#1}\@@href}%
\providecommand \@@href[1]{\endgroup#1\@@endlink}%
\providecommand \@sanitize@url [0]{\catcode `\\12\catcode `\$12\catcode `\&12\catcode `\#12\catcode `\^12\catcode `\_12\catcode `\%12\relax}%
\providecommand \@@startlink[1]{}%
\providecommand \@@endlink[0]{}%
\providecommand \url  [0]{\begingroup\@sanitize@url \@url }%
\providecommand \@url [1]{\endgroup\@href {#1}{\urlprefix }}%
\providecommand \urlprefix  [0]{URL }%
\providecommand \Eprint [0]{\href }%
\providecommand \doibase [0]{http://dx.doi.org/}%
\providecommand \selectlanguage [0]{\@gobble}%
\providecommand \bibinfo  [0]{\@secondoftwo}%
\providecommand \bibfield  [0]{\@secondoftwo}%
\providecommand \translation [1]{[#1]}%
\providecommand \BibitemOpen [0]{}%
\providecommand \bibitemStop [0]{}%
\providecommand \bibitemNoStop [0]{.\EOS\space}%
\providecommand \EOS [0]{\spacefactor3000\relax}%
\providecommand \BibitemShut  [1]{\csname bibitem#1\endcsname}%
\let\auto@bib@innerbib\@empty
%</preamble>
\bibitem [{\citenamefont {Abbott}\ \emph {et~al.}(2016{\natexlab{a}})\citenamefont {Abbott} \emph {et~al.}}]{LIGOScientific:2016vpg}%
  \BibitemOpen
  \bibfield  {author} {\bibinfo {author} {\bibfnamefont {B.~P.}\ \bibnamefont {Abbott}} \emph {et~al.} (\bibinfo {collaboration} {LIGO Scientific, Virgo}),\ }\href {\doibase 10.3847/2041-8205/818/2/L22} {\bibfield  {journal} {\bibinfo  {journal} {Astrophys. J. Lett.}\ }\textbf {\bibinfo {volume} {818}},\ \bibinfo {pages} {L22} (\bibinfo {year} {2016}{\natexlab{a}})},\ \Eprint {http://arxiv.org/abs/1602.03846} {arXiv:1602.03846 [astro-ph.HE]} \BibitemShut {NoStop}%
\bibitem [{\citenamefont {Abbott}\ \emph {et~al.}(2023)\citenamefont {Abbott} \emph {et~al.}}]{KAGRA:2021duu}%
  \BibitemOpen
  \bibfield  {author} {\bibinfo {author} {\bibfnamefont {R.}~\bibnamefont {Abbott}} \emph {et~al.} (\bibinfo {collaboration} {KAGRA, VIRGO, LIGO Scientific}),\ }\href {\doibase 10.1103/PhysRevX.13.011048} {\bibfield  {journal} {\bibinfo  {journal} {Phys. Rev. X}\ }\textbf {\bibinfo {volume} {13}},\ \bibinfo {pages} {011048} (\bibinfo {year} {2023})},\ \Eprint {http://arxiv.org/abs/2111.03634} {arXiv:2111.03634 [astro-ph.HE]} \BibitemShut {NoStop}%
\bibitem [{\citenamefont {Libanore}\ \emph {et~al.}(2023)\citenamefont {Libanore}, \citenamefont {Liguori},\ and\ \citenamefont {Raccanelli}}]{Libanore:2023ovr}%
  \BibitemOpen
  \bibfield  {author} {\bibinfo {author} {\bibfnamefont {S.}~\bibnamefont {Libanore}}, \bibinfo {author} {\bibfnamefont {M.}~\bibnamefont {Liguori}}, \ and\ \bibinfo {author} {\bibfnamefont {A.}~\bibnamefont {Raccanelli}},\ }\href@noop {} {\  (\bibinfo {year} {2023})},\ \Eprint {http://arxiv.org/abs/2306.03087} {arXiv:2306.03087 [astro-ph.CO]} \BibitemShut {NoStop}%
\bibitem [{\citenamefont {Nitz}\ \emph {et~al.}(2023)\citenamefont {Nitz}, \citenamefont {Kumar}, \citenamefont {Wang}, \citenamefont {Kastha}, \citenamefont {Wu}, \citenamefont {Sch\"afer}, \citenamefont {Dhurkunde},\ and\ \citenamefont {Capano}}]{Nitz:2021zwj}%
  \BibitemOpen
  \bibfield  {author} {\bibinfo {author} {\bibfnamefont {A.~H.}\ \bibnamefont {Nitz}}, \bibinfo {author} {\bibfnamefont {S.}~\bibnamefont {Kumar}}, \bibinfo {author} {\bibfnamefont {Y.-F.}\ \bibnamefont {Wang}}, \bibinfo {author} {\bibfnamefont {S.}~\bibnamefont {Kastha}}, \bibinfo {author} {\bibfnamefont {S.}~\bibnamefont {Wu}}, \bibinfo {author} {\bibfnamefont {M.}~\bibnamefont {Sch\"afer}}, \bibinfo {author} {\bibfnamefont {R.}~\bibnamefont {Dhurkunde}}, \ and\ \bibinfo {author} {\bibfnamefont {C.~D.}\ \bibnamefont {Capano}},\ }\href {\doibase 10.3847/1538-4357/aca591} {\bibfield  {journal} {\bibinfo  {journal} {Astrophys. J.}\ }\textbf {\bibinfo {volume} {946}},\ \bibinfo {pages} {59} (\bibinfo {year} {2023})},\ \Eprint {http://arxiv.org/abs/2112.06878} {arXiv:2112.06878 [astro-ph.HE]} \BibitemShut {NoStop}%
\bibitem [{\citenamefont {Diehl}\ \emph {et~al.}(2022)\citenamefont {Diehl}, \citenamefont {Korn}, \citenamefont {Leibundgut}, \citenamefont {Lugaro},\ and\ \citenamefont {Wallner}}]{Diehl:2022jnq}%
  \BibitemOpen
  \bibfield  {author} {\bibinfo {author} {\bibfnamefont {R.}~\bibnamefont {Diehl}}, \bibinfo {author} {\bibfnamefont {A.~J.}\ \bibnamefont {Korn}}, \bibinfo {author} {\bibfnamefont {B.}~\bibnamefont {Leibundgut}}, \bibinfo {author} {\bibfnamefont {M.}~\bibnamefont {Lugaro}}, \ and\ \bibinfo {author} {\bibfnamefont {A.}~\bibnamefont {Wallner}},\ }\href {\doibase 10.1016/j.ppnp.2022.103983} {\bibfield  {journal} {\bibinfo  {journal} {Prog. Part. Nucl. Phys.}\ }\textbf {\bibinfo {volume} {127}},\ \bibinfo {pages} {103983} (\bibinfo {year} {2022})},\ \Eprint {http://arxiv.org/abs/2206.12246} {arXiv:2206.12246 [astro-ph.HE]} \BibitemShut {NoStop}%
\bibitem [{\citenamefont {Berti}\ \emph {et~al.}(2018)\citenamefont {Berti}, \citenamefont {Yagi}, \citenamefont {Yang},\ and\ \citenamefont {Yunes}}]{Berti:2018vdi}%
  \BibitemOpen
  \bibfield  {author} {\bibinfo {author} {\bibfnamefont {E.}~\bibnamefont {Berti}}, \bibinfo {author} {\bibfnamefont {K.}~\bibnamefont {Yagi}}, \bibinfo {author} {\bibfnamefont {H.}~\bibnamefont {Yang}}, \ and\ \bibinfo {author} {\bibfnamefont {N.}~\bibnamefont {Yunes}},\ }\href {\doibase 10.1007/s10714-018-2372-6} {\bibfield  {journal} {\bibinfo  {journal} {Gen. Rel. Grav.}\ }\textbf {\bibinfo {volume} {50}},\ \bibinfo {pages} {49} (\bibinfo {year} {2018})},\ \Eprint {http://arxiv.org/abs/1801.03587} {arXiv:1801.03587 [gr-qc]} \BibitemShut {NoStop}%
\bibitem [{\citenamefont {Abbott}\ \emph {et~al.}(2021)\citenamefont {Abbott} \emph {et~al.}}]{LIGOScientific:2021sio}%
  \BibitemOpen
  \bibfield  {author} {\bibinfo {author} {\bibfnamefont {R.}~\bibnamefont {Abbott}} \emph {et~al.} (\bibinfo {collaboration} {LIGO Scientific, VIRGO, KAGRA}),\ }\href@noop {} {\  (\bibinfo {year} {2021})},\ \Eprint {http://arxiv.org/abs/2112.06861} {arXiv:2112.06861 [gr-qc]} \BibitemShut {NoStop}%
\bibitem [{\citenamefont {Abbott}\ \emph {et~al.}(2019)\citenamefont {Abbott} \emph {et~al.}}]{LIGOScientific:2019fpa}%
  \BibitemOpen
  \bibfield  {author} {\bibinfo {author} {\bibfnamefont {B.~P.}\ \bibnamefont {Abbott}} \emph {et~al.} (\bibinfo {collaboration} {LIGO Scientific, Virgo}),\ }\href {\doibase 10.1103/PhysRevD.100.104036} {\bibfield  {journal} {\bibinfo  {journal} {Phys. Rev. D}\ }\textbf {\bibinfo {volume} {100}},\ \bibinfo {pages} {104036} (\bibinfo {year} {2019})},\ \Eprint {http://arxiv.org/abs/1903.04467} {arXiv:1903.04467 [gr-qc]} \BibitemShut {NoStop}%
\bibitem [{\citenamefont {Kastha}\ \emph {et~al.}(2022)\citenamefont {Kastha}, \citenamefont {Capano}, \citenamefont {Westerweck}, \citenamefont {Cabero}, \citenamefont {Krishnan},\ and\ \citenamefont {Nielsen}}]{Kastha:2021chr}%
  \BibitemOpen
  \bibfield  {author} {\bibinfo {author} {\bibfnamefont {S.}~\bibnamefont {Kastha}}, \bibinfo {author} {\bibfnamefont {C.~D.}\ \bibnamefont {Capano}}, \bibinfo {author} {\bibfnamefont {J.}~\bibnamefont {Westerweck}}, \bibinfo {author} {\bibfnamefont {M.}~\bibnamefont {Cabero}}, \bibinfo {author} {\bibfnamefont {B.}~\bibnamefont {Krishnan}}, \ and\ \bibinfo {author} {\bibfnamefont {A.~B.}\ \bibnamefont {Nielsen}},\ }\href {\doibase 10.1103/PhysRevD.105.064042} {\bibfield  {journal} {\bibinfo  {journal} {Phys. Rev. D}\ }\textbf {\bibinfo {volume} {105}},\ \bibinfo {pages} {064042} (\bibinfo {year} {2022})},\ \Eprint {http://arxiv.org/abs/2111.13664} {arXiv:2111.13664 [gr-qc]} \BibitemShut {NoStop}%
\bibitem [{\citenamefont {Isi}\ \emph {et~al.}(2019)\citenamefont {Isi}, \citenamefont {Giesler}, \citenamefont {Farr}, \citenamefont {Scheel},\ and\ \citenamefont {Teukolsky}}]{Isi:2019aib}%
  \BibitemOpen
  \bibfield  {author} {\bibinfo {author} {\bibfnamefont {M.}~\bibnamefont {Isi}}, \bibinfo {author} {\bibfnamefont {M.}~\bibnamefont {Giesler}}, \bibinfo {author} {\bibfnamefont {W.~M.}\ \bibnamefont {Farr}}, \bibinfo {author} {\bibfnamefont {M.~A.}\ \bibnamefont {Scheel}}, \ and\ \bibinfo {author} {\bibfnamefont {S.~A.}\ \bibnamefont {Teukolsky}},\ }\href {\doibase 10.1103/PhysRevLett.123.111102} {\bibfield  {journal} {\bibinfo  {journal} {Phys. Rev. Lett.}\ }\textbf {\bibinfo {volume} {123}},\ \bibinfo {pages} {111102} (\bibinfo {year} {2019})},\ \Eprint {http://arxiv.org/abs/1905.00869} {arXiv:1905.00869 [gr-qc]} \BibitemShut {NoStop}%
\bibitem [{\citenamefont {Isi}\ \emph {et~al.}(2021)\citenamefont {Isi}, \citenamefont {Farr}, \citenamefont {Giesler}, \citenamefont {Scheel},\ and\ \citenamefont {Teukolsky}}]{Isi:2020tac}%
  \BibitemOpen
  \bibfield  {author} {\bibinfo {author} {\bibfnamefont {M.}~\bibnamefont {Isi}}, \bibinfo {author} {\bibfnamefont {W.~M.}\ \bibnamefont {Farr}}, \bibinfo {author} {\bibfnamefont {M.}~\bibnamefont {Giesler}}, \bibinfo {author} {\bibfnamefont {M.~A.}\ \bibnamefont {Scheel}}, \ and\ \bibinfo {author} {\bibfnamefont {S.~A.}\ \bibnamefont {Teukolsky}},\ }\href {\doibase 10.1103/PhysRevLett.127.011103} {\bibfield  {journal} {\bibinfo  {journal} {Phys. Rev. Lett.}\ }\textbf {\bibinfo {volume} {127}},\ \bibinfo {pages} {011103} (\bibinfo {year} {2021})},\ \Eprint {http://arxiv.org/abs/2012.04486} {arXiv:2012.04486 [gr-qc]} \BibitemShut {NoStop}%
\bibitem [{\citenamefont {Forteza}\ \emph {et~al.}(2023)\citenamefont {Forteza}, \citenamefont {Bhagwat}, \citenamefont {Kumar},\ and\ \citenamefont {Pani}}]{Forteza:2022tgq}%
  \BibitemOpen
  \bibfield  {author} {\bibinfo {author} {\bibfnamefont {X.~J.}\ \bibnamefont {Forteza}}, \bibinfo {author} {\bibfnamefont {S.}~\bibnamefont {Bhagwat}}, \bibinfo {author} {\bibfnamefont {S.}~\bibnamefont {Kumar}}, \ and\ \bibinfo {author} {\bibfnamefont {P.}~\bibnamefont {Pani}},\ }\href {\doibase 10.1103/PhysRevLett.130.021001} {\bibfield  {journal} {\bibinfo  {journal} {Phys. Rev. Lett.}\ }\textbf {\bibinfo {volume} {130}},\ \bibinfo {pages} {021001} (\bibinfo {year} {2023})},\ \Eprint {http://arxiv.org/abs/2205.14910} {arXiv:2205.14910 [gr-qc]} \BibitemShut {NoStop}%
\bibitem [{\citenamefont {Berti}\ \emph {et~al.}(2016)\citenamefont {Berti}, \citenamefont {Sesana}, \citenamefont {Barausse}, \citenamefont {Cardoso},\ and\ \citenamefont {Belczynski}}]{Berti:2016lat}%
  \BibitemOpen
  \bibfield  {author} {\bibinfo {author} {\bibfnamefont {E.}~\bibnamefont {Berti}}, \bibinfo {author} {\bibfnamefont {A.}~\bibnamefont {Sesana}}, \bibinfo {author} {\bibfnamefont {E.}~\bibnamefont {Barausse}}, \bibinfo {author} {\bibfnamefont {V.}~\bibnamefont {Cardoso}}, \ and\ \bibinfo {author} {\bibfnamefont {K.}~\bibnamefont {Belczynski}},\ }\href {\doibase 10.1103/PhysRevLett.117.101102} {\bibfield  {journal} {\bibinfo  {journal} {Phys. Rev. Lett.}\ }\textbf {\bibinfo {volume} {117}},\ \bibinfo {pages} {101102} (\bibinfo {year} {2016})},\ \Eprint {http://arxiv.org/abs/1605.09286} {arXiv:1605.09286 [gr-qc]} \BibitemShut {NoStop}%
\bibitem [{\citenamefont {Yang}\ \emph {et~al.}(2017)\citenamefont {Yang}, \citenamefont {Yagi}, \citenamefont {Blackman}, \citenamefont {Lehner}, \citenamefont {Paschalidis}, \citenamefont {Pretorius},\ and\ \citenamefont {Yunes}}]{Yang:2017zxs}%
  \BibitemOpen
  \bibfield  {author} {\bibinfo {author} {\bibfnamefont {H.}~\bibnamefont {Yang}}, \bibinfo {author} {\bibfnamefont {K.}~\bibnamefont {Yagi}}, \bibinfo {author} {\bibfnamefont {J.}~\bibnamefont {Blackman}}, \bibinfo {author} {\bibfnamefont {L.}~\bibnamefont {Lehner}}, \bibinfo {author} {\bibfnamefont {V.}~\bibnamefont {Paschalidis}}, \bibinfo {author} {\bibfnamefont {F.}~\bibnamefont {Pretorius}}, \ and\ \bibinfo {author} {\bibfnamefont {N.}~\bibnamefont {Yunes}},\ }\href {\doibase 10.1103/PhysRevLett.118.161101} {\bibfield  {journal} {\bibinfo  {journal} {Phys. Rev. Lett.}\ }\textbf {\bibinfo {volume} {118}},\ \bibinfo {pages} {161101} (\bibinfo {year} {2017})},\ \Eprint {http://arxiv.org/abs/1701.05808} {arXiv:1701.05808 [gr-qc]} \BibitemShut {NoStop}%
\bibitem [{\citenamefont {Ma}\ \emph {et~al.}(2023)\citenamefont {Ma}, \citenamefont {Sun},\ and\ \citenamefont {Chen}}]{Ma:2023cwe}%
  \BibitemOpen
  \bibfield  {author} {\bibinfo {author} {\bibfnamefont {S.}~\bibnamefont {Ma}}, \bibinfo {author} {\bibfnamefont {L.}~\bibnamefont {Sun}}, \ and\ \bibinfo {author} {\bibfnamefont {Y.}~\bibnamefont {Chen}},\ }\href {\doibase 10.1103/PhysRevLett.130.141401} {\bibfield  {journal} {\bibinfo  {journal} {Phys. Rev. Lett.}\ }\textbf {\bibinfo {volume} {130}},\ \bibinfo {pages} {141401} (\bibinfo {year} {2023})},\ \Eprint {http://arxiv.org/abs/2301.06705} {arXiv:2301.06705 [gr-qc]} \BibitemShut {NoStop}%
\bibitem [{\citenamefont {Detweiler}(1980)}]{Detweiler:1980gk}%
  \BibitemOpen
  \bibfield  {author} {\bibinfo {author} {\bibfnamefont {S.~L.}\ \bibnamefont {Detweiler}},\ }\href {\doibase 10.1086/158109} {\bibfield  {journal} {\bibinfo  {journal} {Astrophys. J.}\ }\textbf {\bibinfo {volume} {239}},\ \bibinfo {pages} {292} (\bibinfo {year} {1980})}\BibitemShut {NoStop}%
\bibitem [{\citenamefont {Dreyer}\ \emph {et~al.}(2004)\citenamefont {Dreyer}, \citenamefont {Kelly}, \citenamefont {Krishnan}, \citenamefont {Finn}, \citenamefont {Garrison},\ and\ \citenamefont {Lopez-Aleman}}]{Dreyer:2003bv}%
  \BibitemOpen
  \bibfield  {author} {\bibinfo {author} {\bibfnamefont {O.}~\bibnamefont {Dreyer}}, \bibinfo {author} {\bibfnamefont {B.~J.}\ \bibnamefont {Kelly}}, \bibinfo {author} {\bibfnamefont {B.}~\bibnamefont {Krishnan}}, \bibinfo {author} {\bibfnamefont {L.~S.}\ \bibnamefont {Finn}}, \bibinfo {author} {\bibfnamefont {D.}~\bibnamefont {Garrison}}, \ and\ \bibinfo {author} {\bibfnamefont {R.}~\bibnamefont {Lopez-Aleman}},\ }\href {\doibase 10.1088/0264-9381/21/4/003} {\bibfield  {journal} {\bibinfo  {journal} {Class. Quant. Grav.}\ }\textbf {\bibinfo {volume} {21}},\ \bibinfo {pages} {787} (\bibinfo {year} {2004})},\ \Eprint {http://arxiv.org/abs/gr-qc/0309007} {arXiv:gr-qc/0309007} \BibitemShut {NoStop}%
\bibitem [{\citenamefont {Berti}\ \emph {et~al.}(2006)\citenamefont {Berti}, \citenamefont {Cardoso},\ and\ \citenamefont {Will}}]{Berti:2006cc}%
  \BibitemOpen
  \bibfield  {author} {\bibinfo {author} {\bibfnamefont {E.}~\bibnamefont {Berti}}, \bibinfo {author} {\bibfnamefont {V.}~\bibnamefont {Cardoso}}, \ and\ \bibinfo {author} {\bibfnamefont {C.~M.}\ \bibnamefont {Will}},\ }\href {\doibase 10.1063/1.2405024} {\bibfield  {journal} {\bibinfo  {journal} {AIP Conf. Proc.}\ }\textbf {\bibinfo {volume} {873}},\ \bibinfo {pages} {82} (\bibinfo {year} {2006})}\BibitemShut {NoStop}%
\bibitem [{\citenamefont {Nee}\ \emph {et~al.}(2023)\citenamefont {Nee}, \citenamefont {V\"olkel},\ and\ \citenamefont {Pfeiffer}}]{Nee:2023osy}%
  \BibitemOpen
  \bibfield  {author} {\bibinfo {author} {\bibfnamefont {P.~J.}\ \bibnamefont {Nee}}, \bibinfo {author} {\bibfnamefont {S.~H.}\ \bibnamefont {V\"olkel}}, \ and\ \bibinfo {author} {\bibfnamefont {H.~P.}\ \bibnamefont {Pfeiffer}},\ }\href@noop {} {\  (\bibinfo {year} {2023})},\ \Eprint {http://arxiv.org/abs/2302.06634} {arXiv:2302.06634 [gr-qc]} \BibitemShut {NoStop}%
\bibitem [{\citenamefont {Baibhav}\ \emph {et~al.}(2023)\citenamefont {Baibhav}, \citenamefont {Cheung}, \citenamefont {Berti}, \citenamefont {Cardoso}, \citenamefont {Carullo}, \citenamefont {Cotesta}, \citenamefont {Del~Pozzo},\ and\ \citenamefont {Duque}}]{Baibhav:2023clw}%
  \BibitemOpen
  \bibfield  {author} {\bibinfo {author} {\bibfnamefont {V.}~\bibnamefont {Baibhav}}, \bibinfo {author} {\bibfnamefont {M.~H.-Y.}\ \bibnamefont {Cheung}}, \bibinfo {author} {\bibfnamefont {E.}~\bibnamefont {Berti}}, \bibinfo {author} {\bibfnamefont {V.}~\bibnamefont {Cardoso}}, \bibinfo {author} {\bibfnamefont {G.}~\bibnamefont {Carullo}}, \bibinfo {author} {\bibfnamefont {R.}~\bibnamefont {Cotesta}}, \bibinfo {author} {\bibfnamefont {W.}~\bibnamefont {Del~Pozzo}}, \ and\ \bibinfo {author} {\bibfnamefont {F.}~\bibnamefont {Duque}},\ }\href@noop {} {\  (\bibinfo {year} {2023})},\ \Eprint {http://arxiv.org/abs/2302.03050} {arXiv:2302.03050 [gr-qc]} \BibitemShut {NoStop}%
\bibitem [{\citenamefont {Teukolsky}(1973)}]{Teukolsky:1973ha}%
  \BibitemOpen
  \bibfield  {author} {\bibinfo {author} {\bibfnamefont {S.~A.}\ \bibnamefont {Teukolsky}},\ }\href {\doibase 10.1086/152444} {\bibfield  {journal} {\bibinfo  {journal} {Astrophys. J.}\ }\textbf {\bibinfo {volume} {185}},\ \bibinfo {pages} {635} (\bibinfo {year} {1973})}\BibitemShut {NoStop}%
\bibitem [{\citenamefont {Leaver}(1985)}]{Leaver:1985ax}%
  \BibitemOpen
  \bibfield  {author} {\bibinfo {author} {\bibfnamefont {E.~W.}\ \bibnamefont {Leaver}},\ }\href {\doibase 10.1098/rspa.1985.0119} {\bibfield  {journal} {\bibinfo  {journal} {Proc. Roy. Soc. Lond. A}\ }\textbf {\bibinfo {volume} {402}},\ \bibinfo {pages} {285} (\bibinfo {year} {1985})}\BibitemShut {NoStop}%
\bibitem [{\citenamefont {Berti}\ \emph {et~al.}(2009)\citenamefont {Berti}, \citenamefont {Cardoso},\ and\ \citenamefont {Starinets}}]{Berti:2009kk}%
  \BibitemOpen
  \bibfield  {author} {\bibinfo {author} {\bibfnamefont {E.}~\bibnamefont {Berti}}, \bibinfo {author} {\bibfnamefont {V.}~\bibnamefont {Cardoso}}, \ and\ \bibinfo {author} {\bibfnamefont {A.~O.}\ \bibnamefont {Starinets}},\ }\href {\doibase 10.1088/0264-9381/26/16/163001} {\bibfield  {journal} {\bibinfo  {journal} {Class. Quant. Grav.}\ }\textbf {\bibinfo {volume} {26}},\ \bibinfo {pages} {163001} (\bibinfo {year} {2009})},\ \Eprint {http://arxiv.org/abs/0905.2975} {arXiv:0905.2975 [gr-qc]} \BibitemShut {NoStop}%
\bibitem [{\citenamefont {Yang}\ \emph {et~al.}(2012)\citenamefont {Yang}, \citenamefont {Nichols}, \citenamefont {Zhang}, \citenamefont {Zimmerman}, \citenamefont {Zhang},\ and\ \citenamefont {Chen}}]{Yang:2012he}%
  \BibitemOpen
  \bibfield  {author} {\bibinfo {author} {\bibfnamefont {H.}~\bibnamefont {Yang}}, \bibinfo {author} {\bibfnamefont {D.~A.}\ \bibnamefont {Nichols}}, \bibinfo {author} {\bibfnamefont {F.}~\bibnamefont {Zhang}}, \bibinfo {author} {\bibfnamefont {A.}~\bibnamefont {Zimmerman}}, \bibinfo {author} {\bibfnamefont {Z.}~\bibnamefont {Zhang}}, \ and\ \bibinfo {author} {\bibfnamefont {Y.}~\bibnamefont {Chen}},\ }\href {\doibase 10.1103/PhysRevD.86.104006} {\bibfield  {journal} {\bibinfo  {journal} {Phys. Rev. D}\ }\textbf {\bibinfo {volume} {86}},\ \bibinfo {pages} {104006} (\bibinfo {year} {2012})},\ \Eprint {http://arxiv.org/abs/1207.4253} {arXiv:1207.4253 [gr-qc]} \BibitemShut {NoStop}%
\bibitem [{\citenamefont {Yang}\ \emph {et~al.}(2013)\citenamefont {Yang}, \citenamefont {Zimmerman}, \citenamefont {Zengino\u{g}lu}, \citenamefont {Zhang}, \citenamefont {Berti},\ and\ \citenamefont {Chen}}]{Yang:2013uba}%
  \BibitemOpen
  \bibfield  {author} {\bibinfo {author} {\bibfnamefont {H.}~\bibnamefont {Yang}}, \bibinfo {author} {\bibfnamefont {A.}~\bibnamefont {Zimmerman}}, \bibinfo {author} {\bibfnamefont {A.}~\bibnamefont {Zengino\u{g}lu}}, \bibinfo {author} {\bibfnamefont {F.}~\bibnamefont {Zhang}}, \bibinfo {author} {\bibfnamefont {E.}~\bibnamefont {Berti}}, \ and\ \bibinfo {author} {\bibfnamefont {Y.}~\bibnamefont {Chen}},\ }\href {\doibase 10.1103/PhysRevD.88.044047} {\bibfield  {journal} {\bibinfo  {journal} {Phys. Rev. D}\ }\textbf {\bibinfo {volume} {88}},\ \bibinfo {pages} {044047} (\bibinfo {year} {2013})},\ \Eprint {http://arxiv.org/abs/1307.8086} {arXiv:1307.8086 [gr-qc]} \BibitemShut {NoStop}%
\bibitem [{\citenamefont {Chandrasekhar}\ and\ \citenamefont {Detweiler}(1975)}]{Chandrasekhar:1975zza}%
  \BibitemOpen
  \bibfield  {author} {\bibinfo {author} {\bibfnamefont {S.}~\bibnamefont {Chandrasekhar}}\ and\ \bibinfo {author} {\bibfnamefont {S.~L.}\ \bibnamefont {Detweiler}},\ }\href {\doibase 10.1098/rspa.1975.0112} {\bibfield  {journal} {\bibinfo  {journal} {Proc. Roy. Soc. Lond. A}\ }\textbf {\bibinfo {volume} {344}},\ \bibinfo {pages} {441} (\bibinfo {year} {1975})}\BibitemShut {NoStop}%
\bibitem [{\citenamefont {Abbott}\ \emph {et~al.}(2016{\natexlab{b}})\citenamefont {Abbott} \emph {et~al.}}]{LIGOScientific:2016lio}%
  \BibitemOpen
  \bibfield  {author} {\bibinfo {author} {\bibfnamefont {B.~P.}\ \bibnamefont {Abbott}} \emph {et~al.} (\bibinfo {collaboration} {LIGO Scientific, Virgo}),\ }\href {\doibase 10.1103/PhysRevLett.116.221101} {\bibfield  {journal} {\bibinfo  {journal} {Phys. Rev. Lett.}\ }\textbf {\bibinfo {volume} {116}},\ \bibinfo {pages} {221101} (\bibinfo {year} {2016}{\natexlab{b}})},\ \bibinfo {note} {[Erratum: Phys.Rev.Lett. 121, 129902 (2018)]},\ \Eprint {http://arxiv.org/abs/1602.03841} {arXiv:1602.03841 [gr-qc]} \BibitemShut {NoStop}%
\bibitem [{\citenamefont {Capano}\ \emph {et~al.}(2021)\citenamefont {Capano}, \citenamefont {Cabero}, \citenamefont {Westerweck}, \citenamefont {Abedi}, \citenamefont {Kastha}, \citenamefont {Nitz}, \citenamefont {Wang}, \citenamefont {Nielsen},\ and\ \citenamefont {Krishnan}}]{Capano:2021etf}%
  \BibitemOpen
  \bibfield  {author} {\bibinfo {author} {\bibfnamefont {C.~D.}\ \bibnamefont {Capano}}, \bibinfo {author} {\bibfnamefont {M.}~\bibnamefont {Cabero}}, \bibinfo {author} {\bibfnamefont {J.}~\bibnamefont {Westerweck}}, \bibinfo {author} {\bibfnamefont {J.}~\bibnamefont {Abedi}}, \bibinfo {author} {\bibfnamefont {S.}~\bibnamefont {Kastha}}, \bibinfo {author} {\bibfnamefont {A.~H.}\ \bibnamefont {Nitz}}, \bibinfo {author} {\bibfnamefont {Y.-F.}\ \bibnamefont {Wang}}, \bibinfo {author} {\bibfnamefont {A.~B.}\ \bibnamefont {Nielsen}}, \ and\ \bibinfo {author} {\bibfnamefont {B.}~\bibnamefont {Krishnan}},\ }\href@noop {} {\  (\bibinfo {year} {2021})},\ \Eprint {http://arxiv.org/abs/2105.05238} {arXiv:2105.05238 [gr-qc]} \BibitemShut {NoStop}%
\bibitem [{\citenamefont {Giesler}\ \emph {et~al.}(2019)\citenamefont {Giesler}, \citenamefont {Isi}, \citenamefont {Scheel},\ and\ \citenamefont {Teukolsky}}]{Giesler_2019}%
  \BibitemOpen
  \bibfield  {author} {\bibinfo {author} {\bibfnamefont {M.}~\bibnamefont {Giesler}}, \bibinfo {author} {\bibfnamefont {M.}~\bibnamefont {Isi}}, \bibinfo {author} {\bibfnamefont {M.~A.}\ \bibnamefont {Scheel}}, \ and\ \bibinfo {author} {\bibfnamefont {S.~A.}\ \bibnamefont {Teukolsky}},\ }\href {\doibase 10.1103/physrevx.9.041060} {\bibfield  {journal} {\bibinfo  {journal} {Physical Review X}\ }\textbf {\bibinfo {volume} {9}} (\bibinfo {year} {2019}),\ 10.1103/physrevx.9.041060}\BibitemShut {NoStop}%
\bibitem [{\citenamefont {Dhani}(2021)}]{Dhani:2020nik}%
  \BibitemOpen
  \bibfield  {author} {\bibinfo {author} {\bibfnamefont {A.}~\bibnamefont {Dhani}},\ }\href {\doibase 10.1103/PhysRevD.103.104048} {\bibfield  {journal} {\bibinfo  {journal} {Phys. Rev. D}\ }\textbf {\bibinfo {volume} {103}},\ \bibinfo {pages} {104048} (\bibinfo {year} {2021})},\ \Eprint {http://arxiv.org/abs/2010.08602} {arXiv:2010.08602 [gr-qc]} \BibitemShut {NoStop}%
\bibitem [{\citenamefont {Li}\ \emph {et~al.}(2022)\citenamefont {Li}, \citenamefont {Sun}, \citenamefont {Lo}, \citenamefont {Payne},\ and\ \citenamefont {Chen}}]{Li:2021wgz}%
  \BibitemOpen
  \bibfield  {author} {\bibinfo {author} {\bibfnamefont {X.}~\bibnamefont {Li}}, \bibinfo {author} {\bibfnamefont {L.}~\bibnamefont {Sun}}, \bibinfo {author} {\bibfnamefont {R.~K.~L.}\ \bibnamefont {Lo}}, \bibinfo {author} {\bibfnamefont {E.}~\bibnamefont {Payne}}, \ and\ \bibinfo {author} {\bibfnamefont {Y.}~\bibnamefont {Chen}},\ }\href {\doibase 10.1103/PhysRevD.105.024016} {\bibfield  {journal} {\bibinfo  {journal} {Phys. Rev. D}\ }\textbf {\bibinfo {volume} {105}},\ \bibinfo {pages} {024016} (\bibinfo {year} {2022})},\ \Eprint {http://arxiv.org/abs/2110.03116} {arXiv:2110.03116 [gr-qc]} \BibitemShut {NoStop}%
\bibitem [{\citenamefont {Zertuche}\ \emph {et~al.}(2022)\citenamefont {Zertuche}, \citenamefont {Mitman}, \citenamefont {Khera}, \citenamefont {Stein}, \citenamefont {Boyle}, \citenamefont {Deppe}, \citenamefont {H{\'e}bert}, \citenamefont {Iozzo}, \citenamefont {Kidder}, \citenamefont {Moxon} \emph {et~al.}}]{MaganaZertuche:2021syq}%
  \BibitemOpen
  \bibfield  {author} {\bibinfo {author} {\bibfnamefont {L.~M.}\ \bibnamefont {Zertuche}}, \bibinfo {author} {\bibfnamefont {K.}~\bibnamefont {Mitman}}, \bibinfo {author} {\bibfnamefont {N.}~\bibnamefont {Khera}}, \bibinfo {author} {\bibfnamefont {L.~C.}\ \bibnamefont {Stein}}, \bibinfo {author} {\bibfnamefont {M.}~\bibnamefont {Boyle}}, \bibinfo {author} {\bibfnamefont {N.}~\bibnamefont {Deppe}}, \bibinfo {author} {\bibfnamefont {F.}~\bibnamefont {H{\'e}bert}}, \bibinfo {author} {\bibfnamefont {D.~A.}\ \bibnamefont {Iozzo}}, \bibinfo {author} {\bibfnamefont {L.~E.}\ \bibnamefont {Kidder}}, \bibinfo {author} {\bibfnamefont {J.}~\bibnamefont {Moxon}},  \emph {et~al.},\ }\href {\doibase 10.1103/PhysRevD.105.104015} {\bibfield  {journal} {\bibinfo  {journal} {Phys. Rev. D}\ }\textbf {\bibinfo {volume} {105}},\ \bibinfo {pages} {104015} (\bibinfo {year} {2022})},\ \Eprint {http://arxiv.org/abs/2110.15922} {arXiv:2110.15922 [gr-qc]} \BibitemShut {NoStop}%
\bibitem [{\citenamefont {Campanelli}\ and\ \citenamefont {Lousto}(1999)}]{Campanelli:1998jv}%
  \BibitemOpen
  \bibfield  {author} {\bibinfo {author} {\bibfnamefont {M.}~\bibnamefont {Campanelli}}\ and\ \bibinfo {author} {\bibfnamefont {C.~O.}\ \bibnamefont {Lousto}},\ }\href {\doibase 10.1103/PhysRevD.59.124022} {\bibfield  {journal} {\bibinfo  {journal} {Phys. Rev. D}\ }\textbf {\bibinfo {volume} {59}},\ \bibinfo {pages} {124022} (\bibinfo {year} {1999})},\ \Eprint {http://arxiv.org/abs/gr-qc/9811019} {arXiv:gr-qc/9811019} \BibitemShut {NoStop}%
\bibitem [{\citenamefont {Yang}\ \emph {et~al.}(2015{\natexlab{a}})\citenamefont {Yang}, \citenamefont {Zimmerman},\ and\ \citenamefont {Lehner}}]{Yang:2014tla}%
  \BibitemOpen
  \bibfield  {author} {\bibinfo {author} {\bibfnamefont {H.}~\bibnamefont {Yang}}, \bibinfo {author} {\bibfnamefont {A.}~\bibnamefont {Zimmerman}}, \ and\ \bibinfo {author} {\bibfnamefont {L.}~\bibnamefont {Lehner}},\ }\href {\doibase 10.1103/PhysRevLett.114.081101} {\bibfield  {journal} {\bibinfo  {journal} {Phys. Rev. Lett.}\ }\textbf {\bibinfo {volume} {114}},\ \bibinfo {pages} {081101} (\bibinfo {year} {2015}{\natexlab{a}})},\ \Eprint {http://arxiv.org/abs/1402.4859} {arXiv:1402.4859 [gr-qc]} \BibitemShut {NoStop}%
\bibitem [{\citenamefont {Yang}\ \emph {et~al.}(2015{\natexlab{b}})\citenamefont {Yang}, \citenamefont {Zhang}, \citenamefont {Green},\ and\ \citenamefont {Lehner}}]{Yang:2015jja}%
  \BibitemOpen
  \bibfield  {author} {\bibinfo {author} {\bibfnamefont {H.}~\bibnamefont {Yang}}, \bibinfo {author} {\bibfnamefont {F.}~\bibnamefont {Zhang}}, \bibinfo {author} {\bibfnamefont {S.~R.}\ \bibnamefont {Green}}, \ and\ \bibinfo {author} {\bibfnamefont {L.}~\bibnamefont {Lehner}},\ }\href {\doibase 10.1103/PhysRevD.91.084007} {\bibfield  {journal} {\bibinfo  {journal} {Phys. Rev. D}\ }\textbf {\bibinfo {volume} {91}},\ \bibinfo {pages} {084007} (\bibinfo {year} {2015}{\natexlab{b}})},\ \Eprint {http://arxiv.org/abs/1502.08051} {arXiv:1502.08051 [gr-qc]} \BibitemShut {NoStop}%
\bibitem [{\citenamefont {Ripley}\ \emph {et~al.}(2021)\citenamefont {Ripley}, \citenamefont {Loutrel}, \citenamefont {Giorgi},\ and\ \citenamefont {Pretorius}}]{Ripley:2020xby}%
  \BibitemOpen
  \bibfield  {author} {\bibinfo {author} {\bibfnamefont {J.~L.}\ \bibnamefont {Ripley}}, \bibinfo {author} {\bibfnamefont {N.}~\bibnamefont {Loutrel}}, \bibinfo {author} {\bibfnamefont {E.}~\bibnamefont {Giorgi}}, \ and\ \bibinfo {author} {\bibfnamefont {F.}~\bibnamefont {Pretorius}},\ }\href {\doibase 10.1103/PhysRevD.103.104018} {\bibfield  {journal} {\bibinfo  {journal} {Phys. Rev. D}\ }\textbf {\bibinfo {volume} {103}},\ \bibinfo {pages} {104018} (\bibinfo {year} {2021})},\ \Eprint {http://arxiv.org/abs/2010.00162} {arXiv:2010.00162 [gr-qc]} \BibitemShut {NoStop}%
\bibitem [{\citenamefont {Sberna}\ \emph {et~al.}(2022)\citenamefont {Sberna}, \citenamefont {Bosch}, \citenamefont {East}, \citenamefont {Green},\ and\ \citenamefont {Lehner}}]{Sberna:2021eui}%
  \BibitemOpen
  \bibfield  {author} {\bibinfo {author} {\bibfnamefont {L.}~\bibnamefont {Sberna}}, \bibinfo {author} {\bibfnamefont {P.}~\bibnamefont {Bosch}}, \bibinfo {author} {\bibfnamefont {W.~E.}\ \bibnamefont {East}}, \bibinfo {author} {\bibfnamefont {S.~R.}\ \bibnamefont {Green}}, \ and\ \bibinfo {author} {\bibfnamefont {L.}~\bibnamefont {Lehner}},\ }\href {\doibase 10.1103/PhysRevD.105.064046} {\bibfield  {journal} {\bibinfo  {journal} {Phys. Rev. D}\ }\textbf {\bibinfo {volume} {105}},\ \bibinfo {pages} {064046} (\bibinfo {year} {2022})},\ \Eprint {http://arxiv.org/abs/2112.11168} {arXiv:2112.11168 [gr-qc]} \BibitemShut {NoStop}%
\bibitem [{\citenamefont {Redondo-Yuste}\ and\ \citenamefont {Lehner}(2023)}]{Redondo-Yuste:2022czg}%
  \BibitemOpen
  \bibfield  {author} {\bibinfo {author} {\bibfnamefont {J.}~\bibnamefont {Redondo-Yuste}}\ and\ \bibinfo {author} {\bibfnamefont {L.}~\bibnamefont {Lehner}},\ }\href {\doibase 10.1007/JHEP02(2023)240} {\bibfield  {journal} {\bibinfo  {journal} {JHEP}\ }\textbf {\bibinfo {volume} {02}},\ \bibinfo {pages} {240} (\bibinfo {year} {2023})},\ \Eprint {http://arxiv.org/abs/2212.06175} {arXiv:2212.06175 [gr-qc]} \BibitemShut {NoStop}%
\bibitem [{\citenamefont {Cheung}\ \emph {et~al.}(2023)\citenamefont {Cheung} \emph {et~al.}}]{Cheung:2022rbm}%
  \BibitemOpen
  \bibfield  {author} {\bibinfo {author} {\bibfnamefont {M.~H.-Y.}\ \bibnamefont {Cheung}} \emph {et~al.},\ }\href {\doibase 10.1103/PhysRevLett.130.081401} {\bibfield  {journal} {\bibinfo  {journal} {Phys. Rev. Lett.}\ }\textbf {\bibinfo {volume} {130}},\ \bibinfo {pages} {081401} (\bibinfo {year} {2023})},\ \Eprint {http://arxiv.org/abs/2208.07374} {arXiv:2208.07374 [gr-qc]} \BibitemShut {NoStop}%
\bibitem [{\citenamefont {Mitman}\ \emph {et~al.}(2023)\citenamefont {Mitman} \emph {et~al.}}]{Mitman:2022qdl}%
  \BibitemOpen
  \bibfield  {author} {\bibinfo {author} {\bibfnamefont {K.}~\bibnamefont {Mitman}} \emph {et~al.},\ }\href {\doibase 10.1103/PhysRevLett.130.081402} {\bibfield  {journal} {\bibinfo  {journal} {Phys. Rev. Lett.}\ }\textbf {\bibinfo {volume} {130}},\ \bibinfo {pages} {081402} (\bibinfo {year} {2023})},\ \Eprint {http://arxiv.org/abs/2208.07380} {arXiv:2208.07380 [gr-qc]} \BibitemShut {NoStop}%
\bibitem [{\citenamefont {London}\ \emph {et~al.}(2014)\citenamefont {London}, \citenamefont {Shoemaker},\ and\ \citenamefont {Healy}}]{London:2014cma}%
  \BibitemOpen
  \bibfield  {author} {\bibinfo {author} {\bibfnamefont {L.}~\bibnamefont {London}}, \bibinfo {author} {\bibfnamefont {D.}~\bibnamefont {Shoemaker}}, \ and\ \bibinfo {author} {\bibfnamefont {J.}~\bibnamefont {Healy}},\ }\href {\doibase 10.1103/PhysRevD.90.124032} {\bibfield  {journal} {\bibinfo  {journal} {Phys. Rev. D}\ }\textbf {\bibinfo {volume} {90}},\ \bibinfo {pages} {124032} (\bibinfo {year} {2014})},\ \bibinfo {note} {[Erratum: Phys.Rev.D 94, 069902 (2016)]},\ \Eprint {http://arxiv.org/abs/1404.3197} {arXiv:1404.3197 [gr-qc]} \BibitemShut {NoStop}%
\bibitem [{\citenamefont {Mourier}\ \emph {et~al.}(2021)\citenamefont {Mourier}, \citenamefont {Jim\'enez~Forteza}, \citenamefont {Pook-Kolb}, \citenamefont {Krishnan},\ and\ \citenamefont {Schnetter}}]{Mourier:2020mwa}%
  \BibitemOpen
  \bibfield  {author} {\bibinfo {author} {\bibfnamefont {P.}~\bibnamefont {Mourier}}, \bibinfo {author} {\bibfnamefont {X.}~\bibnamefont {Jim\'enez~Forteza}}, \bibinfo {author} {\bibfnamefont {D.}~\bibnamefont {Pook-Kolb}}, \bibinfo {author} {\bibfnamefont {B.}~\bibnamefont {Krishnan}}, \ and\ \bibinfo {author} {\bibfnamefont {E.}~\bibnamefont {Schnetter}},\ }\href {\doibase 10.1103/PhysRevD.103.044054} {\bibfield  {journal} {\bibinfo  {journal} {Phys. Rev. D}\ }\textbf {\bibinfo {volume} {103}},\ \bibinfo {pages} {044054} (\bibinfo {year} {2021})},\ \Eprint {http://arxiv.org/abs/2010.15186} {arXiv:2010.15186 [gr-qc]} \BibitemShut {NoStop}%
\bibitem [{\citenamefont {Prasad}\ \emph {et~al.}(2020)\citenamefont {Prasad}, \citenamefont {Gupta}, \citenamefont {Bose}, \citenamefont {Krishnan},\ and\ \citenamefont {Schnetter}}]{Prasad_2020}%
  \BibitemOpen
  \bibfield  {author} {\bibinfo {author} {\bibfnamefont {V.}~\bibnamefont {Prasad}}, \bibinfo {author} {\bibfnamefont {A.}~\bibnamefont {Gupta}}, \bibinfo {author} {\bibfnamefont {S.}~\bibnamefont {Bose}}, \bibinfo {author} {\bibfnamefont {B.}~\bibnamefont {Krishnan}}, \ and\ \bibinfo {author} {\bibfnamefont {E.}~\bibnamefont {Schnetter}},\ }\href {\doibase 10.1103/physrevlett.125.121101} {\bibfield  {journal} {\bibinfo  {journal} {Physical Review Letters}\ }\textbf {\bibinfo {volume} {125}} (\bibinfo {year} {2020}),\ 10.1103/physrevlett.125.121101}\BibitemShut {NoStop}%
\bibitem [{\citenamefont {Okounkova}(2020)}]{okounkova2020revisiting}%
  \BibitemOpen
  \bibfield  {author} {\bibinfo {author} {\bibfnamefont {M.}~\bibnamefont {Okounkova}},\ }\href@noop {} {\enquote {\bibinfo {title} {Revisiting non-linearity in binary black hole mergers},}\ } (\bibinfo {year} {2020}),\ \Eprint {http://arxiv.org/abs/2004.00671} {arXiv:2004.00671 [gr-qc]} \BibitemShut {NoStop}%
\bibitem [{\citenamefont {Jaramillo}\ and\ \citenamefont {Krishnan}(2022)}]{jaramillo2022airyfunction}%
  \BibitemOpen
  \bibfield  {author} {\bibinfo {author} {\bibfnamefont {J.~L.}\ \bibnamefont {Jaramillo}}\ and\ \bibinfo {author} {\bibfnamefont {B.}~\bibnamefont {Krishnan}},\ }\href@noop {} {\enquote {\bibinfo {title} {Airy-function approach to binary black hole merger waveforms: The fold-caustic diffraction model},}\ } (\bibinfo {year} {2022}),\ \Eprint {http://arxiv.org/abs/2206.02117} {arXiv:2206.02117 [gr-qc]} \BibitemShut {NoStop}%
\bibitem [{\citenamefont {Chen}\ \emph {et~al.}(2022)\citenamefont {Chen}, \citenamefont {Kumar}, \citenamefont {Khera}, \citenamefont {Deppe}, \citenamefont {Dhani}, \citenamefont {Boyle}, \citenamefont {Giesler}, \citenamefont {Kidder}, \citenamefont {Pfeiffer}, \citenamefont {Scheel},\ and\ \citenamefont {Teukolsky}}]{Chen_2022}%
  \BibitemOpen
  \bibfield  {author} {\bibinfo {author} {\bibfnamefont {Y.}~\bibnamefont {Chen}}, \bibinfo {author} {\bibfnamefont {P.}~\bibnamefont {Kumar}}, \bibinfo {author} {\bibfnamefont {N.}~\bibnamefont {Khera}}, \bibinfo {author} {\bibfnamefont {N.}~\bibnamefont {Deppe}}, \bibinfo {author} {\bibfnamefont {A.}~\bibnamefont {Dhani}}, \bibinfo {author} {\bibfnamefont {M.}~\bibnamefont {Boyle}}, \bibinfo {author} {\bibfnamefont {M.}~\bibnamefont {Giesler}}, \bibinfo {author} {\bibfnamefont {L.~E.}\ \bibnamefont {Kidder}}, \bibinfo {author} {\bibfnamefont {H.~P.}\ \bibnamefont {Pfeiffer}}, \bibinfo {author} {\bibfnamefont {M.~A.}\ \bibnamefont {Scheel}}, \ and\ \bibinfo {author} {\bibfnamefont {S.~A.}\ \bibnamefont {Teukolsky}},\ }\href {\doibase 10.1103/physrevd.106.124045} {\bibfield  {journal} {\bibinfo  {journal} {Physical Review D}\ }\textbf {\bibinfo {volume} {106}} (\bibinfo {year} {2022}),\ 10.1103/physrevd.106.124045}\BibitemShut {NoStop}%
\bibitem [{\citenamefont {Brill}\ and\ \citenamefont {Lindquist}(1963)}]{BrillLindquist}%
  \BibitemOpen
  \bibfield  {author} {\bibinfo {author} {\bibfnamefont {D.~R.}\ \bibnamefont {Brill}}\ and\ \bibinfo {author} {\bibfnamefont {R.~W.}\ \bibnamefont {Lindquist}},\ }\href {\doibase 10.1103/PhysRev.131.471} {\bibfield  {journal} {\bibinfo  {journal} {Phys. Rev.}\ }\textbf {\bibinfo {volume} {131}},\ \bibinfo {pages} {471} (\bibinfo {year} {1963})}\BibitemShut {NoStop}%
\bibitem [{\citenamefont {Bowen}\ and\ \citenamefont {York}(1980)}]{BowenYork}%
  \BibitemOpen
  \bibfield  {author} {\bibinfo {author} {\bibfnamefont {J.~M.}\ \bibnamefont {Bowen}}\ and\ \bibinfo {author} {\bibfnamefont {J.~W.}\ \bibnamefont {York}},\ }\href {\doibase 10.1103/PhysRevD.21.2047} {\bibfield  {journal} {\bibinfo  {journal} {Phys. Rev. D}\ }\textbf {\bibinfo {volume} {21}},\ \bibinfo {pages} {2047} (\bibinfo {year} {1980})}\BibitemShut {NoStop}%
\bibitem [{Note1()}]{Note1}%
  \BibitemOpen
  \bibinfo {note} {For initial data corresponding to a large separation of the black holes, this parameter can be interpreted as the individual momenta of the black holes. However, these simulations have an initial separation of $0.4 M_\circ $, so this interpretation is not applicable. Nonetheless, increasing $P$ corresponds to increasing the coordinate velocities of the black holes.}\BibitemShut {Stop}%
\bibitem [{\citenamefont {Hawking}\ and\ \citenamefont {Hartle}(1972)}]{HawkingHartle72}%
  \BibitemOpen
  \bibfield  {author} {\bibinfo {author} {\bibfnamefont {S.~W.}\ \bibnamefont {Hawking}}\ and\ \bibinfo {author} {\bibfnamefont {J.~B.}\ \bibnamefont {Hartle}},\ }\href {\doibase 10.1007/BF01645515} {\bibfield  {journal} {\bibinfo  {journal} {Communications in Mathematical Physics}\ }\textbf {\bibinfo {volume} {27}},\ \bibinfo {pages} {283} (\bibinfo {year} {1972})}\BibitemShut {NoStop}%
\bibitem [{\citenamefont {Stein}(2019)}]{Stein:2019mop}%
  \BibitemOpen
  \bibfield  {author} {\bibinfo {author} {\bibfnamefont {L.~C.}\ \bibnamefont {Stein}},\ }\href {\doibase 10.21105/joss.01683} {\bibfield  {journal} {\bibinfo  {journal} {J. Open Source Softw.}\ }\textbf {\bibinfo {volume} {4}},\ \bibinfo {pages} {1683} (\bibinfo {year} {2019})},\ \Eprint {http://arxiv.org/abs/1908.10377} {arXiv:1908.10377 [gr-qc]} \BibitemShut {NoStop}%
\bibitem [{\citenamefont {Brizuela}\ \emph {et~al.}(2006)\citenamefont {Brizuela}, \citenamefont {Martin-Garcia},\ and\ \citenamefont {Mena~Marugan}}]{Brizuela:2006ne}%
  \BibitemOpen
  \bibfield  {author} {\bibinfo {author} {\bibfnamefont {D.}~\bibnamefont {Brizuela}}, \bibinfo {author} {\bibfnamefont {J.~M.}\ \bibnamefont {Martin-Garcia}}, \ and\ \bibinfo {author} {\bibfnamefont {G.~A.}\ \bibnamefont {Mena~Marugan}},\ }\href {\doibase 10.1103/PhysRevD.74.044039} {\bibfield  {journal} {\bibinfo  {journal} {Phys. Rev. D}\ }\textbf {\bibinfo {volume} {74}},\ \bibinfo {pages} {044039} (\bibinfo {year} {2006})},\ \Eprint {http://arxiv.org/abs/gr-qc/0607025} {arXiv:gr-qc/0607025} \BibitemShut {NoStop}%
\bibitem [{\citenamefont {Brizuela}\ \emph {et~al.}(2007)\citenamefont {Brizuela}, \citenamefont {Martin-Garcia},\ and\ \citenamefont {Marugan}}]{Brizuela:2007zza}%
  \BibitemOpen
  \bibfield  {author} {\bibinfo {author} {\bibfnamefont {D.}~\bibnamefont {Brizuela}}, \bibinfo {author} {\bibfnamefont {J.~M.}\ \bibnamefont {Martin-Garcia}}, \ and\ \bibinfo {author} {\bibfnamefont {G.~A.~M.}\ \bibnamefont {Marugan}},\ }\href {\doibase 10.1103/PhysRevD.76.024004} {\bibfield  {journal} {\bibinfo  {journal} {Phys. Rev. D}\ }\textbf {\bibinfo {volume} {76}},\ \bibinfo {pages} {024004} (\bibinfo {year} {2007})},\ \Eprint {http://arxiv.org/abs/gr-qc/0703069} {arXiv:gr-qc/0703069} \BibitemShut {NoStop}%
\bibitem [{\citenamefont {Brizuela}\ \emph {et~al.}(2009)\citenamefont {Brizuela}, \citenamefont {Martin-Garcia},\ and\ \citenamefont {Tiglio}}]{Brizuela:2009qd}%
  \BibitemOpen
  \bibfield  {author} {\bibinfo {author} {\bibfnamefont {D.}~\bibnamefont {Brizuela}}, \bibinfo {author} {\bibfnamefont {J.~M.}\ \bibnamefont {Martin-Garcia}}, \ and\ \bibinfo {author} {\bibfnamefont {M.}~\bibnamefont {Tiglio}},\ }\href {\doibase 10.1103/PhysRevD.80.024021} {\bibfield  {journal} {\bibinfo  {journal} {Phys. Rev. D}\ }\textbf {\bibinfo {volume} {80}},\ \bibinfo {pages} {024021} (\bibinfo {year} {2009})},\ \Eprint {http://arxiv.org/abs/0903.1134} {arXiv:0903.1134 [gr-qc]} \BibitemShut {NoStop}%
\bibitem [{\citenamefont {Spiers}\ \emph {et~al.}(2023)\citenamefont {Spiers}, \citenamefont {Pound},\ and\ \citenamefont {Wardell}}]{spiers2023secondorder}%
  \BibitemOpen
  \bibfield  {author} {\bibinfo {author} {\bibfnamefont {A.}~\bibnamefont {Spiers}}, \bibinfo {author} {\bibfnamefont {A.}~\bibnamefont {Pound}}, \ and\ \bibinfo {author} {\bibfnamefont {B.}~\bibnamefont {Wardell}},\ }\href@noop {} {\enquote {\bibinfo {title} {Second-order perturbations of the schwarzschild spacetime: practical, covariant and gauge-invariant formalisms},}\ } (\bibinfo {year} {2023}),\ \Eprint {http://arxiv.org/abs/2306.17847} {arXiv:2306.17847 [gr-qc]} \BibitemShut {NoStop}%
\bibitem [{\citenamefont {Ioka}\ and\ \citenamefont {Nakano}(2007)}]{Ioka:2007ak}%
  \BibitemOpen
  \bibfield  {author} {\bibinfo {author} {\bibfnamefont {K.}~\bibnamefont {Ioka}}\ and\ \bibinfo {author} {\bibfnamefont {H.}~\bibnamefont {Nakano}},\ }\href {\doibase 10.1103/PhysRevD.76.061503} {\bibfield  {journal} {\bibinfo  {journal} {Phys. Rev. D}\ }\textbf {\bibinfo {volume} {76}},\ \bibinfo {pages} {061503} (\bibinfo {year} {2007})},\ \Eprint {http://arxiv.org/abs/0704.3467} {arXiv:0704.3467 [astro-ph]} \BibitemShut {NoStop}%
\bibitem [{\citenamefont {Nakano}\ and\ \citenamefont {Ioka}(2007)}]{Nakano:2007cj}%
  \BibitemOpen
  \bibfield  {author} {\bibinfo {author} {\bibfnamefont {H.}~\bibnamefont {Nakano}}\ and\ \bibinfo {author} {\bibfnamefont {K.}~\bibnamefont {Ioka}},\ }\href {\doibase 10.1103/PhysRevD.76.084007} {\bibfield  {journal} {\bibinfo  {journal} {Phys. Rev. D}\ }\textbf {\bibinfo {volume} {76}},\ \bibinfo {pages} {084007} (\bibinfo {year} {2007})},\ \Eprint {http://arxiv.org/abs/0708.0450} {arXiv:0708.0450 [gr-qc]} \BibitemShut {NoStop}%
\bibitem [{\citenamefont {Forteza}\ and\ \citenamefont {Mourier}(2021)}]{Forteza:2021wfq}%
  \BibitemOpen
  \bibfield  {author} {\bibinfo {author} {\bibfnamefont {X.~J.}\ \bibnamefont {Forteza}}\ and\ \bibinfo {author} {\bibfnamefont {P.}~\bibnamefont {Mourier}},\ }\href {\doibase 10.1103/PhysRevD.104.124072} {\bibfield  {journal} {\bibinfo  {journal} {Phys. Rev. D}\ }\textbf {\bibinfo {volume} {104}},\ \bibinfo {pages} {124072} (\bibinfo {year} {2021})},\ \Eprint {http://arxiv.org/abs/2107.11829} {arXiv:2107.11829 [gr-qc]} \BibitemShut {NoStop}%
\bibitem [{Note2()}]{Note2}%
  \BibitemOpen
  \bibinfo {note} {We follow a similar 'bootstrap' strategy as in~\cite {Nee:2023osy}, where we first confirm the presence of the fundamental tone at late times, and the first fundamental tone, before searching for higher tones.}\BibitemShut {Stop}%
\bibitem [{Note3()}]{Note3}%
  \BibitemOpen
  \bibinfo {note} {We evaluate the amplitudes on an interval of $4M$ centered around the time of the fit $t_0$. The range of amplitudes in the interval is the uncertainty bars in Fig.~\ref {fig:Ampl2}.}\BibitemShut {Stop}%
\bibitem [{\citenamefont {Kehagias}\ \emph {et~al.}(2023)\citenamefont {Kehagias}, \citenamefont {Perrone}, \citenamefont {Riotto},\ and\ \citenamefont {Riva}}]{Kehagias:2023ctr}%
  \BibitemOpen
  \bibfield  {author} {\bibinfo {author} {\bibfnamefont {A.}~\bibnamefont {Kehagias}}, \bibinfo {author} {\bibfnamefont {D.}~\bibnamefont {Perrone}}, \bibinfo {author} {\bibfnamefont {A.}~\bibnamefont {Riotto}}, \ and\ \bibinfo {author} {\bibfnamefont {F.}~\bibnamefont {Riva}},\ }\href@noop {} {\  (\bibinfo {year} {2023})},\ \Eprint {http://arxiv.org/abs/2301.09345} {arXiv:2301.09345 [gr-qc]} \BibitemShut {NoStop}%
\bibitem [{\citenamefont {L{\"{o}}ffler}\ \emph {et~al.}(2012)\citenamefont {L{\"{o}}ffler}, \citenamefont {Faber}, \citenamefont {Bentivegna}, \citenamefont {Bode}, \citenamefont {Diener}, \citenamefont {Haas}, \citenamefont {Hinder}, \citenamefont {Mundim}, \citenamefont {Ott}, \citenamefont {Schnetter}, \citenamefont {Allen}, \citenamefont {Campanelli},\ and\ \citenamefont {Laguna}}]{Loffler:2011ay}%
  \BibitemOpen
  \bibfield  {author} {\bibinfo {author} {\bibfnamefont {F.}~\bibnamefont {L{\"{o}}ffler}}, \bibinfo {author} {\bibfnamefont {J.}~\bibnamefont {Faber}}, \bibinfo {author} {\bibfnamefont {E.}~\bibnamefont {Bentivegna}}, \bibinfo {author} {\bibfnamefont {T.}~\bibnamefont {Bode}}, \bibinfo {author} {\bibfnamefont {P.}~\bibnamefont {Diener}}, \bibinfo {author} {\bibfnamefont {R.}~\bibnamefont {Haas}}, \bibinfo {author} {\bibfnamefont {I.}~\bibnamefont {Hinder}}, \bibinfo {author} {\bibfnamefont {B.~C.}\ \bibnamefont {Mundim}}, \bibinfo {author} {\bibfnamefont {C.~D.}\ \bibnamefont {Ott}}, \bibinfo {author} {\bibfnamefont {E.}~\bibnamefont {Schnetter}}, \bibinfo {author} {\bibfnamefont {G.}~\bibnamefont {Allen}}, \bibinfo {author} {\bibfnamefont {M.}~\bibnamefont {Campanelli}}, \ and\ \bibinfo {author} {\bibfnamefont {P.}~\bibnamefont {Laguna}},\ }\href {\doibase doi:10.1088/0264-9381/29/11/115001} {\bibfield  {journal} {\bibinfo  {journal} {Class. Quantum Grav.}\ }\textbf {\bibinfo {volume} {29}},\ \bibinfo {pages} {115001} (\bibinfo {year} {2012})},\ \Eprint {http://arxiv.org/abs/arXiv:1111.3344 [gr-qc]} {arXiv:1111.3344 [gr-qc]} \BibitemShut {NoStop}%
\bibitem [{\citenamefont {Zlochower}\ \emph {et~al.}(2022)\citenamefont {Zlochower}, \citenamefont {Brandt}, \citenamefont {Diener}, \citenamefont {Gabella}, \citenamefont {Gracia-Linares}, \citenamefont {Haas}, \citenamefont {Kedia}, \citenamefont {Alcubierre}, \citenamefont {Alic}, \citenamefont {Allen}, \citenamefont {Ansorg}, \citenamefont {Babiuc-Hamilton}, \citenamefont {Baiotti}, \citenamefont {Benger}, \citenamefont {Bentivegna}, \citenamefont {Bernuzzi}, \citenamefont {Bode}, \citenamefont {Bozzola}, \citenamefont {Brendal}, \citenamefont {Bruegmann}, \citenamefont {Campanelli}, \citenamefont {Cipolletta}, \citenamefont {Corvino}, \citenamefont {Cupp}, \citenamefont {Pietri}, \citenamefont {Dimmelmeier}, \citenamefont {Dooley}, \citenamefont {Dorband}, \citenamefont {Elley}, \citenamefont {Khamra}, \citenamefont {Etienne}, \citenamefont {Faber}, \citenamefont {Font}, \citenamefont {Frieben}, \citenamefont {Giacomazzo}, \citenamefont {Goodale}, \citenamefont {Gundlach}, \citenamefont {Hawke}, \citenamefont {Hawley}, \citenamefont {Hinder}, \citenamefont {Huerta}, \citenamefont {Husa}, \citenamefont {Iyer}, \citenamefont {Johnson}, \citenamefont {Joshi}, \citenamefont {Kastaun}, \citenamefont {Kellermann}, \citenamefont {Knapp}, \citenamefont {Koppitz}, \citenamefont {Laguna}, \citenamefont {Lanferman}, \citenamefont {L{\"o}ffler}, \citenamefont {Masso}, \citenamefont {Menger}, \citenamefont {Merzky}, \citenamefont {Miller}, \citenamefont {Miller}, \citenamefont {Moesta}, \citenamefont {Montero}, \citenamefont {Mundim}, \citenamefont {Nelson}, \citenamefont {Nerozzi}, \citenamefont {Noble}, \citenamefont {Ott}, \citenamefont {Paruchuri}, \citenamefont {Pollney}, \citenamefont {Radice}, \citenamefont {Radke}, \citenamefont {Reisswig}, \citenamefont {Rezzolla}, \citenamefont {Rideout}, \citenamefont {Ripeanu}, \citenamefont {Sala}, \citenamefont {Schewtschenko}, \citenamefont {Schnetter}, \citenamefont {Schutz}, \citenamefont {Seidel}, \citenamefont {Seidel}, \citenamefont {Shalf}, \citenamefont {Sible}, \citenamefont {Sperhake}, \citenamefont {Stergioulas}, \citenamefont {Suen}, \citenamefont {Szilagyi}, \citenamefont {Takahashi}, \citenamefont {Thomas}, \citenamefont {Thornburg}, \citenamefont {Tobias}, \citenamefont {Tonita}, \citenamefont {Walker}, \citenamefont {Wan}, \citenamefont {Wardell}, \citenamefont {Werneck}, \citenamefont {Witek}, \citenamefont {Zilh{\~a}o},\ and\ \citenamefont {Zink}}]{EinsteinToolkit:2022_05}%
  \BibitemOpen
  \bibfield  {author} {\bibinfo {author} {\bibfnamefont {Y.}~\bibnamefont {Zlochower}}, \bibinfo {author} {\bibfnamefont {S.~R.}\ \bibnamefont {Brandt}}, \bibinfo {author} {\bibfnamefont {P.}~\bibnamefont {Diener}}, \bibinfo {author} {\bibfnamefont {W.~E.}\ \bibnamefont {Gabella}}, \bibinfo {author} {\bibfnamefont {M.}~\bibnamefont {Gracia-Linares}}, \bibinfo {author} {\bibfnamefont {R.}~\bibnamefont {Haas}}, \bibinfo {author} {\bibfnamefont {A.}~\bibnamefont {Kedia}}, \bibinfo {author} {\bibfnamefont {M.}~\bibnamefont {Alcubierre}}, \bibinfo {author} {\bibfnamefont {D.}~\bibnamefont {Alic}}, \bibinfo {author} {\bibfnamefont {G.}~\bibnamefont {Allen}}, \bibinfo {author} {\bibfnamefont {M.}~\bibnamefont {Ansorg}}, \bibinfo {author} {\bibfnamefont {M.}~\bibnamefont {Babiuc-Hamilton}}, \bibinfo {author} {\bibfnamefont {L.}~\bibnamefont {Baiotti}}, \bibinfo {author} {\bibfnamefont {W.}~\bibnamefont {Benger}}, \bibinfo {author} {\bibfnamefont {E.}~\bibnamefont {Bentivegna}}, \bibinfo {author} {\bibfnamefont {S.}~\bibnamefont {Bernuzzi}}, \bibinfo {author} {\bibfnamefont {T.}~\bibnamefont {Bode}}, \bibinfo {author} {\bibfnamefont {G.}~\bibnamefont {Bozzola}}, \bibinfo {author} {\bibfnamefont {B.}~\bibnamefont {Brendal}}, \bibinfo {author} {\bibfnamefont {B.}~\bibnamefont {Bruegmann}}, \bibinfo {author} {\bibfnamefont {M.}~\bibnamefont {Campanelli}}, \bibinfo {author} {\bibfnamefont {F.}~\bibnamefont {Cipolletta}}, \bibinfo {author} {\bibfnamefont {G.}~\bibnamefont {Corvino}}, \bibinfo {author} {\bibfnamefont {S.}~\bibnamefont {Cupp}}, \bibinfo {author} {\bibfnamefont {R.~D.}\ \bibnamefont {Pietri}}, \bibinfo {author} {\bibfnamefont {H.}~\bibnamefont {Dimmelmeier}}, \bibinfo {author} {\bibfnamefont {R.}~\bibnamefont {Dooley}}, \bibinfo {author} {\bibfnamefont {N.}~\bibnamefont {Dorband}}, \bibinfo {author} {\bibfnamefont {M.}~\bibnamefont {Elley}}, \bibinfo {author} {\bibfnamefont {Y.~E.}\ \bibnamefont {Khamra}}, \bibinfo {author} {\bibfnamefont {Z.}~\bibnamefont {Etienne}}, \bibinfo {author} {\bibfnamefont {J.}~\bibnamefont {Faber}}, \bibinfo {author} {\bibfnamefont {T.}~\bibnamefont {Font}}, \bibinfo {author} {\bibfnamefont {J.}~\bibnamefont {Frieben}}, \bibinfo {author} {\bibfnamefont {B.}~\bibnamefont {Giacomazzo}}, \bibinfo {author} {\bibfnamefont {T.}~\bibnamefont {Goodale}}, \bibinfo {author} {\bibfnamefont {C.}~\bibnamefont {Gundlach}}, \bibinfo {author} {\bibfnamefont {I.}~\bibnamefont {Hawke}}, \bibinfo {author} {\bibfnamefont {S.}~\bibnamefont {Hawley}}, \bibinfo {author} {\bibfnamefont {I.}~\bibnamefont {Hinder}}, \bibinfo {author} {\bibfnamefont {E.~A.}\ \bibnamefont {Huerta}}, \bibinfo {author} {\bibfnamefont {S.}~\bibnamefont {Husa}}, \bibinfo {author} {\bibfnamefont {S.}~\bibnamefont {Iyer}}, \bibinfo {author} {\bibfnamefont {D.}~\bibnamefont {Johnson}}, \bibinfo {author} {\bibfnamefont {A.~V.}\ \bibnamefont {Joshi}}, \bibinfo {author} {\bibfnamefont {W.}~\bibnamefont {Kastaun}}, \bibinfo {author} {\bibfnamefont {T.}~\bibnamefont {Kellermann}}, \bibinfo {author} {\bibfnamefont {A.}~\bibnamefont {Knapp}}, \bibinfo {author} {\bibfnamefont {M.}~\bibnamefont {Koppitz}}, \bibinfo {author} {\bibfnamefont {P.}~\bibnamefont {Laguna}}, \bibinfo {author} {\bibfnamefont {G.}~\bibnamefont {Lanferman}}, \bibinfo {author} {\bibfnamefont {F.}~\bibnamefont {L{\"o}ffler}}, \bibinfo {author} {\bibfnamefont {J.}~\bibnamefont {Masso}}, \bibinfo {author} {\bibfnamefont {L.}~\bibnamefont {Menger}}, \bibinfo {author} {\bibfnamefont {A.}~\bibnamefont {Merzky}}, \bibinfo {author} {\bibfnamefont {J.~M.}\ \bibnamefont {Miller}}, \bibinfo {author} {\bibfnamefont {M.}~\bibnamefont {Miller}}, \bibinfo {author} {\bibfnamefont {P.}~\bibnamefont {Moesta}}, \bibinfo {author} {\bibfnamefont {P.}~\bibnamefont {Montero}}, \bibinfo {author} {\bibfnamefont {B.}~\bibnamefont {Mundim}}, \bibinfo {author} {\bibfnamefont {P.}~\bibnamefont {Nelson}}, \bibinfo {author} {\bibfnamefont {A.}~\bibnamefont {Nerozzi}}, \bibinfo {author} {\bibfnamefont {S.~C.}\ \bibnamefont {Noble}}, \bibinfo {author} {\bibfnamefont {C.}~\bibnamefont {Ott}}, \bibinfo {author} {\bibfnamefont {R.}~\bibnamefont {Paruchuri}}, \bibinfo {author} {\bibfnamefont {D.}~\bibnamefont {Pollney}}, \bibinfo {author} {\bibfnamefont {D.}~\bibnamefont {Radice}}, \bibinfo {author} {\bibfnamefont {T.}~\bibnamefont {Radke}}, \bibinfo {author} {\bibfnamefont {C.}~\bibnamefont {Reisswig}}, \bibinfo {author} {\bibfnamefont {L.}~\bibnamefont {Rezzolla}}, \bibinfo {author} {\bibfnamefont {D.}~\bibnamefont {Rideout}}, \bibinfo {author} {\bibfnamefont {M.}~\bibnamefont {Ripeanu}}, \bibinfo {author} {\bibfnamefont {L.}~\bibnamefont {Sala}}, \bibinfo {author} {\bibfnamefont {J.~A.}\ \bibnamefont {Schewtschenko}}, \bibinfo {author} {\bibfnamefont {E.}~\bibnamefont {Schnetter}}, \bibinfo {author} {\bibfnamefont {B.}~\bibnamefont {Schutz}}, \bibinfo {author} {\bibfnamefont {E.}~\bibnamefont {Seidel}}, \bibinfo {author} {\bibfnamefont {E.}~\bibnamefont {Seidel}}, \bibinfo {author} {\bibfnamefont {J.}~\bibnamefont {Shalf}}, \bibinfo {author} {\bibfnamefont {K.}~\bibnamefont {Sible}}, \bibinfo {author} {\bibfnamefont {U.}~\bibnamefont {Sperhake}}, \bibinfo {author} {\bibfnamefont {N.}~\bibnamefont {Stergioulas}}, \bibinfo {author} {\bibfnamefont {W.-M.}\ \bibnamefont {Suen}}, \bibinfo {author} {\bibfnamefont {B.}~\bibnamefont {Szilagyi}}, \bibinfo {author} {\bibfnamefont {R.}~\bibnamefont {Takahashi}}, \bibinfo {author} {\bibfnamefont {M.}~\bibnamefont {Thomas}}, \bibinfo {author} {\bibfnamefont {J.}~\bibnamefont {Thornburg}}, \bibinfo {author} {\bibfnamefont {M.}~\bibnamefont {Tobias}}, \bibinfo {author} {\bibfnamefont {A.}~\bibnamefont {Tonita}}, \bibinfo {author} {\bibfnamefont {P.}~\bibnamefont {Walker}}, \bibinfo {author} {\bibfnamefont {M.-B.}\ \bibnamefont {Wan}}, \bibinfo {author} {\bibfnamefont {B.}~\bibnamefont {Wardell}}, \bibinfo {author} {\bibfnamefont {L.}~\bibnamefont {Werneck}}, \bibinfo {author} {\bibfnamefont {H.}~\bibnamefont {Witek}}, \bibinfo {author} {\bibfnamefont {M.}~\bibnamefont {Zilh{\~a}o}}, \ and\ \bibinfo {author} {\bibfnamefont {B.}~\bibnamefont {Zink}},\ }\href {\doibase 10.5281/zenodo.6588641} {\enquote {\bibinfo {title} {The {E}instein {T}oolkit},}\ } (\bibinfo {year} {2022}),\ \bibinfo {note} {to find out more, visit http://einsteintoolkit.org}\BibitemShut {NoStop}%
\bibitem [{EinsteinToolkit()}]{EinsteinToolkit:web}%
  \BibitemOpen
  EinsteinToolkit,\ \href {http://einsteintoolkit.org/} {\enquote {\bibinfo {title} {{Einstein Toolkit}: Open software for relativistic astrophysics},}\ }\BibitemShut {NoStop}%
\bibitem [{\citenamefont {Ansorg}\ \emph {et~al.}(2004)\citenamefont {Ansorg}, \citenamefont {Br{\"u}gmann},\ and\ \citenamefont {Tichy}}]{Ansorg:2004ds}%
  \BibitemOpen
  \bibfield  {author} {\bibinfo {author} {\bibfnamefont {M.}~\bibnamefont {Ansorg}}, \bibinfo {author} {\bibfnamefont {B.}~\bibnamefont {Br{\"u}gmann}}, \ and\ \bibinfo {author} {\bibfnamefont {W.}~\bibnamefont {Tichy}},\ }\href {\doibase 10.1103/PhysRevD.70.064011} {\bibfield  {journal} {\bibinfo  {journal} {Phys. Rev. D}\ }\textbf {\bibinfo {volume} {70}},\ \bibinfo {pages} {064011} (\bibinfo {year} {2004})},\ \Eprint {http://arxiv.org/abs/arXiv:gr-qc/0404056} {arXiv:gr-qc/0404056} \BibitemShut {NoStop}%
%%CITATION = GR-QC/0404056;%%
\bibitem [{\citenamefont {Brown}\ \emph {et~al.}(2009)\citenamefont {Brown}, \citenamefont {Diener}, \citenamefont {Sarbach}, \citenamefont {Schnetter},\ and\ \citenamefont {Tiglio}}]{Brown:2008sb}%
  \BibitemOpen
  \bibfield  {author} {\bibinfo {author} {\bibfnamefont {J.~D.}\ \bibnamefont {Brown}}, \bibinfo {author} {\bibfnamefont {P.}~\bibnamefont {Diener}}, \bibinfo {author} {\bibfnamefont {O.}~\bibnamefont {Sarbach}}, \bibinfo {author} {\bibfnamefont {E.}~\bibnamefont {Schnetter}}, \ and\ \bibinfo {author} {\bibfnamefont {M.}~\bibnamefont {Tiglio}},\ }\href {\doibase 10.1103/PhysRevD.79.044023} {\bibfield  {journal} {\bibinfo  {journal} {Phys. Rev. D}\ }\textbf {\bibinfo {volume} {79}},\ \bibinfo {pages} {044023} (\bibinfo {year} {2009})},\ \Eprint {http://arxiv.org/abs/arXiv:0809.3533 [gr-qc]} {arXiv:0809.3533 [gr-qc]} \BibitemShut {NoStop}%
%%CITATION = 0809.3533;%%
\bibitem [{\citenamefont {Pook-Kolb}\ \emph {et~al.}(2019)\citenamefont {Pook-Kolb}, \citenamefont {Birnholtz}, \citenamefont {Krishnan},\ and\ \citenamefont {Schnetter}}]{Pook-Kolb:2018igu}%
  \BibitemOpen
  \bibfield  {author} {\bibinfo {author} {\bibfnamefont {D.}~\bibnamefont {Pook-Kolb}}, \bibinfo {author} {\bibfnamefont {O.}~\bibnamefont {Birnholtz}}, \bibinfo {author} {\bibfnamefont {B.}~\bibnamefont {Krishnan}}, \ and\ \bibinfo {author} {\bibfnamefont {E.}~\bibnamefont {Schnetter}},\ }\href {\doibase 10.1103/PhysRevD.99.064005} {\bibfield  {journal} {\bibinfo  {journal} {Phys. Rev. D}\ }\textbf {\bibinfo {volume} {99}},\ \bibinfo {pages} {064005} (\bibinfo {year} {2019})},\ \Eprint {http://arxiv.org/abs/1811.10405} {arXiv:1811.10405 [gr-qc]} \BibitemShut {NoStop}%
\bibitem [{\citenamefont {Gupta}\ \emph {et~al.}(2018)\citenamefont {Gupta}, \citenamefont {Krishnan}, \citenamefont {Nielsen},\ and\ \citenamefont {Schnetter}}]{Gupta:2018znn}%
  \BibitemOpen
  \bibfield  {author} {\bibinfo {author} {\bibfnamefont {A.}~\bibnamefont {Gupta}}, \bibinfo {author} {\bibfnamefont {B.}~\bibnamefont {Krishnan}}, \bibinfo {author} {\bibfnamefont {A.}~\bibnamefont {Nielsen}}, \ and\ \bibinfo {author} {\bibfnamefont {E.}~\bibnamefont {Schnetter}},\ }\href {\doibase 10.1103/PhysRevD.97.084028} {\bibfield  {journal} {\bibinfo  {journal} {Phys. Rev. D}\ }\textbf {\bibinfo {volume} {97}},\ \bibinfo {pages} {084028} (\bibinfo {year} {2018})},\ \Eprint {http://arxiv.org/abs/1801.07048} {arXiv:1801.07048 [gr-qc]} \BibitemShut {NoStop}%
\bibitem [{Note4()}]{Note4}%
  \BibitemOpen
  \bibinfo {note} {If $t_{\protect \rm simulation} = t_{\protect \rm Sch} + f(r,\theta , \varphi )$, then the wave equation remains separable---we have in mind a more complicated relation such as $t_{\protect \rm simulation} = f(t_{\protect \rm Sch},r,\theta , \varphi )$.}\BibitemShut {Stop}%
\end{thebibliography}%
